\documentclass[prd,aps,twocolumn,a4paper,showkeys,nofootinbib]{revtex4-1}

\usepackage[colorlinks=true,backref=true,linkcolor=blue,linktocpage=true,anchorcolor=black,citecolor=blue,filecolor=black,menucolor=black,pagecolor=black,urlcolor=blue]{hyperref}
\usepackage{graphicx,psfrag}
\usepackage{mathrsfs}
\usepackage{amsmath,amsfonts,amssymb}
\usepackage{multirow}
\usepackage{comment}
\usepackage{bm}
\usepackage{ulem}
\usepackage{hyperref}
\usepackage{enumitem}

\usepackage{pifont} 
\newcommand{\cmark}{\ding{51}}%
\newcommand{\xmark}{\ding{55}}%

\newcommand{\m}{-}
\newcommand{\be}{\begin{equation}}
\newcommand{\ee}{\end{equation}}
\newcommand{\bea}{\begin{eqnarray}}
\newcommand{\eea}{\end{eqnarray}}
\newcommand{\bel}{\begin{align}}
\newcommand{\eel}{\end{align}}

\newcommand{\ord}{\mathcal{O}}
\newcommand{\f}{\frac}
\newcommand{\gsf}{\text{1GSF}}
\newcommand{\bk}{\text{0GSF}}
\newcommand{\el}{\ell}
\newcommand{\nn}{\nonumber}

\def\GMc2{G M_{\odot} c^{-2}}

\newcommand\B{\rule[-1.2ex]{0pt}{0pt}} 
\newcommand\T{\rule{0pt}{2.6ex}}       
\newcommand\BAM{{\tt BAM}}
\newcommand\BAMc{{\tt BAM:}}

\def\TEOBResum{\texttt{TEOBResum}}
\def\TEOBResumS{\texttt{TEOBResumS}}

\def\BAM{{\texttt{BAM}}}

\newcommand\PN[1]{PN${}^{\rm (#1)}$}
\newcommand\GSF[2]{GSF{#1}${}^{\rm (#2)}$}
\newcommand\GSFp[2]{GSF{#1}${}^{\rm (#2)}_{4.5}$}

\DeclareSymbolFontAlphabet{\mathrsfs}{rsfs}
\DeclareMathAlphabet{\mathcal}{OMS}{cmsy}{m}{n}
\DeclareSymbolFontAlphabet{\mathrsfs}{rsfs}
\DeclareMathAlphabet\mathbfcal{OMS}{cmsy}{b}{n}

\usepackage{color}
\definecolor{cyan}{rgb}{0,0.9,0.9}
\definecolor{orange}{rgb}{0.9,0.5,0}
\definecolor{magenta}{rgb}{1,0,1}
\definecolor{purple}{rgb}{0.8,0.4,0.8}
\definecolor{gray}{rgb}{0.8242,0.8242,0.8242}
\definecolor{dodgerblue}{rgb}{0.12, 0.56, 1.0}

\begin{document}

\title{Effective-one-body multipolar waveform for \\ tidally interacting binary neutron stars up to merger}

\author{Sarp \surname{Akcay}$^{1}$}
\author{Sebastiano \surname{Bernuzzi}${}^{1}$}
\author{Francesco \surname{Messina}$^{2,3}$}
\author{Alessandro \surname{Nagar}$^{4,5,6}$}
\author{N\'estor \surname{Ortiz}$^{1}$}
\author{Piero \surname{Rettegno}$^{5,7}$}
 
\affiliation{${}^1$Theoretisch-Physikalisches Institut, Friedrich-Schiller-Universit{\"a}t Jena, 07743, Jena, Germany}
\affiliation{${}^{2}$Dipartimento di Fisica, Universit\`a degli studi di Milano Bicocca, Piazza della Scienza 3, 20126 Milano, Italy}
\affiliation{${}^{3}$INFN, Sezione di Milano Bicocca, Piazza della Scienza 3, 20126 Milano, Italy}
\affiliation{${}^4$Centro Fermi - Museo Storico della Fisica e Centro Studi e Ricerche Enrico Fermi, Rome, Italy}
\affiliation{${}^5$INFN Sezione di Torino, Via P. Giuria 1, 10125 Torino, Italy}
\affiliation{${}^6$Institut des Hautes Etudes Scientifiques, 91440 Bures-sur-Yvette, France}
\affiliation{${}^{7}$ Dipartimento di Fisica, Universit\`a di Torino, via P. Giuria 1, I-10125 Torino, Italy}


\begin{abstract}
  Gravitational-wave astronomy with coalescing binary neutron star sources
  requires the availability of gravitational waveforms with tidal effects accurate
  up to merger. 
  This article
  presents an improved version of \TEOBResum, a nonspinning effective-one-body (EOB)
  waveform model with enhanced analytical information in its tidal sector.
  The tidal potential governing the conservative dynamics employs 
  resummed expressions based on post-Newtonian (PN) and
  gravitational self-force (GSF) information.
  In particular, we compute a GSF-resummed expression for the
  leading-order octupolar gravitoelectric term and
  incorporate the leading-order gravitomagnetic term
  (either in PN-expanded or GSF-resummed form).
  The multipolar waveform and fluxes are augmented with
  gravitoelectric and magnetic terms recently obtained in PN. 
  %
  The new analytical information enhances tidal effects toward 
  merger accelerating the coalescence. We quantify the impact
  on the gravitational-wave phasing of each physical effect.
  The most important contribution  is given  by the resummed
  gravitoelectric octupolar  term entering the  EOB interaction
  potential, that can yield up to 1~rad of dephasing
  (depending on the NS model) with respect to its nonresummed
  version.
  %
  The model's energetics and the gravitational wave
  phasing are validated with eccentricity-reduced and multi-resolution
  numerical relativity simulations with different equations of
  state and mass ratios.
  We also present EOB-NR waveform comparisons
  for higher multipolar modes beyond the dominant quadrupole one.
\end{abstract}


\maketitle


\section{Introduction}

The analysis of gravitational waves (GW) from binary neutron star 
events requires detailed waveform models that include
tidal effects \cite{TheLIGOScientific:2017qsa,Abbott:2018exr,Abbott:2018wiz}.
Semi-analytical inspiral waveforms with tidal effects valid up to merger have been
constructed to date only in a few works \cite{Baiotti:2010xh,Bernuzzi:2012ci,Bernuzzi:2014owa,Hinderer:2016eia}.
These models build on the effective-one-body (EOB) formalism for the
general-relativistic two-body problem~\cite{Buonanno:1998gg,Buonanno:2000ef} and its
extension to include tidal interactions~\cite{Damour:2009wj}.
Their common starting point is the general-relativistic theory of tidal
properties of neutron stars (NSs)~\cite{Damour:1983a,Hinderer:2007mb,Flanagan:2007ix,Damour:2009vw,Binnington:2009bb} and 
a post-Newtonian (PN) expression for
the EOB potential based on the calculations of Refs.~\cite{Hinderer:2009ca,Damour:2009wj,Vines:2010ca,Vines:2011ud,Damour:2012yf,Bini:2012gu,Bini:2014zxa}. 
The conservative part of the dynamics of circularized binaries is currently known
at next-to-next-to-leading order (NNLO), i.e., formal 7PN level~\cite{Bini:2012gu}
(or 2PN, since the Newtonian contribution starts in fact at 5PN~\cite{Damour:1983a}). 
On the other hand, for generic, noncircular motion, the conservative dynamics
is fully known only at 6PN~\cite{Vines:2010ca}, since Ref.~\cite{Bini:2012gu} only
focused on circular motion.
In addition, the tidal correction to the waveform amplitude is analytically known at 
6PN~\cite{Vines:2010ca}, including gravitomagnetic and subdominant gravitoelectric
multipolar contributions~\cite{Banihashemi:2018xfb}.
%
%
Note that  waveform amplitude 
corrections due to tidal-tail terms 
are also exactly known analytically up to relative 
2.5PN (i.e., global 7.5PN order\,\footnote{We recall that the tidal waveform information is 
only lacking the knowledge of the 2PN (7PN) quadrupolar term, though, as argued in Ref.~\cite{Damour:2012yf},
its effect is expected to be small. Once this term becomes available, one will automatically have access 
to 3.5PN tail terms in the the tidal waveform amplitude.}) thanks to the analytical knowledge of 
the resummed tail factor that enters  the factorized EOB waveform~\cite{Damour:2008gu,Faye:2014fra}.  

Such a large amount of analytical information has been compared over time 
with numerical relativity (NR) simulations of inspiralling and coalescing 
neutron stars of increased accuracy~\cite{Damour:2009wj,Baiotti:2010xh,Bernuzzi:2012ci,Hotokezaka:2013mm}.
It was pointed out as early as in Ref.~\cite{Damour:2009wj} that the EOB treatment
of tidal effects (at the time just at 1PN level) seemed prone to underestimating their
actual magnitude in the last few inspiral orbits up to merger. 
This fact became progressively apparent as the reliability of
NR simulations increased, with improved handling of the error 
budget~\cite{Bernuzzi:2011aq,Bernuzzi:2012ci,Hotokezaka:2013mm,Bernuzzi:2014owa},
clearly pointing out that the gravitational attraction yielded by the EOB interaction 
potential based on PN-expanded NNLO tidal information was not sufficiently 
strong so as to match the NR predictions within their error bars. 
Bini and Damour~\cite{Bini:2014zxa} proposed to blend together the 
aforementioned NNLO tidal information with gravitational-self-force (GSF)~\cite{Dolan:2014pja} 
information in a special resummed expression for the (gravitoelectric) 
potential which enhanced the tidal attraction due to the presence of a pole at 
the Schwarzschild light-ring.
Such a potential was incorporated (with a modification concerning the light-ring
location, see below) in the (nonspinning) \TEOBResum{} model~\cite{Bernuzzi:2014owa},
that is built upon of the point-mass, nonspinning, EOB dynamics 
of  Refs.~\cite{Damour:2014sva,Nagar:2017jdw,Nagar:2018zoe}.
The key prescription suggested in Ref.~\cite{Bini:2014zxa} and implemented
Ref.~\cite{Bernuzzi:2014owa} is to substitute the test-mass light-ring pole 
$r=3$ (in dimensionless units) with the the light-ring of the NNLO EOB model. 
The pole effectively amplifies tides in a regime in which the two NS cannot be
described as isolated objects. Note that the pole singularity is never reached 
since the EOB dynamics terminates at a larger radius. 
\TEOBResum{} reproduces
NR waveforms within their errors up to merger for a large sample of
binaries, including binaries with nonprecessing spins
\cite{Bernuzzi:2014owa,Nagar:2018zoe,Dietrich:2017feu}. 
To date, \TEOBResum{} has been tested against the 
largest sample of NR data available~\cite{Dietrich:2018phi}.
Some phase differences with respect to the NR data are however present
for binaries with large mass ratio and/or for NS with large tidal
polarizability parameters, thus indicating that reproducing the GW 
from last few orbits using EOB requires even stronger tides
\cite{Bernuzzi:2014owa,Dietrich:2017feu,Nagar:2018zoe, Hotokezaka:2015xka, Hotokezaka:2016bzh}. 

A possible mechanism leading to an effective amplification of 
tidal effects close to merger is the resonance between the NS $f$-mode
and the orbital frequency, cf., e.g., Refs.~\cite{Kokkotas:1995xe,Ho:1998hq}. 
This idea has been implemented in the
EOB formalism in Refs.~\cite{Hinderer:2016eia,Steinhoff:2016rfi}, and there it is referred to
as ``dynamical tides''. The point-mass EOB baseline used in those works is the
one developed in
Refs.~\cite{Pan:2011gk,Pan:2013tva,Pan:2013rra,Taracchini:2013rva} in
combination with the PN tidal NNLO EOB potential. When compared to NR data, the model has performances
very similar to the GSF resummation approach. Notably, both methods
either reproduce the data within their errors or slightly underestimate 
the GW phase near merger \cite{Dietrich:2017feu}.

In this work we incorporate in \TEOBResum{} all the analytical tidal information
that is currently available: (i) the $\ell=3$ GSF-resummed contribution to the EOB $A$ 
potential, that is computed in this paper for the first time; (ii) the gravitomagnetic tidal potential;
(iii) the tidal contributions to the EOB $B$ potential of Ref.~\cite{Vines:2010ca},
and (iv) the full 1PN tidal corrections to the multipolar waveform~\cite{Banihashemi:2018xfb}.
We then compare the performance of the model against long-end, error-controlled, NR data computed by the 
computational relativity ({\tt CoRe}) collaboration.

The paper is organized as follows. In Sec.~\ref{sbsec:tidal_potentials}, we compute a 
GSF-resummed expression for the electric $\ell=3$ term of the tidal EOB $A_\text{T}$ 
potential [cf. Eq.~(\ref{eq:A_tidal})\,]. We also include the LO gravitomagnetic, $(2-)$, 
term either in PN series or in GSF-resummed form.  We additionally incorporate the leading 
order tidal correction to the $B$ potential [cf. Eq.~(\ref{eq:Vines_Flan_BT_term})\,], 
as computed in Ref.~\cite{Vines:2010ca}. The gravitoelectric and gravitomagnetic 
corrections to the tidal multipolar waveform computed in Ref.~\cite{Banihashemi:2018xfb}
are also incorporated into the factorized and resummed EOB waveform.
In Sec.~\ref{sec:phasing}, we evaluate the effect of each new term on
the GW phasing for a set of sample binaries.
We find that the largest effect on the tidal phase is generated by the
new GSF-resummed $\ell=3$ electric term, with significantly smaller
contributions from the gravitomagnetic term, the tidal correction to
the $B$ potential and the sub-dominant multipoles.
We also consider the gravitomagnetic contribution parameterized by
static Love numbers \cite{Landry:2015cva} (as opposed to irrotational) 
and find that this gravitomagnetic effect is also very small.
The \TEOBResum{}/NR comparison is driven in Sec.~\ref{sec:nrar} and 
concerns both the energetics (through the gauge-invariant relation between
binding energy and orbital angular momentum) and the phasing, notably 
considering also higher multipolar modes.
In particular, we consider twelve best eccentricity-reduced and multiple-resolution
simulations of irrotational and quasi-circular binary neutron star
mergers computed by the {\tt CoRe} collaboration \cite{Dietrich:2018phi} and
previously presented in Ref.~\cite{Dietrich:2017aum}. 
The high accuracy of these data currently
provides us with the most stringent strong-field constraints available from NR, 
as shown in Fig.~\ref{fig:Phasing_fig1}.
Within this data set, we also consider simulation data with mass ratios other than unity such as  $q\approx(1.5,1.75,2)$ 
computed in Refs.~\cite{Dietrich:2015pxa,Dietrich:2016hky,Dietrich:2017feu}. 
While these data are less accurate, they give some insights on the model
performances in an ``extreme'' region of the parameter space.
We additionally present comparisons of NR and EOB waveforms for modes beyond 
the leading-order quadrupole in Fig.~\ref{fig:EOBNR_phase_higher_modes}.
Conclusions are collected in Sec.~\ref{sec:end}. The paper is then
completed by two technical Appendixes. Appendix~\ref{sec:A3plus} reports
the explicit derivation of the GSF-resummed $\ell=3$ tidal potential.
Appendix~\ref{sec:PA} briefly discusses the numerical implementation of
the model, focusing in particular on the performances yielded by the
use of the post-adiabatic approximation of Ref.~\cite{Nagar:2018gnk}.

We use geometric units $c=G=1$. To convert from geometric to physical units
we recall that $G M_\odot/c^3=4.925490947\times 10^{-6}$~sec.
The $(2,2)$-mode GW frequency, $f$, relates to the dimensionless
$(2,2)$-mode angular frequency $\hat{\omega}$ via 
$ f\approx 32.3125 \,\hat{\omega} \,(M_\odot/M)\,$kHz \cite{Damour:2009wj}.
For example, $\hat{\omega}\approx 8.356\times 10^{-4}$ at $f=10\,$Hz for a typical NS binary with $M =2.7 M_\odot$. For the remainder of this article, we employ dimensionless units rescaled with respect to $M$.

\section{Tidal effects in \TEOBResum{}}
\label{sec:new_tides}
This section summarizes the main analytical results. We use the
following definitions:
\be
q\equiv \f{m_A}{m_B}\geq 1 \ , \quad X_{A}\equiv \f{m_{A}}{M} \label{eq:defs}
\ee
with $A,B$ labelling the stars and $M=m_A+m_B$. Let us also introduce the symmetric mass ratio $\nu \equiv X_A X_B$.

\subsection{Tidal potential: Gravitoelectric and magnetic terms}
\label{sbsec:tidal_potentials}
The key idea of EOB is to map the binary motion to geodesic motion in an effective Schwarzschild spacetime
(or Kerr for binaries with spin). The dynamics are described by the following EOB Hamiltonian 
\be
H_\text{EOB} = M \sqrt{1+2\nu (\hat{H}_\text{eff}-1)} \label{eq:H_EOB},
\ee
which is given by
\be
\hat{H}_\text{eff} = \sqrt{p_{r_\ast}^2+ A(r)\left(1 + \f{p_\varphi^2}{r^2}+ 2\nu (4-3\nu)\f{p_{r_\ast}^4}{r^2}\right)} \label{eq:Heff}
\ee
in polar coordinates $(r,\varphi)$ and per unit mass conjugate momenta $(p_{r_\ast},p_\varphi)$ for planar 
motion~\cite{Buonanno:1998gg,Buonanno:2000ef, Damour:2001tu}.
It has been shown that the point-mass dynamics is well described by a Pad\'{e} resummation
of the 5PN expression for the radial potential $A(r)$ \cite{Damour:2009kr} 
(henceforth the point-mass potential $A_0$).

In EOB, the tidal interaction for quasicircular inspiral dynamics is incorporated
by augmenting the point-mass potential as follows~\cite{Damour:2009wj}
\be
A = A_0 + A_\text{T} \label{eq:A0_plus_AT},
\ee
where
\begin{align}
A_\text{T}(u) =  \sum_{l\ge 2}& A^{(\ell +)\text{LO}}_A(u)\hat{A}^{(\ell+)}_A(u)\nn \\
+& A^{(\el -)\text{LO}}_A(u)\hat{A}^{(\ell-)}_A(u) + \left(A\leftrightarrow B\right)\label{eq:A_tidal}
\ ,
\end{align}
where the signs $\pm$ correspond to gravitoelectric and gravitomagnetic terms, respectively,
and $u=M/r$ is the inverse of the dimensionless EOB radial coordinate. 
The leading-order (LO) terms are given by
\begin{subequations}
\begin{align}
  A^{(\el +)\text{LO}}_A(u) &=- \kappa^{(\el +)}_A u^{2\el+2}, \label{eq:AplusLO}\\
  A^{(\el -)\text{LO}}_A(u) &=- \kappa^{(\el -)}_A u^{2\el+3}, \label{eq:AminusLO}
\end{align}
\end{subequations}
where
\begin{subequations}
\begin{align}
  \kappa^{(\el+)}_A &= 2 {k_A^{(\el)}}\f{X_B}{X_A}\f{X_A^{2\el+1} }{\mathcal{C}_A^{2\el+1}}\label{eq:kappaplus_general}
\end{align}
%
For the $(\el-)$ sector, we currently have
\be
\kappa^{(2-)}_A = \f{1}{2} {j_A^{(2)}}\f{X_B}{X_A}\f{X_A^5}{\mathcal{C}_A^{5}}\label{eq:kappaminus_general}.
\ee
\end{subequations}
$k_A^{(\el)}$ and $j_A^{(\el)}$ are the dimensionless gravitoelectric
and gravitomagnetic Love numbers \cite{Damour:2009vw}, and $\mathcal{C}_A \equiv m_A/R_A$ is the compactness parameter.
$k^{(\el)}_A$ is often denoted as $k_\el$ in the literature and relates to the other 
commonly used Love number (polarizability) $\bar\lambda_\el$ via $ k_\el = (2\el-1)!!\,
\mathcal{C}^{2\el+1} \bar\lambda_\el/2$ \cite{Yagi:2013sva}
which, in our notation, translates to
\begin{subequations}
\be
\Lambda^{(\el)}_A \equiv \frac{2}{(2\el-1)!!}\, \mathcal{C}^{-(2\el+1)}_A k^{(\el)}_A 
\ .
\ee
Similarly, for the gravitomagnetic sector, we have
\be
\Sigma^{(\el)}_A\equiv\f{\el-1}{4(\el+2)}\f{1}{(2\el-1)!!}\,
\mathcal{C}_A^{-(2\el+1)} j^{(\el)}_A 
\label{eq:sigma_bar}
\ee
\end{subequations}
which is denoted by $\bar\sigma^{(\el)}$, e.g., in Ref.~\cite{Yagi:2013sva}.
For $\el=2$, our gravitomagnetic Love number $j^{(2)}$ relates to the $k_2^\text{mag}$ of
Ref.~\cite{Landry:2015cva} via $k_2^\text{mag}=j^{(2)}/(24\mathcal{C})$ \cite{Banihashemi:2018xfb,Pani:2018inf}.
We use quasi-universal fitting relations to obtain $\Sigma$ from $\Lambda$
\cite{Jimenez-Forteza:2018buh, Yagi:2013sva, Yagi:2013bca}, specifically the fits of Ref.~\cite{Jimenez-Forteza:2018buh}.

Following Ref.~\cite{Damour:2009vw}, we introduce their $\kappa^T_{2}\equiv \kappa^{(2+)}_A + \kappa^{(2+)}_B$
and, similarly, $\kappa^T_{2-}\equiv \kappa^{(2-)}_A + \kappa^{(2-)}_B$.
For $q=1$, we use $\Lambda\equiv\Lambda^{(2)}_A=\Lambda^{(2)}_B$
and $\Sigma\equiv\Sigma^{(2)}_A=\Sigma^{(2)}_B$.
These relations yield $\kappa^T_{2}= 3\Lambda/16$ and $\kappa^T_{2-}= 3\Sigma/2$. 
We will employ $\kappa^T_2$ and $\Lambda$ interchangeably to quantify the strength of the  tidal interactions
including gravitomagnetic cases as $|\Sigma|$ grows monotonically with $\Lambda$.

The potentials $\hat{A}_A^{(\el\pm)}(u)$ contain the terms beyond LO. 
In particular, the $(\el+)$ contributions are known up
to $\el=3$ as a series in $u$
%
\be
\hat{A}^{(\el +)}_{A}(u) = 1 + \alpha^{(\ell +)}_{1 A} u + \alpha^{(\ell +)}_{2A} u^2 \ .\label{eq:A_PN}
\ee
with
\begin{align}
  \alpha^{(2 +)}_{1 A} & =\f{5}{2} X_A,\\
  \alpha^{(2 +)}_{2 A} &= 3+\f{1}{8}X_A+\f{337}{28}X_A^2,\label{eq:alpha_2s}\\
  \alpha^{(3 +)}_{1 A} & =-2+\f{15}{2} X_A,\\
  \alpha^{(3 +)}_{2 A} &= \f{8}{3}-\f{311}{24}X_A+\f{110}{3}X_A^2 \label{eq:alpha_3s}.
\end{align}
For $\ell=4$, we are currently limited to the LO term, thus $\hat{A}_A^{(4+)}(u)=1$.

In the $(\el-)$ sector, only the gravitomagnetic NLO term is known:
\begin{align}
 \alpha^{(2 -)}_{1 A} &= 1+\f{11}{6}X_A +X_A^2 \label{eq:alpha_2minus}.
\end{align}

Ref.~\cite{Bini:2014zxa} offered an alternative series representation for the tidal potentials
$\hat{A}^{(\el\pm)}_A(u)$ in terms of the mass ratio $X_A$ as a consequence of a resummation
procedure done using results from first-order GSF approach.
Using $X_A=m_A/M\ll 1$ as an expansion parameter, they wrote
\be
	\hat{A}^{(\el\pm)}_A(u) = \hat{A}^{(\el\pm)\text{0GSF}}+X_A\hat{A}^{(\el\pm)\text{1GSF}}+X_A^2\hat{A}^{(\el\pm)\text{2GSF}} + \ldots \label{eq:Ahat_SF_expansion} 
\ee
%
For the 1GSF terms, Ref.~\cite{Bini:2014zxa} introduced light-ring
(LR) singularity factorized potentials
$\tilde{A}^{(2\pm)}(u)\equiv (1-3u)^{7/2} \hat{A}^{(2\pm)1\text{GSF}} $.
Using Ref.~\cite{Dolan:2014pja}'s numerical GSF data, they
constructed a global four-parameter fit to
$\tilde{A}^{(2\pm)}(u)$ and explicitly displayed
the fit parameters for the $(2+)$ potential.
As Ref.~\cite{Dolan:2014pja}'s numerical data received
a minor, $\sim \mathcal{O}(10^{-5})$,
correction after the publication of Ref.~\cite{Bini:2014zxa},
we repeated their fit to 
\be
	\tilde{A}^{(2+)}(u)
	\approx
	\f{5}{2} u\, (1-a_1 u)(1-a_2 u)\, \f{1+n_1 u}{1+d_2 u^2} \label{eq:Ahat2_plus_fit}
\ee
and obtained the following minor changes to their fit parameters 
\begin{align}
a_1 &=8.53352,\quad a_2  = 3.04309, \nn\\ 
n_1 & =  0.840064,\quad d_2  =  17.7324. \label{eq:BD_old_params}
\end{align}
These should be compared with Eq.~(7.27) of Ref.~\cite{Bini:2014zxa}.

For the $(2-)$ potential, we employ a similar fit using Ref.~\cite{Dolan:2014pja}'s updated data:
\be
	\tilde{A}^{(2-)}(u) 
	\approx 
	\f{11}{6} u\, (1-a^\m_1 u)(1-a^\m_2 u)\, \f{1+n_1^- u}{1+d_2^- u^2} \label{eq:Ahat2_minus_fit}
\ee
with 
\begin{align}
  \label{eq:Ahat2_minus_fit_coefs}
  a_1^- &= \ 0.728591,\\
  a_2^- &= 3.10037,\\ 
  n_1^- &= -15.0442,\\
  d_2^- &= 12.5523. 
\end{align}
For the 0GSF, 2GSF potentials,
from Ref.~\cite{Bini:2014zxa} we have
\begin{align}
 \hat{A}^{(2+)0\text{GSF}}&= 1+\f{3u^2}{1-3u}\\
 \hat{A}^{(2+)2\text{GSF}}&=\f{337}{28}\f{u^2}{(1-3u)^p}\\
\hat{A}^{(2-)0\text{GSF}}&=\f{1-2u}{1-3u}\label{eq:Ahat_zeros},\\
\hat{A}^{(2-)2\text{GSF}}&= \f{u}{(1-3u)^{p_{2-}}}\label{eq:Ahat_two},
\end{align}
where the values of $p, p_{2-}$ are currently unknown due to
lack of second-order GSF results. However, Sec.~VIID
of Ref.~\cite{Bini:2014zxa} provided a proof
that $p,p_{2-}\ge 4$ and a further argument that $p \le 6$.

%
We now wish to resum the $(3+)$ tidal potential in the
same fashion as was done for the $(2\pm)$ tidal
potentials. 
To this end, we introduce the following GSF series for $\hat{A}^{(3+)}_A(u)$ 
\begin{align}
        \label{eq:A3plus_GSF_all}
	\hat{A}_A^{(3+)}(u)=& (1-2u) \left( 1 + \f{8}{3} \f{u^2}{(1-3u)}\right)+ X_A\,\f{\tilde{A}^{(3+)}}{(1-3u)^{7/2}} \nn\\
	 &+ X_A^2\, \f{110}{3}\f{u^2}{(1-3u)^{p_{3+}}},
\end{align}
where $p_{3+} \ge 4$ \cite{Bini:2014zxa}.
Next, using Refs.~\cite{Dolan:2014pja,Nolan:2015vpa}'s numerical data, 
we construct a global fit for the LR factorized 1GSF potential:
\begin{align}
\tilde{A}^{(3+)}(u) &\equiv (1-3u)^{7/2} \hat{A}^{(3+)\gsf}  \label{eq:A3tilde_fit}\\
\approx& \frac{15}{2}u(1+ C_1 u + C_2 u^2+C_3 u^3)\, \f{1+C_4 u+C_5 u^2}{1+C_6 u^2} ,\nn
\end{align}
with
\begin{align}
	C_1 &= -3.68210, \	C_2 = 5.17100,\	C_3 = -7.63916,\nn\\
	C_4 &= -8.63278, \	C_5 = 16.3601, \	C_6 = \ \ 12.3197 \, . \label{eq:A3tilde_fit_coefs}
\end{align}
The details of this derivation are collected in Appendix~\ref{sec:A3plus}.

To pragmatically reduce the number of unknowns here we set
$p_{2-}=p_{3+}=p$ and we mostly stick to the (conservative)
value $p=4$, as in Ref.~\cite{Bernuzzi:2014owa}. However,
to get an idea of the sensitivity of our results to the
changes in $p$, we shall also show some results obtained
using $p=9/2$. In principle, since the complete tidal potential
is analytically known only at 2PN relative order, one may think
to transform the parameters $\{p, p_{2-},p_{3+}\}$ into effective
functions (that may depend on EOS and mass ratio) to be determined
by comparisons with highly accurate NR simulations. Consistently
with Ref.~\cite{Nagar:2018zoe} (see Sec.~IIIC and notably Fig.~12),
the NR phasing error of (some) NR simulations of the {\tt CoRe} catalog,
that we shall also use here, is {\it smaller} than the EOB/NR
phase difference towards merger. This thus suggests that state-of-the-art
NR simulations might be used to meaningfully inform the tidal
sector of the EOB model towards merger. However, to do so consistently
all over the BNS parameter space we would need a few dozen of high-quality
numerical BNS simulations with error budget of the order of (at least)
0.2~rad up to merger. This is currently not the case when $\kappa_2^{\rm T}$
is of the order of (or larger than) 150, so that this kind of tuning
is postponed to future work. In any case, at least for $\kappa_2^T\simeq 100$,
we shall confirm that the simplifying choice $p=4$ yields a good representation
of the tidal interaction; similarly, the value  $p\gtrsim 5$ seems to
universally overestimate the strength of the tidal forces in the last
few orbits up to merger. 
 \begin{figure}[t]
  \centering
  \includegraphics[width=0.99\linewidth]{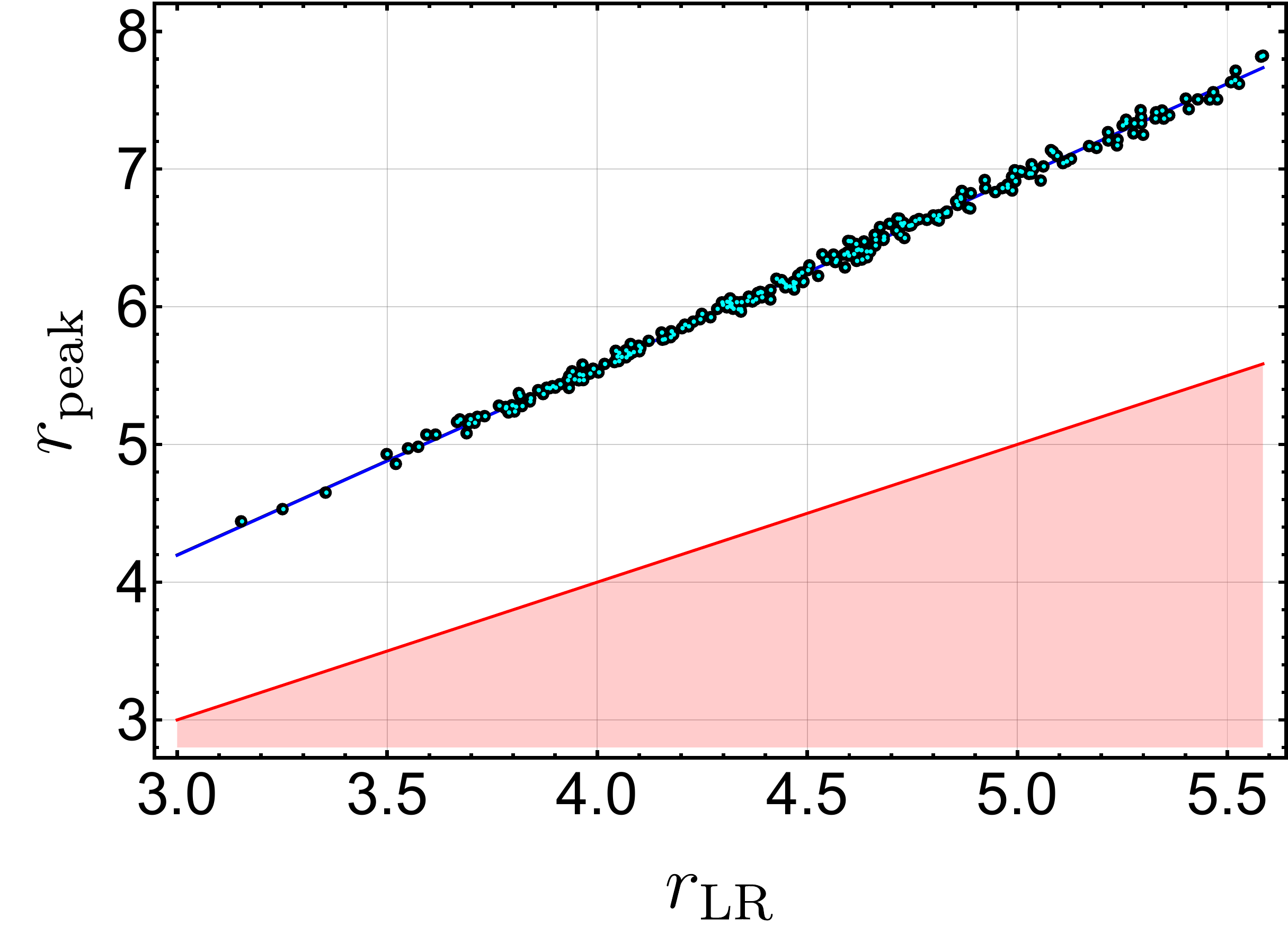}
   \caption{
   Distribution of values for $r_\text{peak}$ vs. $r_\text{LR}$ for 250 points chosen
   from the $\{q,\Lambda_A,\Lambda_B\}$ parameter space. 
   The black dots represent data obtained using the \TEOBResum{} model of Ref.~\cite{Nagar:2018zoe} dubbed \GSF{2}{+}nm in Table~\ref{table:opt}.
   The data for the cyan (light colored) points are obtained by augmenting this \TEOBResum{} with
   $\hat{A}^{(3+)}+\hat{A}^{(2-)}$ (\GSF{23}{+}\PN{-}).
      As $r_\text{LR}$ is determined by finding the maximum of $r^{-2}A^\text{NNLO}(r)$ (see text)
   the resulting values for $r_\text{LR}$ are the same regardless of how we augment \TEOBResum.
   However, the values of $r_\text{peak}$ do differ slightly, but this is not easily discernible in this
   plot, which is why the 250 cyan points appear to lie exactly on top of the 250 black points.
   The linear fit is given by $r_\text{peak}\approx 1.37 r_\text{LR} +0.09$.
   The red region is the forbidden zone corresponding to $r_\text{peak} \le r_\text{LR}$.}
   \label{fig:rPeak_rLR}
  \end{figure}
The \TEOBResum{} model of Ref.~\cite{Bernuzzi:2014owa} employs PN series
for all the tidal potentials with the exception of $(2+)$
for which the GSF series of Ref.~\cite{Bini:2014zxa} is adopted with $p=4$. 
Additionally, as explained in Ref.~\cite{Bernuzzi:2014owa},
\TEOBResum{} replaces the Schwarzschild LR, $u=1/3$, 
with the maximum of $u^2 A^\text{NNLO}$,
i.e., the EOB effective photon potential.
$A^\text{NNLO}(u)$ is the EOB potential in which the
point-mass $A_0$ potential is added to
the tidal $A_\text{T}$ potential containing only
the PN series for the $(2+),(3+),(4+)$ tidal terms
(see Ref.~\cite{Bernuzzi:2014owa} and Sec.~IIIA
of Ref.~\cite{Nagar:2018zoe}).

\begin{figure}[t!]
  \centering
  \includegraphics[width=0.45\textwidth]{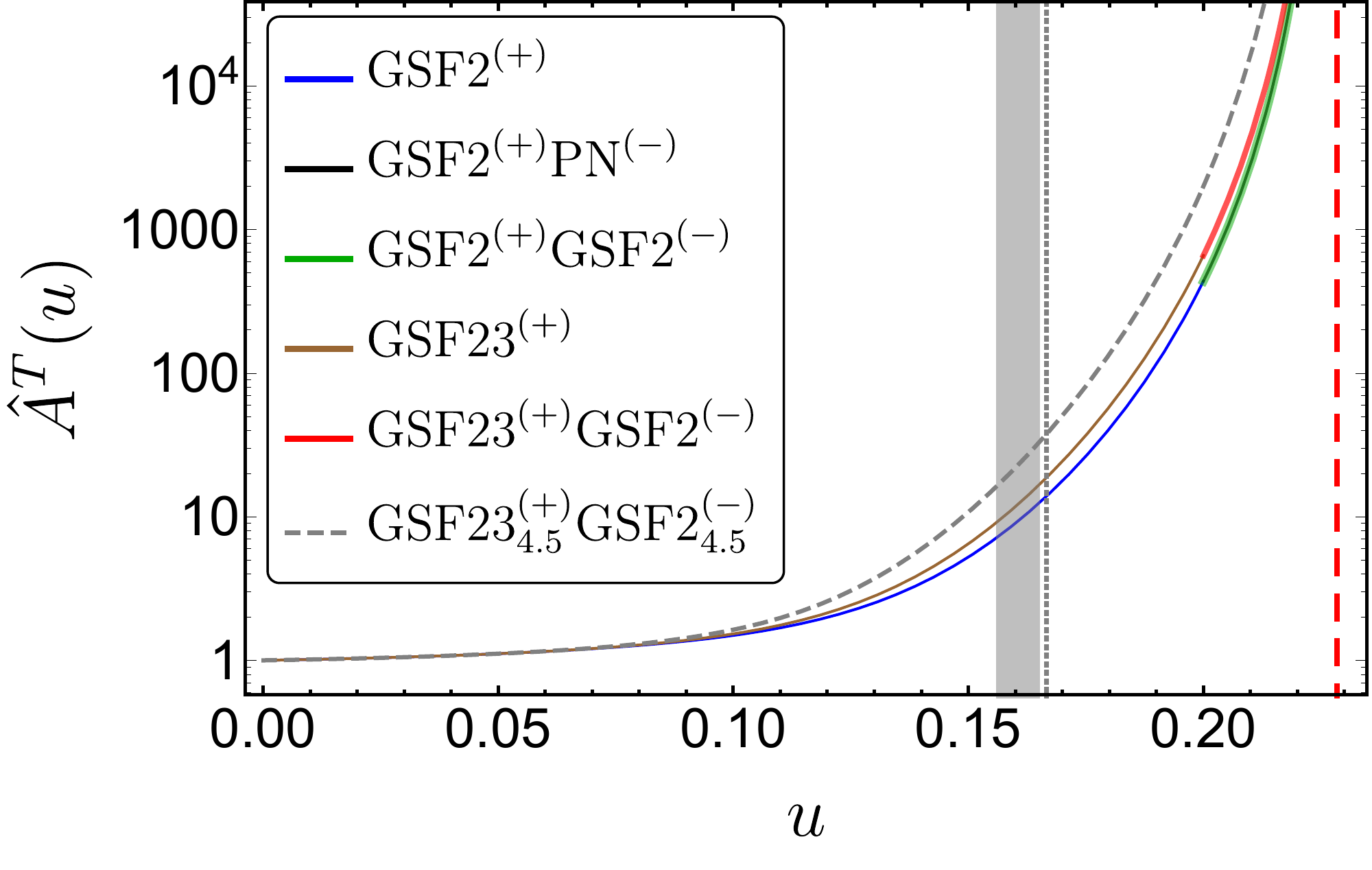} 
  \caption{A sample of tidal EOB potentials at our disposal shown against each
    other for $q=1$, $\Lambda=1531.34$ corresponding to $\kappa_2^T=287.126$. 
     The vertical gray region marks the various values for $r_\text{peak}$ at which the orbital
     frequency peaks for each EOB variant plotted here. See Table~\ref{table:opt} for
     explanation of the legend labels. The vertical red dashed line marks the
     location of the NNLO effective light ring, $u_\text{LR}^{\rm NNLO}\approx 0.228$,
     for this set of parameters. The vertical gray dotted line marks the Schwarzschild
     LSO at $u=1/6$. As the relative contribution of the $(2-)$ tides
     is $\lesssim \ord(10^{-2})$, the potentials with $(2-)$ tides overlap rather well
     with their no-$(2-)$ counterparts. Therefore, to distinguish these, we opted to plot
     them over limited domains as shown by the solid black, green,
     and red curves with the black curve under the green one. Note that the dashed,
     gray, line corresponds to the potential with $p=4$ replaced by $p=4.5$,
     in order to appreciate the sensitivity to this parameter.}
   \label{fig:A_of_u}
  \end{figure}

Following then Ref.~\cite{Bernuzzi:2014owa} to obtain the
complete tidal potential we have to finally replace the
denominators $(1-3u)$ in Eq.~\eqref{eq:A3plus_GSF_all}
with $(1-r_\text{LR}^{\rm NNLO} u)$, where  $r^{\rm NNLO}_\text{LR}$
corresponds to the peak of  $u^2 (A_0+A_T^\text{NNLO}(u))$.
Such a new GSF-resummed potential will then yield a different
effective light-ring, defined this time as the peak
of $u^2 (A_0+A_T^{\text{GSF*}}(u))$, where $A_T^{\text{GSF}*}$
indicates any tidal potential with GSF-resummed information.
Clearly, one has to
a posteriori check that the so constructed dynamics
never passes through $r_\text{LR}^{\rm NNLO}$ in the
physically meaningful region. To do so easily, we can
monitor the behavior of the orbital frequency and
identify the radius where it peaks. This point  $r_\text{peak}$, is
close to the peak of the $\ell=m=2$ waveform amplitude that we
conventionally identify as the merger point.
In Fig.~\ref{fig:rPeak_rLR} we plot $r_\text{peak}$ vs. $r_\text{LR}$
for 250 points in the $\{q,\Lambda_A,\Lambda_B\}$ parameter space for
\TEOBResum{} of Ref.~\cite{Bernuzzi:2014owa} and \TEOBResum{}
supplied with (3+) and (2-) tides as GSF series.  The figure
illustrates that the EOB radial separation never hits
the $r_{\rm LR}^{\rm NNLO}$ effective light ring location.
With our new GSF series for the $(2-),(3+)$ potentials
we now have several different options to flex the original \TEOBResumS{}
model. We show some of our main choices in Fig.~\ref{fig:A_of_u},
where the legend is explained in Table~\ref{table:opt}.

Our final addition to \TEOBResum{} regards the tidal contribution $B_\text{T}(u)$
to the EOB $B$ potential in the EOB Hamiltonian. From Ref.~\cite{Vines:2010ca},
one has that the contribution that is added to the PN-expanded point-mass
part of the potential $B_0$ is 
\be
B_\text{T}(u) = 3 \kappa_2^T (3-5\nu) u^6 \label{eq:Vines_Flan_BT_term}\ .
\ee
To incorporate this information within \TEOBResum{} we need first to review
the choices previously made. In particular, let us remember that the current
$B$ function is defined as $B\equiv D/A$, where the $D$ function is the 3PN-accurate
one that is resummed as a Pad\'e (0,3) approximant as $D\equiv (1+6\nu u^2-2(3\nu-26)\nu u^3)^{-1}$,
and $A\equiv A_0+A_T$, i.e., the total potential as a sum of the point-mass with the tidal part.
As a consequence, the $B$ function obtained in this way already incorporates
the tidal contribution, that is, however, inconsistent, once PN-expanded, with
Eq.~\eqref{eq:Vines_Flan_BT_term}. There are several ways to overcome this difficulty and
have the correct PN-expansion of the tidal $B$ potential. The simplest is just to
add to the current $B$ potential a term $B'_{\rm T}(u)$ such that the term proportional
to $\kappa_2^T$ of the PN-expanded $B+B'_{\rm T}$ coincides with
Eq.~\eqref{eq:Vines_Flan_BT_term}. This condition yields
\be
\label{eq:BTprime}
B_{\rm T}'(u) = \kappa_2^T(8-15\nu)u^6  \, .
\ee
We shall investigate the effect of this additional term on
phasing in Sec.~\ref{sec:phasing} below.
\vspace{1mm}
%

\subsection{Tidal Waveform}
\label{sbsec:new_multipoles}
When including the effects of the tides on the waveform, the point-mass waveform $h^0_{\el m}$ is augmented via \cite{Damour:2012yf}
\be
h_{\el m}= h^0_{\el m}+h^T_{\el m}= h^{\text{Newt}}_{\el m}\,(\hat{h}^0_{\el m} + \hat{h}_{\el m}^T), \label{eq:h_lm}
\ee
%
%
with the general expression for $h^{\text{Newt}}_{\el m}$ given, e.g., by Eq.~(18) of Ref.~\cite{Damour:2012ky} modulo normalization and sign conventions.
Until recently, only the $(2+)$ NLO contribution to $\hat{h}^T_{22}$ was known \cite{Vines:2011ud}, but thanks to Ref.~\cite{Banihashemi:2018xfb}, we now have access to all the NLO information for the $(2+)$ contributions to $\hat{h}^T_{21},\hat{h}^T_{31},\hat{h}^T_{33}$ as well as the $(2+)$ LO contribution to $\hat{h}^T_{32}$, and the LO $(2-)$ contributions for $\el \le 3, m\le \el$.
We adopt all of this new information to all our tidal choices for \TEOBResum{} with one exception which we label by ``nm'' (no multipoles) in Table~\ref{table:opt} and Fig.~\ref{fig:DeltaPhi_EOB}.

Rewriting the results for $\hat{h}^T_{\el m}$ from Appendix~A of Ref.~\cite{Banihashemi:2018xfb} in our own notation, and using $X_B=1-X_A$, we obtain
\begin{widetext}
\begin{align}
\hat{h}_{22}^{\rm T}&=\kappa_A^{(2+)}\left(\f{3-2X_A}{1-X_A} \right)x^5+\biggl[\f{14}{9}\kappa_A^{(2-)}-\kappa_A^{(2+)}\frac{\left(202-560 X_A+340 X_A^2-45 X_A^3\right)}{42 (1-X_A)}\biggr]x^6+(A\leftrightarrow B)\label{eq:h22_tidal},\\
\hat{h}_{21}^{\rm T}&=\biggl[\kappa_A^{(2+)}\left(-\frac{9}{2}+6X_A\right)-\kappa_A^{(2-)}\frac{1}{2(1-X_A)}\biggr]x^5-(A\leftrightarrow B)\label{eq:h21_tidal},\\
\hat{h}_{33}^{\rm T}&=-6\,\kappa_A^{(2+)}(1-X_A)x^5+\biggl[\,\kappa_A^{(2+)}\left(21-\frac{89}{2}X_A+\frac{55}{2}X_A^2-5X_A^3\right)+ \f{1}{2}\kappa_A^{(2-)}(9X_A-5)\biggr]x^6-(A\leftrightarrow B)\label{eq:h33_tidal},\\
\hat{h}_{32}^{\rm T}&=\biggl[4\,\kappa_A^{(2+)}(2-4X_A+3X_A^2)+\f{4}{3}\kappa_A^{(2-)}\biggr]x^5+(A\leftrightarrow B)\label{eq:h32_tidal},\\
\hat{h}_{31}^{\rm T}&=-6\,\kappa_A^{(2+)}(1-X_A)x^5+\biggl[\,\kappa_A^{(2+)}\left(1+\frac{5}{6} X_A-\frac{131}{6} X_A^2+\frac{65}{3} X_A^3\right)+\f{1}{2}\kappa_A^{(2-)}(17X_A-13)\biggr]x^6-(A\leftrightarrow B)\label{eq:h31_tidal}.
\end{align}
\end{widetext}
Note that some of the $(A\leftrightarrow B)$ terms are preceded by a minus sign. 
This PN-expanded tidal part is then incorporated in the \TEOBResum{} following
Appendix A of Ref.~\cite{Damour:2012yf}, in particular with the tail factor factorized
in front of the tidal waveform contribution as above. As usual in EOB models,
the PN variable $x$ is replaced by the EOB velocity variable $v_\Omega=r_\Omega\Omega$,
where $r_\Omega=r\psi^{1/3}$ and $\psi$ is computed using the EOB Hamiltonian~\cite{Damour:2006tr,Damour:2009wj}.

\section{Effect of enhanced analytical information on GW phasing}
\label{sec:phasing}

In this section we evaluate the impact, in terms of accumulated GW phase, 
of the new analytical information discussed above. In particular we separately
focus on the effect of the GSF-resummed $\ell=3$ potential and on all other
contributions (gravitomagnetic effects and additional tidal corrections to waveform
amplitude etc.) that turn out to be largely subdominant.
The key options for the models investigated here are
summarized in Table~\ref{table:opt}. For example, \GSF{23}{+}\GSF{2}{-} 
represents \TEOBResum{} employing $(2\pm),(3+)$ GSF-resummed
tides, with our standard choice $p=4$.
Finally, we also mention the possibility of flexing $p$,
with the subscript $4.5$ representing the choice $p=9/2$.
The default, or baseline, \TEOBResum{} model that is used
as benchmark for our comparisons is \GSF{2}{+}.

\begin{table}[t]
\centering
\begin{tabular}{lcccccccc}
  \hline\hline
  Shortname  & $\hat{A}^{(2+)}$ & $\hat{A}^{(3+)}$  & $\hat{A}^{(2-)}$& $p$ & $\hat{h}^T_{\ell m}$ \\
  \hline
  \PN{+} & PN        & PN & PN  & -   & \cmark \\
  \GSF{2}{+}nm & GSF-R & PN & \xmark & 4 & \xmark \\
  \GSF{2}{+} & GSF-R & PN & \xmark & 4 & \cmark \\
  \GSF{2}{+}\PN{-} & GSF-R & PN & PN     & 4 & \cmark\\
  \GSF{2}{+}\GSF{2}{-} & GSF-R & PN & GSF-R  & 4 & \cmark\\
  \GSF{23}{+} & GSF-R & GSF-R & \xmark & 4 & \cmark \\
  \GSF{23}{+}\PN{-} & GSF-R & GSF-R & PN     & 4 & \cmark\\
  \GSF{23}{+}\GSF{2}{-} & GSF-R & GSF-R & GSF-R  & 4 & \cmark\\
  \GSFp{23}{+}\GSFp{2}{-} & GSF-R & GSF-R & GSF-R  & 4.5 &\cmark \B \\ 
 \hline\hline
\end{tabular}
\caption{Summary of the key analytical terms and components of
  \TEOBResum{} tested in this work. GSF-R and PN stand for ``GSF
  resummed'' and post-Newtonian expressions described in Sec.~\ref{sbsec:tidal_potentials}.
  All models include the $B_\text{T}$ term of Eq.~(\ref{eq:Vines_Flan_BT_term}), which is individually tested in
  Fig.~\ref{fig:Delta_phi_B_tidal}. All models except \GSF{2}{+}nm include the waveform
  multipoles described in Sec.~\ref{sbsec:new_multipoles}.
 For example, \GSF{2}{+}\PN{-} represents the EOB model in which the $(2+)$ tide is modelled as a GSF series
 and the $(2-)$ tide as a PN series. 
 }
\label{table:opt}
\end{table}
%

\subsection{Impact of the $\ell=3$ GSF-resummed potential}
\label{sec:l3}
\begin{figure}[t]
  \centering
  \includegraphics[width=0.99\linewidth]{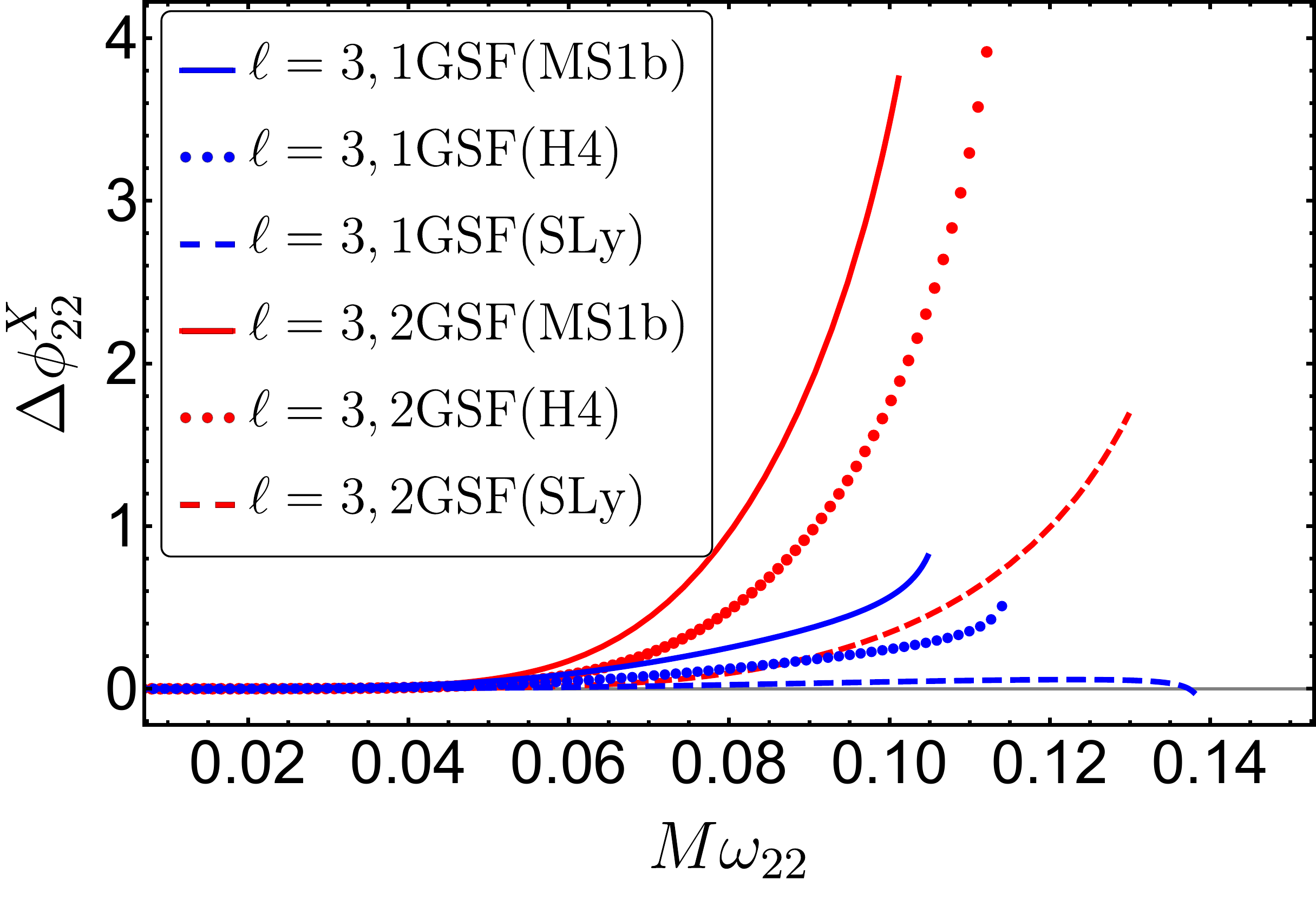} 
   \caption{\label{fig:el3_GSF_phase_contributions}The effect of the 1GSF and 2GSF $\el=3$ terms on the $(2,2)$ mode phase.
   Here, $\Delta\phi^X_{22}\equiv  \phi^b_{22} - \phi^X_{22}$, where $\phi_{22}^b$ is the phase
   of the baseline BNS run that contains only the $\el=2$ 0GSF, 1GSF terms for tides.
   We then add to this base either the 1GSF or 2GSF $\el=3$ tides and re-evolve the inspiral to obtain
   the corresponding $\phi_{22}^X$.
   We used $\kappa_2^T = 73.53, 191.4, 289.6$ for the SLy, H4, MS1b EOS, respectively.   
  }
  \end{figure}

Let us start by investigating the impact of the GSF-resummed $\ell=3$ contribution 
to the tidal potential. Its effect is to make the EOB $A$ potential more negative 
(i.e., more attractive) with respect to the corresponding PN-expanded NNLO 
$\ell=3$ part, so that the binary inspirals faster up to merger.
Figure~\ref{fig:el3_GSF_phase_contributions} shows the effect of 
the $\ell=3$ 1GSF and 2GSF terms individually,
where we employed three equal-mass BNS configurations: 
SLy, H4, and MS1b with $\kappa_2^T = 73.53, 191.4$, and $289.6$, respectively.
We recall that the 1GSF and 2GSF terms come from Eq.~\eqref{eq:A3plus_GSF_all} above, with
$r_{\rm LR}$ of the Schwarzschild geometry replaced by the corresponding EOB one
of the NNLO tidal potential, and having fixed $p=4$. 
The figure shows the phase difference versus GW frequency $M\omega_{22}$. 
Note that the curves end at the peak values of $M\omega_{22}$,
which approximately correspond to the peak of the $(2,2)$ 
waveform mode amplitude that was found to be rather close and consistent
with the merger frequency coming from NR simulations~\cite{Bernuzzi:2014owa}.
The figure illustrates the contribution of each term to the total $(2,2)$-mode phase of a baseline 
tidal model consisting of only the $\el=2$ 0GSF, 1GSF terms (no $\el>2$ tides whatsoever).

Note that, the phase accumulation due to the new terms starts very late in the inspiral, 
$M\omega_{22}\gtrsim 0.06$, consistent with the fact that the $\el=3$ GSF-resummed 
potential becomes distinguishable only in the last few cycles before the merger
as can be seen by comparing the brown and blue curves in Fig.~\ref{fig:A_of_u}.
Overall, we see that the first-order $\el=3$ GSF term contributes up $\ord(1)$ 
radian and the second-order term up to roughly 4 radians.
%
%

\begin{figure}[t]
  \centering
  \includegraphics[width=0.99\linewidth]{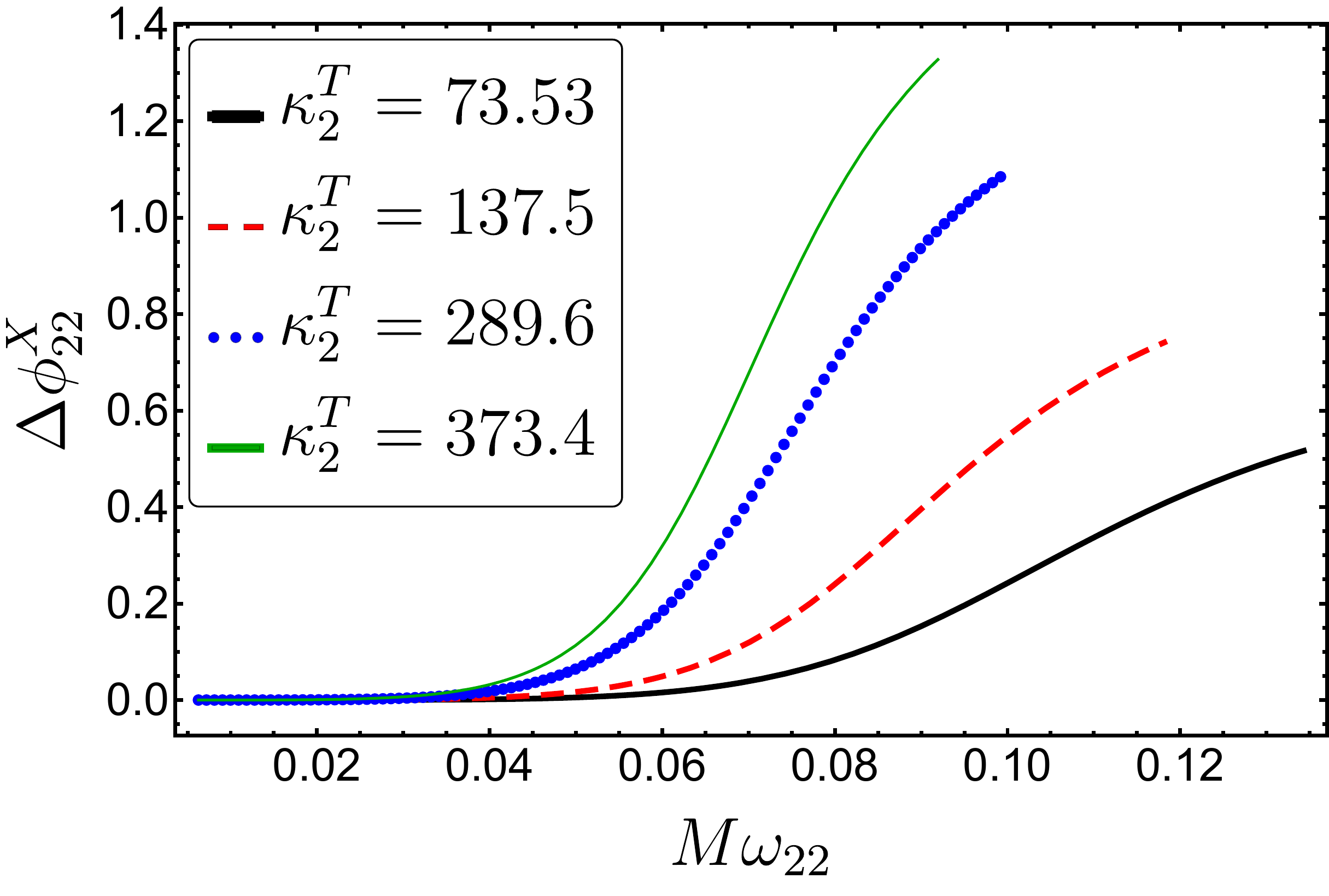} 
   \caption{The effect of the entire GSF-resummed $\el=3$ contribution, model \GSF{23}{+}, 
   on the $(2,2)$-waveform phase as compared with respect to the phase of the baseline \GSF{2}{+} model. 
    Here, $\Delta\phi_{22}^X\equiv \phi_{22}^\text{\GSF{2}{+}}\!\! -\phi^X_{22}$. 
    As $\kappa_2^T$ increases, the phase difference grows corresponding to the tides becoming more attractive, thus the neutron stars merge sooner.
  }\label{fig:Delta_phi_GSF23}
  \end{figure}

Having gained a quantitative understanding of the impact of the separate $\ell=3$ GSF-resummed
contributions to the potential, we incorporate them,  
Eq.~\eqref{eq:A3plus_GSF_all}, into \TEOBResum{}, thus replacing the previously used 
PN series truncated at NNLO. According to the summary of the various terms listed in 
Table~\ref{table:opt}, we name this flavor of the model \GSF{23}{+}.  We gauge its effect 
on the $(2,2)$ phase by comparing it to the phase  resulting from the \GSF{2}{+} 
model which will serve as our standard baseline for the remainder of this article unless 
otherwise noted. We show the resulting phase differences, $\Delta\phi_{22}^X\equiv \phi_{22}^\text{\GSF{2}{+}}\!\! -\phi^X_{22}$, 
in Fig.~\ref{fig:Delta_phi_GSF23} for three equal-mass configurations: $\{\text{EOS},q, \Lambda\}=\{ \text{SLy}, 1, 392.151 \}, \{\text{ALF2},1,733.323\}, \{\text{MS1b},1,1544.53 \}$, and 
MS1b with $q=1.5,\Lambda_A=1099.9, \Lambda_B = 4391.144$
translating to $\kappa_2^T=73.53, 137.5, 289.6$, and $ 373.4$, respectively.
As \GSF{23}{+} is more attractive than \GSF{2}{+}, because of the stronger $\ell=3$ contribution, 
it plunges faster, it accumulates less phase, and therefore $\Delta\phi_{22}^X$ is positive. 
Also note, in passing, that the merger frequency decreases as $\kappa_2^T$ increases 
because of the correspondingly augmented tidal interaction \cite{Bernuzzi:2014kca}.
%
%

\subsection{Impact of all other tidal contributions}
\label{sec:others}
\begin{figure*}[t]
\center
  \includegraphics[width=0.48\textwidth]{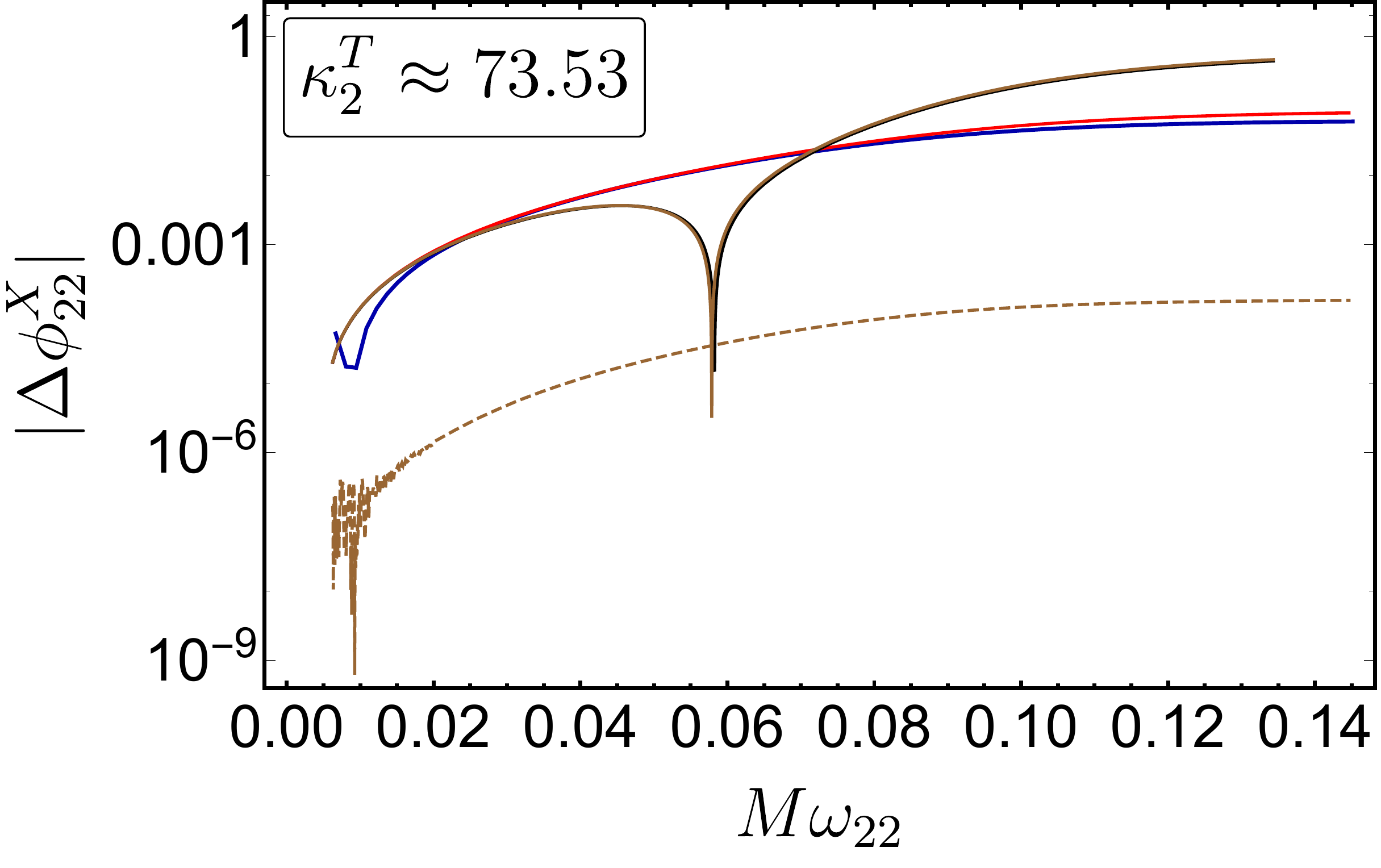}\hspace{5mm}
  \includegraphics[width=0.48\textwidth]{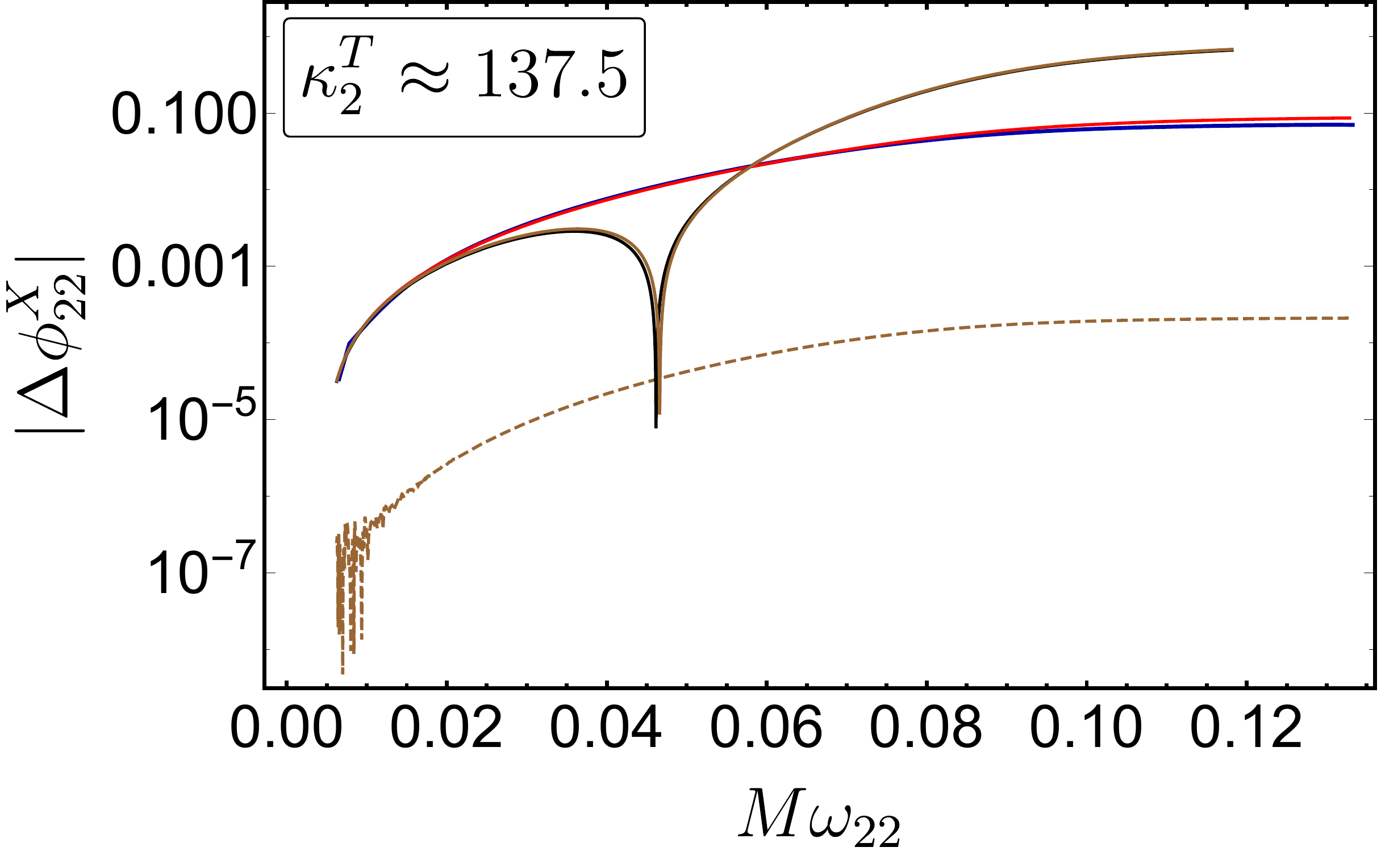}\\
  \includegraphics[width=0.48\textwidth]{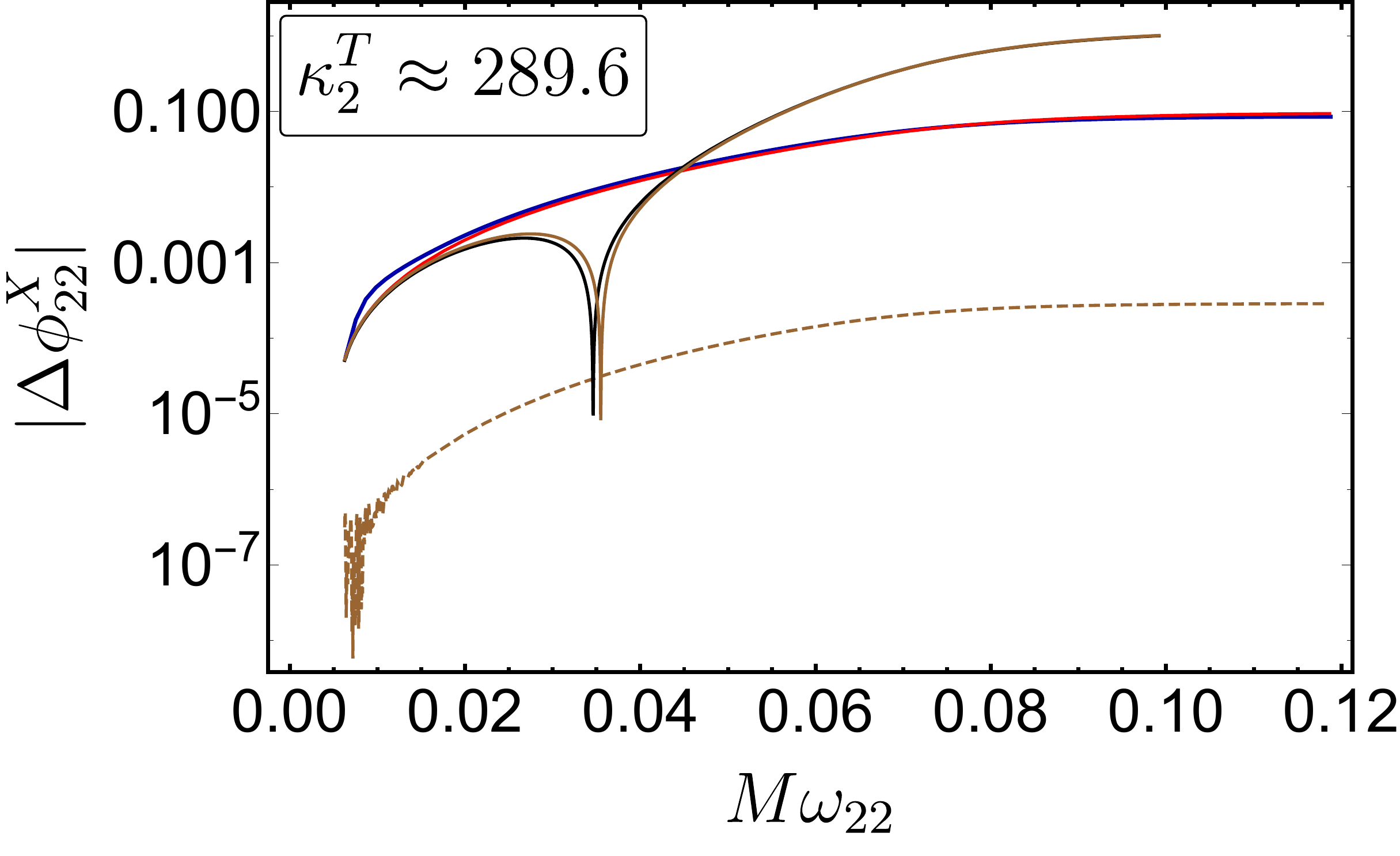}\hspace{5mm}
  \includegraphics[width=0.47\textwidth]{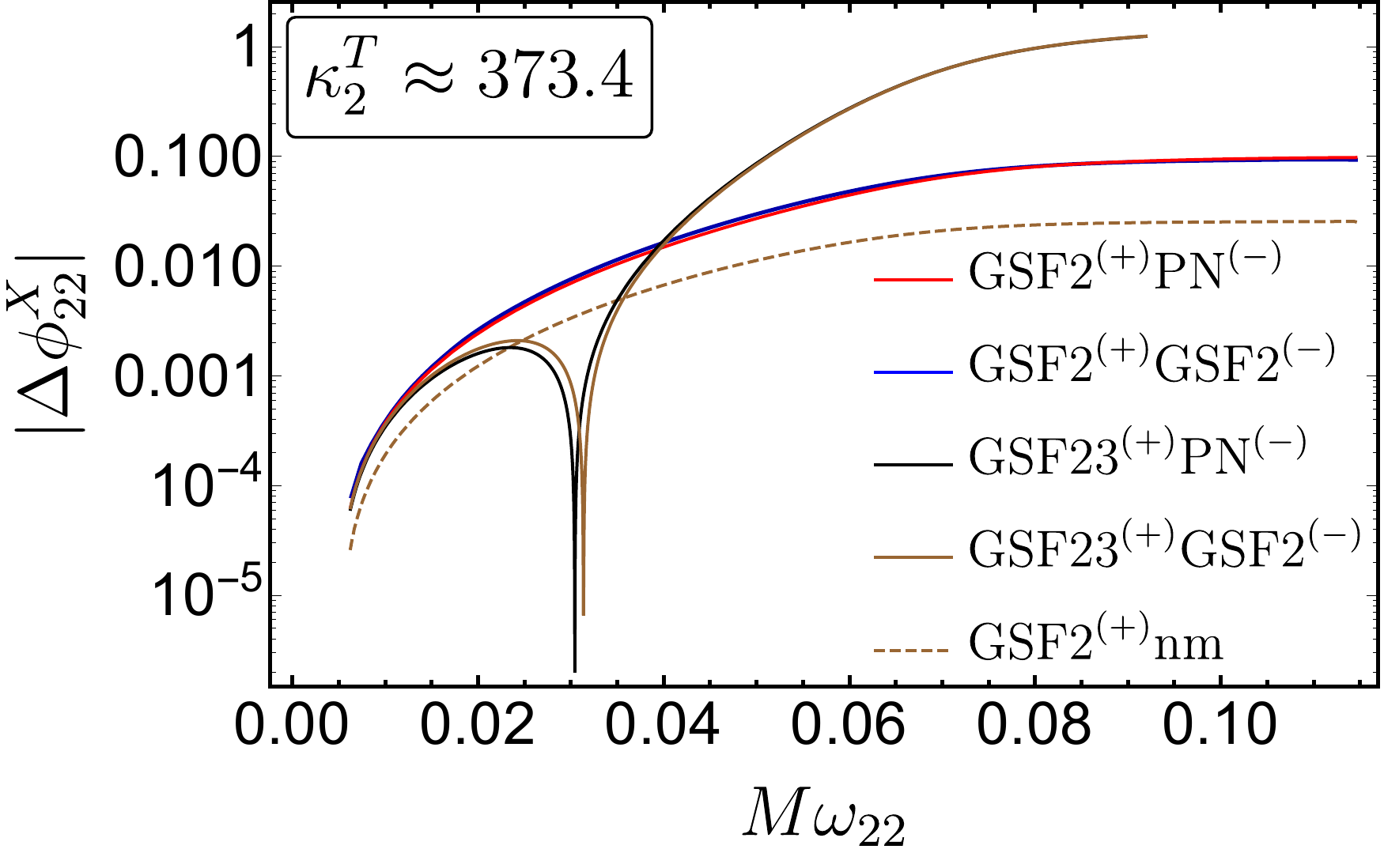}
  \caption{\label{fig:DeltaPhi_EOB} 
  The phase difference $\Delta\phi^X_{22}\equiv \phi_{22}^\text{\GSF{2}{+}}\! \! -\phi^X_{22}$, between the baseline \TEOBResum{} model, \GSF{2}{+}, and various tidally augmented  \TEOBResum{} variants listed in Table~\ref{table:opt}.
  Starting from the upper-left panel and going clockwise, we have $\{\text{EOS},q, \Lambda\}=\{ \text{SLy}, 1, 392.151 \}, \{\text{ALF2},1,733.323\}, \{\text{MS1b},1,1544.53 \}$. 
  The lower-right panel corresponds to MS1b with $q=1.5,\Lambda_A=1099.9, \Lambda_B = 4391.144$. 
  The blue and the red curves are negative because the gravitomagnetic Love number $\Sigma$ is negative for irrotational fluids. 
  The sign change of the \GSF{23}{+} curves (brown and black) is explained in the text.
  }
  \end{figure*}

As detailed in Sec.~\ref{sec:new_tides} we have added the gravitomagnetic tidal interaction to \TEOBResum{}
either as a PN series or a GSF-resummation. 
We have additionally augmented the EOB $B$ potential and the multipolar waveforms with new analytical
tidal information. The contribution of these new terms are subdominant compared to the $\el=3$ GSF-resummed
tide. Their effects on the evolution of the GW phase
is shown in Fig.~\ref{fig:DeltaPhi_EOB} once again in terms of $\Delta\phi_{22}^X\equiv \phi_{22}^\text{\GSF{2}{+}}\!\! -\phi^X_{22}$.
As $\Delta\phi^X_{22}$ varies in sign and over several orders of magnitude, 
we opted to display $|\Delta\phi^X_{22}|$ as semilog plots in the figure, where
the four panels correspond to the same four cases chosen for Fig.~\ref{fig:Delta_phi_GSF23}.
In the following subsections, we discuss the effects of these subdominant terms.
%
%
%

\subsubsection{Gravitomagnetic tides: irrotational fluids}\label{sbsec:irrotational_GM}

Since the gravitomagnetic Love number is negative, 
the contribution of the $(2-)$ tide, whether as a PN or GSF series, 
yields $\Delta\phi_{22}<0$.  This is in concordance with our physical intuition 
if we recall that the overall sign of the tidal potential is negative.
Hence, gravitomagnetic terms make it less negative thus extending 
the inspiral time and increasing the accumulated phase which, when 
subtracted from the smaller phase of \GSF{2}{+}, expectedly 
yields a negative number.

We see in Fig.~\ref{fig:DeltaPhi_EOB} that 
the contribution of the negative gravitomagnetic terms (red, blue curves) 
to the phase is $\lesssim 0.1$ radian up to the EOB mergers given roughly 
by $0.12 \lesssim M\omega_{22} \lesssim 0.14$ depending on $\kappa_2^T$.
Moreover, the difference between using PN vs. GSF series for the $(2-)$ tides is 
almost indistinguishable as can be seen both in the \GSF{2}{+} (red vs. blue curves)
and \GSF{23}{+} cases (black vs. brown curves).
Note that the sign change of the black, brown curves in Fig.~\ref{fig:DeltaPhi_EOB} is due to the sign change in the corresponding tidal potential ${A}^{(2-)}+{A}^{(3+)}$ 
because these terms have opposite signs and different weak-field behavior ($u^7$ vs. $u^{10}$, respectively).
Even less distinguishable than the gravitomagnetic contribution is the effect of augmenting the waveform by adding the $(2+)$NLO and $(2-)$LO terms to $\hat{h}^T_{22}$.
This effect is represented by the dashed brown curves labelled \GSF{2}{+}nm 
and amounts to at most $\sim 0.02\,$radian. 
%

%
%
%
%
%
%
%
%
  \begin{figure}[h!]
  \centering
  \includegraphics[width=0.99\linewidth]{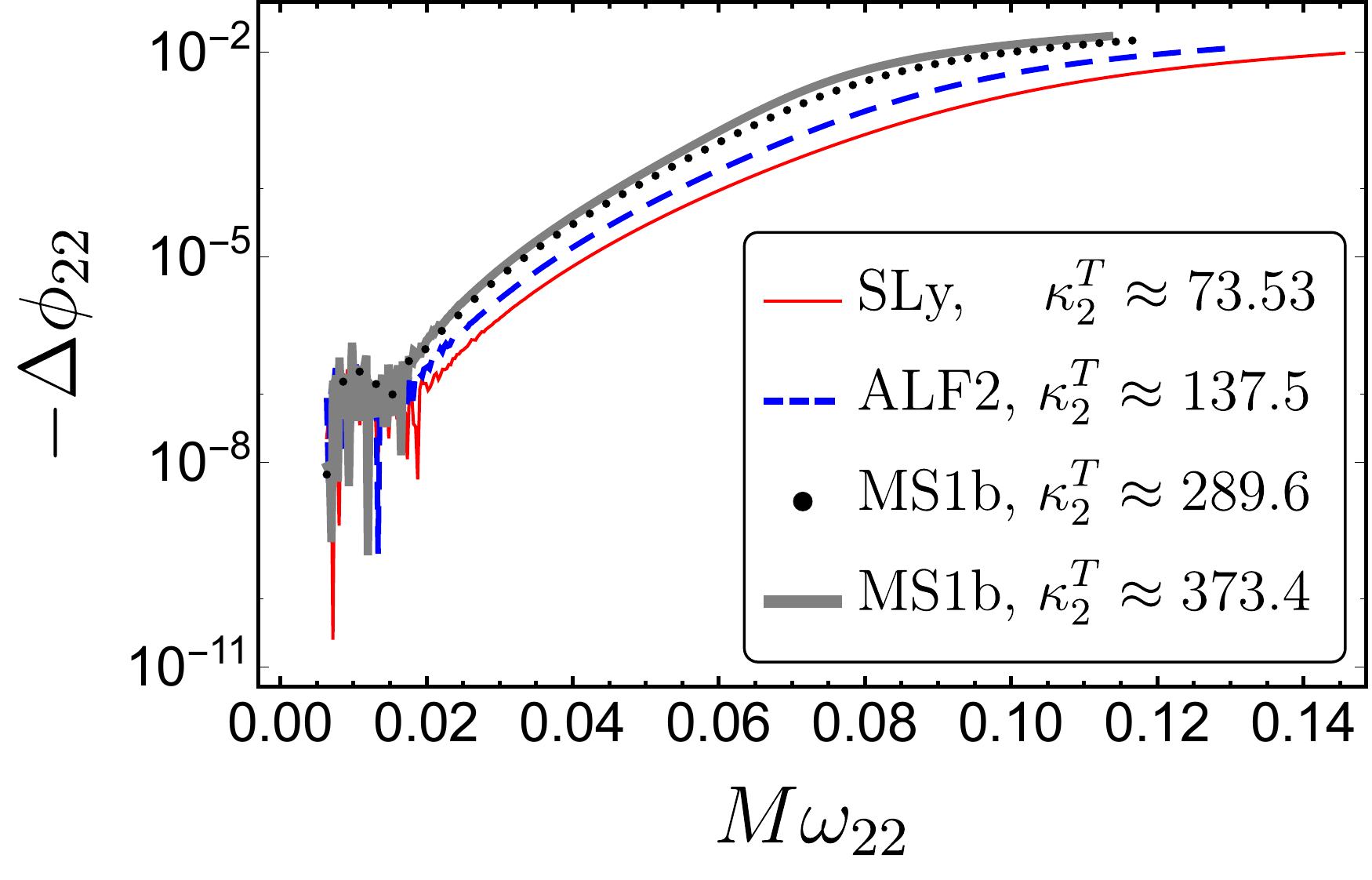}
   \caption{The effect of augmenting the $B$ potential with the tidal term $B'_\text{T}$ of Eq.~(\ref{eq:BTprime}) on the GW phase $\phi_{22}$ for the four cases of Fig.~\ref{fig:DeltaPhi_EOB}.
   In each case, the phase difference is computed with respect to the baseline model \GSF{2}{+} (see Table~\ref{table:opt}).
  Note that we plot $-\Delta\phi_{22}$. See Sec.~\ref{sbsec:BT} for why $\Delta\phi_{22}$ is negative in this comparison.}
   \label{fig:Delta_phi_B_tidal}
  \end{figure}

\begin{table*}[t]
\centering
\begin{tabular}{lcccccccccccccccc}
\hline\hline
 \BAM  & EOS & $\kappa_2^T$ & $ m_A~[M_\odot]$ & $q$  & \hspace{3mm}$\Lambda_A$\hspace{3mm} & \hspace{3mm}$ \Lambda_B$ \hspace{3mm} & \hspace{3mm}$k_A^{(2)}$\hspace{3mm} & \hspace{3mm}$ k_B^{(2)}$ \hspace{3mm} & \hspace{2mm}$\mathcal{C}_A$\hspace{2mm} & \hspace{2mm}$ \mathcal{C}_B$ \hspace{2mm} &  \hspace{2mm}$\Sigma_A$\hspace{2mm} &  \hspace{2mm}$\Sigma_B$\hspace{2mm} & \hspace{2mm}$j_A^{(2)}$\hspace{2mm} & \hspace{2mm}$ j_B^{(2)}$ \hspace{2mm} & \hspace{2mm}  Ref. \T \\
   \hline
  
   {\tt 0011} & ALF2 & 72.12 & 1.500 & 1.000 & 384.7 & 384.7 & 0.1044 & 0.1044 & 0.1784 & 0.1784 & $-3.775$ & $-3.775$ & $-0.03278$ & $-0.03278$ & \cite{Dietrich:2017aum}\\
  {\tt 0095} & SLy & 73.53 & 1.350 & 1.000 & 392.2 & 392.2 & 0.09333 & 0.09334 & 0.1738 & 0.1738 & $-3.823$ & $-3.823$ & $-0.02911$ & $-0.02911$ & \cite{Dietrich:2017aum}\\
{\tt 0127} & SLy & 78.05 & 1.650  & 1.503 & 1371. & 93.45 & 0.1172 & 0.06433 & 0.1416 & 0.2150 & $-8.761$ & $-1.553$ & $-0.02398$ & $-0.03421$ & \cite{Dietrich:2017feu,Dietrich:2016hky}\\
 {\tt 0017} & ALF2 & 132.7 & 1.650  & 1.500 & 2218. & 196.6 & 0.1443 & 0.08692 & 0.1341 & 0.1967 & $-12.17$ & $-2.460$ & $-0.02533$ & $-0.03480$ &  \cite{Dietrich:2017aum}\\
  {\tt 0107} & SLy & 136.6 & 1.354  & 1.224 & 1320. & 383.9 & 0.1174 & 0.09288 & 0.1427 & 0.1744 & $-8.542$ & $-3.770$ & $-0.02430$ & $-0.02919$ & \cite{Dietrich:2017xqb}\\
 {\tt 0021} & ALF2 & 139.6 & 1.750  & 1.750 & 122.3 & 3528. & 0.07505 & 0.1517 & 0.2101 & 0.1234 & $-1.830$ & $-16.78$ & $-0.03596$ & $-0.02309$ & 
  \cite{Dietrich:2017aum}\\
  {\tt 0037} & H4 & 191.4 & 1.372  & 1.000 & 1020. & 1021. & 0.1140 & 0.1140 & 0.1494 & 0.1494 & $-7.181$ & $-7.182$ & $-0.02567$ & $-0.02567$ & \cite{Dietrich:2017aum}\\
{\tt 0048} & H4 & 192.8 & 1.528  & 1.250 & 1990. & 500.2 & 0.1262 & 0.09762 & 0.1334 & 0.1671 & $-11.30$ & $-4.478$ & $-0.02293$ & $-0.02797$ & \cite{Dietrich:2017aum}\\
  {\tt 0058} & MPA1 & 115.3 & 1.350  & 1.000 & 614.9 & 614.9 & 0.1120 & 0.1120 & 0.1648 & 0.1648 & $-5.128$ & $-5.128$ & $-0.02988$ & $-0.02988$ & \cite{Dietrich:2017aum}\\
  {\tt 0094} & MS1b & 250.2 & 1.944 & 2.059 & 9249. & 183.7 & 0.1619 & 0.08698 & 0.1031 & 0.1994 & $-33.14$ & $-2.358$ & $-0.01855$ & $-0.03572$ & \cite{Dietrich:2016hky,Dietrich:2015pxa}\\
  {\tt 0091} & MS1b & 280.4 & 1.650  & 1.500 & 502.2 & 4391. & 0.1099 & 0.1525 & 0.1709 & 0.1183 & $-4.490$ & $-19.56$ & $-0.03143$ & $-0.02173$ & \cite{Dietrich:2017feu,Dietrich:2016hky} \\
  {\tt 0064} & MS1b & 289.6  & 1.350 & 1.000 & 1542. & 1546. & 0.1347 & 0.1347 & 0.1422 & 0.1422 & $-9.492$ & $-9.508$ & $-0.02653$ & $-0.02651$ & \cite{Dietrich:2017aum}\\
   \hline\hline
\end{tabular}
\caption{
The initial configuration for the \BAM{} and EOB runs that we use for the EOBNR comparisons. Note $q\equiv
  m_A/m_B>1$ and the numbers for the \BAM{} NR runs are approximate as they are extracted from NR initial data. See Sec.~\ref{sbsec:tidal_potentials} for the notation.}
\label{table:BAM_runs}
\end{table*}

\begin{table}[t]
 \centering
\begin{tabular}{lcccc}
\hline\hline
 \BAM  & \hspace{3mm} $\hat{\omega}^\text{mrg}$ & \hspace{1mm} $f^\text{mrg} $(Hz) & \hspace{2mm} $j^\text{mrg}$ \hspace{2mm} & \hspace{3mm} $E_b^\text{mrg}$ \hspace{2mm} \\
  \hline
  {\tt0011} & 0.1615 & 1739 & 3.358 & $-0.06302$ \\ 
  {\tt0095} & 0.1711 & 2048 & 3.318 & $-0.06520$ \\ 
  {\tt0127} & 0.1364 & 1604 & 3.404 & $-0.05969$ \\ 
  {\tt0017} & 0.1197 & 1406 & 3.532 & $-0.05388$ \\ 
  {\tt0107} & 0.1333 & 1750 & 3.489 & $-0.05592$ \\ 
  {\tt0021} & 0.1075 & 1263 & 3.584 & $-0.05151$ \\ 
  {\tt0037} & 0.1357 & 1598 & 3.516 & $-0.05467$ \\ 
  {\tt0048} & 0.1168 & 1371 & 3.568 & $-0.05213$ \\ 
  {\tt0058} & 0.1486 & 1778 & 3.451 & $-0.05804$ \\ 
  {\tt0094} & 0.08877 & 993 & 3.700 & $-0.04686$ \\ 
  {\tt0091} & 0.1014 & 1191 & 3.658 & $-0.04864$ \\ 
  {\tt0064} & 0.1234 & 1477 & 3.612 & $-0.05068$ \\ 
 \hline\hline
\end{tabular}
\caption{\BAM{} merger (mrg) data. The merger is taken to occur at the peak 
  of the amplitude of the $\ell=m=2$ mode. $f$ is the corresponding quadrupole GW frequency in Hz.
  $j$ is the angular momentum and $E_b$ the binding energy (\ref{sbsec:Eb_vs_J}).}
\label{table:BAM_merger_data}
\end{table}

\subsubsection{Gravitomagnetic tides: static fluid}\label{sbsec:static_GM}

For the sake of comparison, we also considered gravitomagnetic tides for static fluids. 
As a generic difference, we note that static gravitomagnetic Love numbers are positive as opposed to
irrotational ones, and, for polytropes, their absolute values are about twice those of irrotational 
Love numbers (see Fig.~1 of Ref.~\cite{Landry:2015cva}).
For realistic EOS, we obtain the static Love numbers from the quasi-universal relations of 
Ref.~\cite{Jimenez-Forteza:2018buh} which yield $\Sigma_\text{stat}\sim 2|\Sigma|$, roughly 
in agreement with the polytropic ratio mentioned above. As a result, we would expect static 
Love numbers to result in phase differences that are roughly twice the magnitude of $\Delta\phi^\text{red,blue}_{22}$ 
of Fig.~\ref{fig:DeltaPhi_EOB} and with  a positive sign. Repeating the runs of Fig.~\ref{fig:DeltaPhi_EOB} for \GSF{2}{+}\PN{-} and \GSF{2}{+}\GSF{2}{-} with $\Sigma_\text{stat}$, we indeed find that $\Delta\phi_{22}$ now accumulates up to $\sim 0.2$ radian at the EOB merger ($0.12 \lesssim M\omega_{22} \lesssim 0.14$), but has, 
as expected, the opposite sign to the irrotational case.

We opt for irrotational Love numbers because we think they represent more realistic scenarios:
in Ref.~\cite{Landry:2015cva}, Landry and Poisson studied gravitomagnetic
tidal interactions relaxing the hypothesis that the NS fluid be in hydrostatic equilibrium.
Instead, they considered fluids in an irrotational state, thus allowing for internal currents
induced by gravitomagnetic tidal fields.
It was only recently shown~\cite{Pani:2018inf} that the independent formalism for relativistic tides in
Ref.~\cite{Damour:2009vw} by Damour and Nagar indeed implicitly enforces the fluid to an
irrotational state and is equivalent to the Landry-Poisson formulation.
Here we follow the Damour-Nagar conventions for Love numbers as shown in Sec.~\ref{sbsec:tidal_potentials}.

\subsubsection{Leading-order tidal term in the  EOB $B$ potential}
\label{sbsec:BT}
The consequence of augmenting the $B$ potential by $B_\text{T}'$ of Eq.~\eqref{eq:BTprime}
is shown in Fig.~\ref{fig:Delta_phi_B_tidal}, once again in terms of $\Delta\phi^X_{22}$
 where $X$ now represents \GSF{2}{+} augmented with $B'_\text{T}$.
We show the phase difference again for four points in $\{q,\Lambda_A,\Lambda_B\}$ space with increasing $\kappa_2^T$.
Even for very large $\kappa_2^T$, the effect of the $B'_\text{T}$ term on the phase of the waveform is too small
to matter for the current generation of ground-based detectors.
Note that, unlike in Fig.~\ref{fig:DeltaPhi_EOB}, $\Delta\phi_{22}$ is now negative because $\dot{r} \propto B^{-1/2}$
(cf. Eq.~(6b) of Ref.~\cite{Damour:2012ky}). Hence increasing $B$ decreases $\dot{r}$,
thus lengthening the inspiral time. The $B_\text{T}'$ term has been added to all models
of Table~\ref{table:opt}.
%

\section{EOB/NR comparisons: energetics and waveforms}
\label{sec:nrar}
%
%
%
\begin{figure*}[t]
  \centering
  \includegraphics[width=0.34\linewidth]{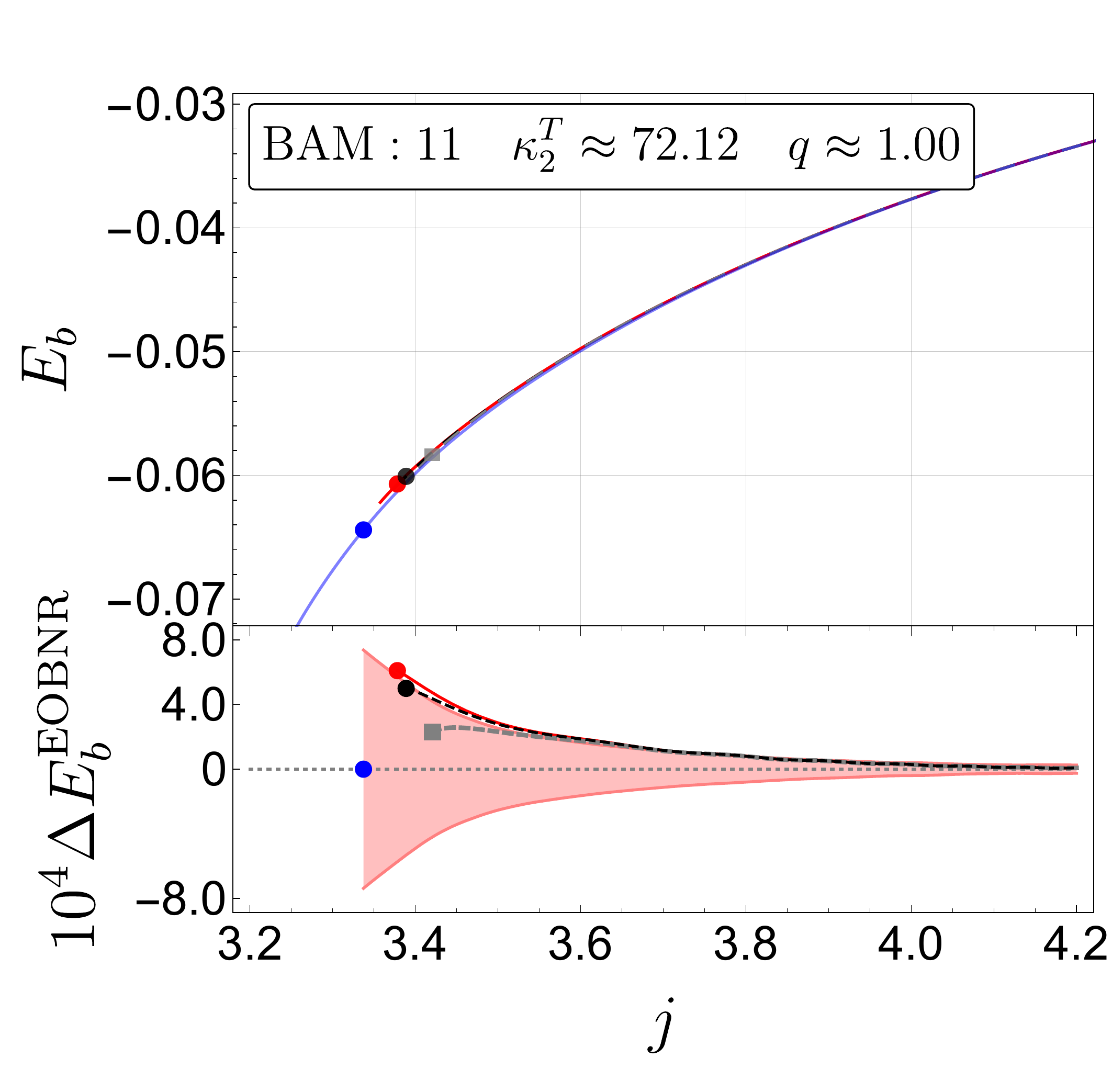}
  \hspace{-1mm}
  \includegraphics[width=0.325\linewidth]{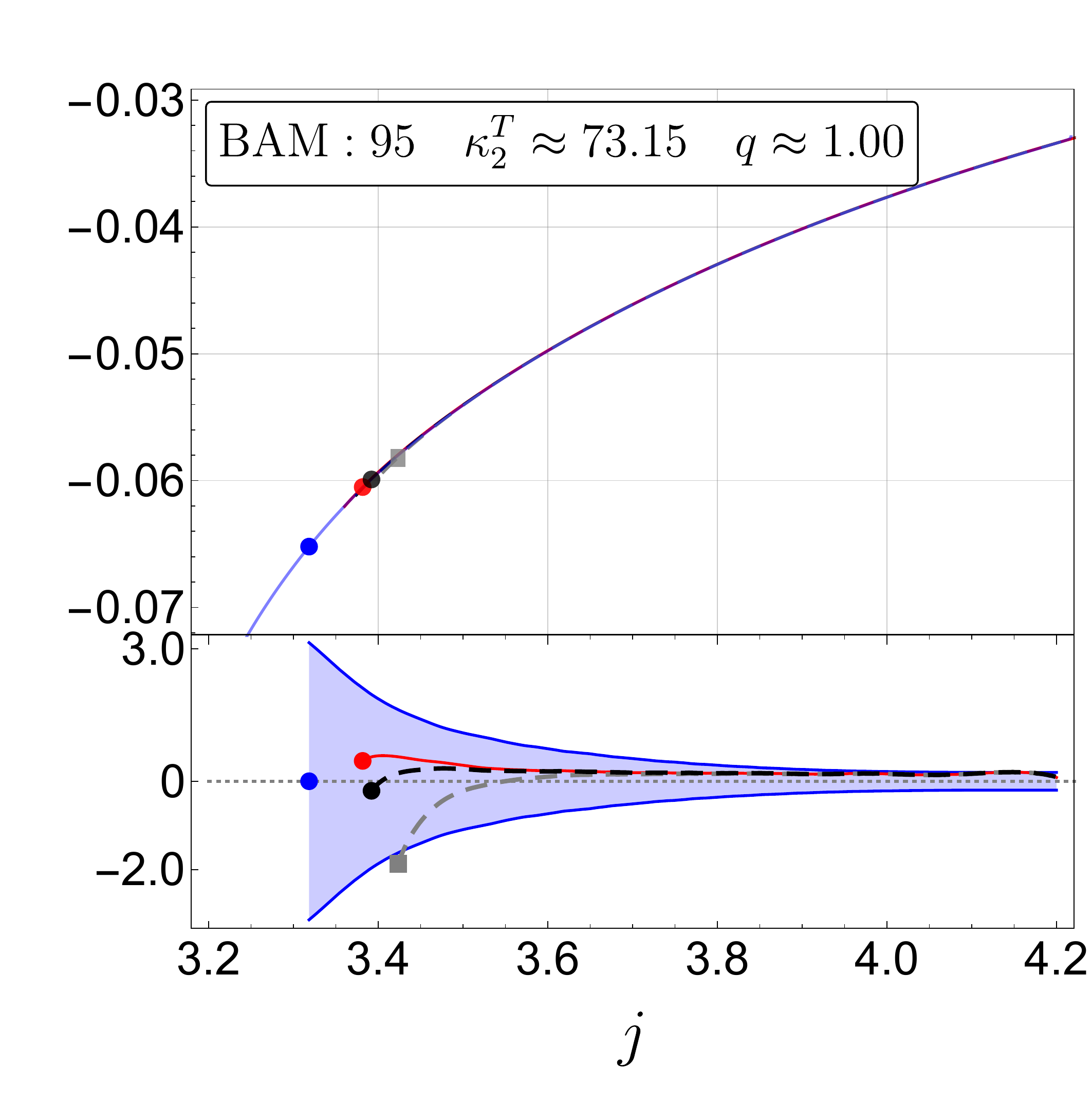}  
  \hspace{-2mm}
  \includegraphics[width=0.325\linewidth]{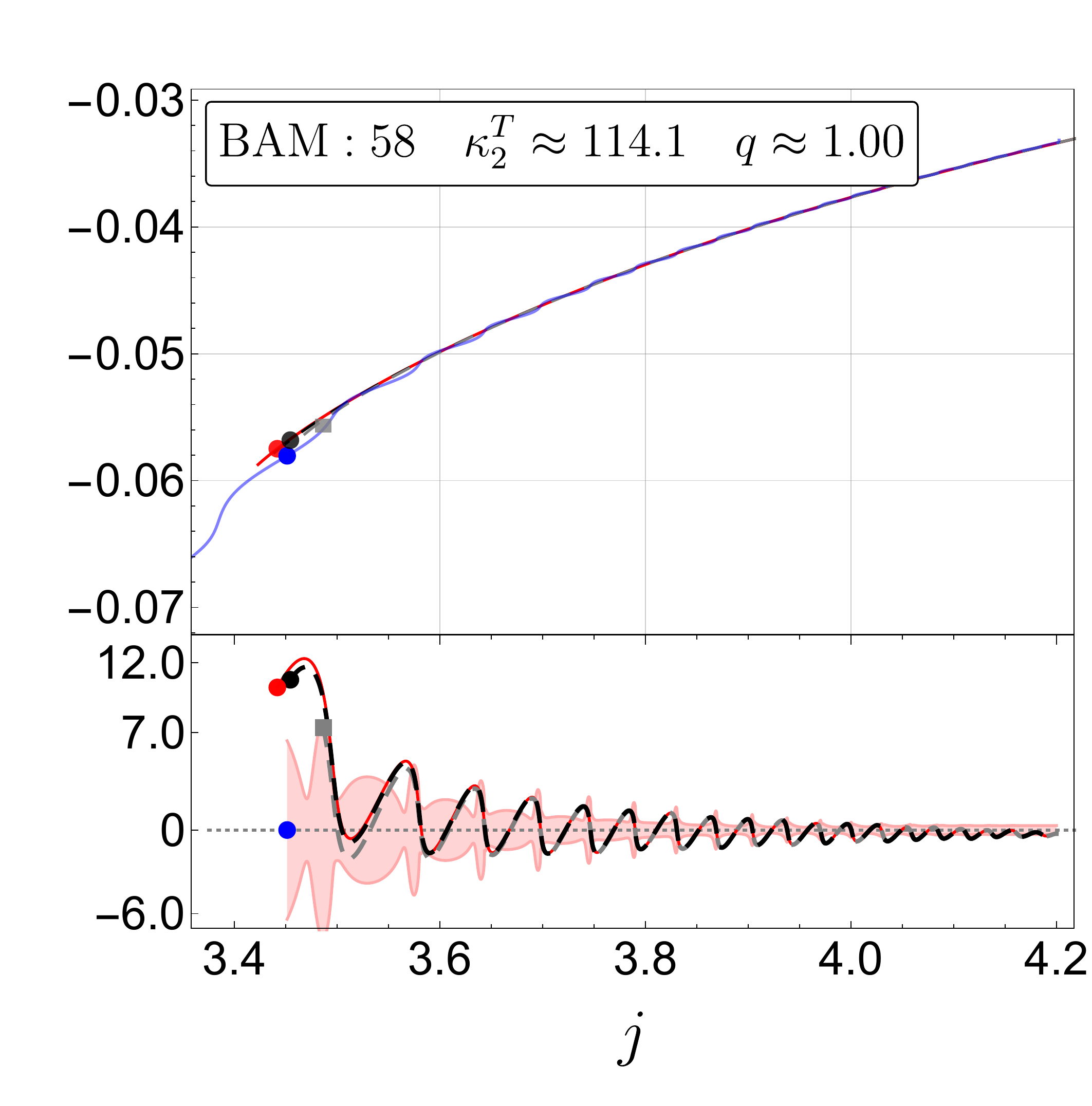}\\
  \vspace{-2mm}
  \includegraphics[width=0.34\linewidth]{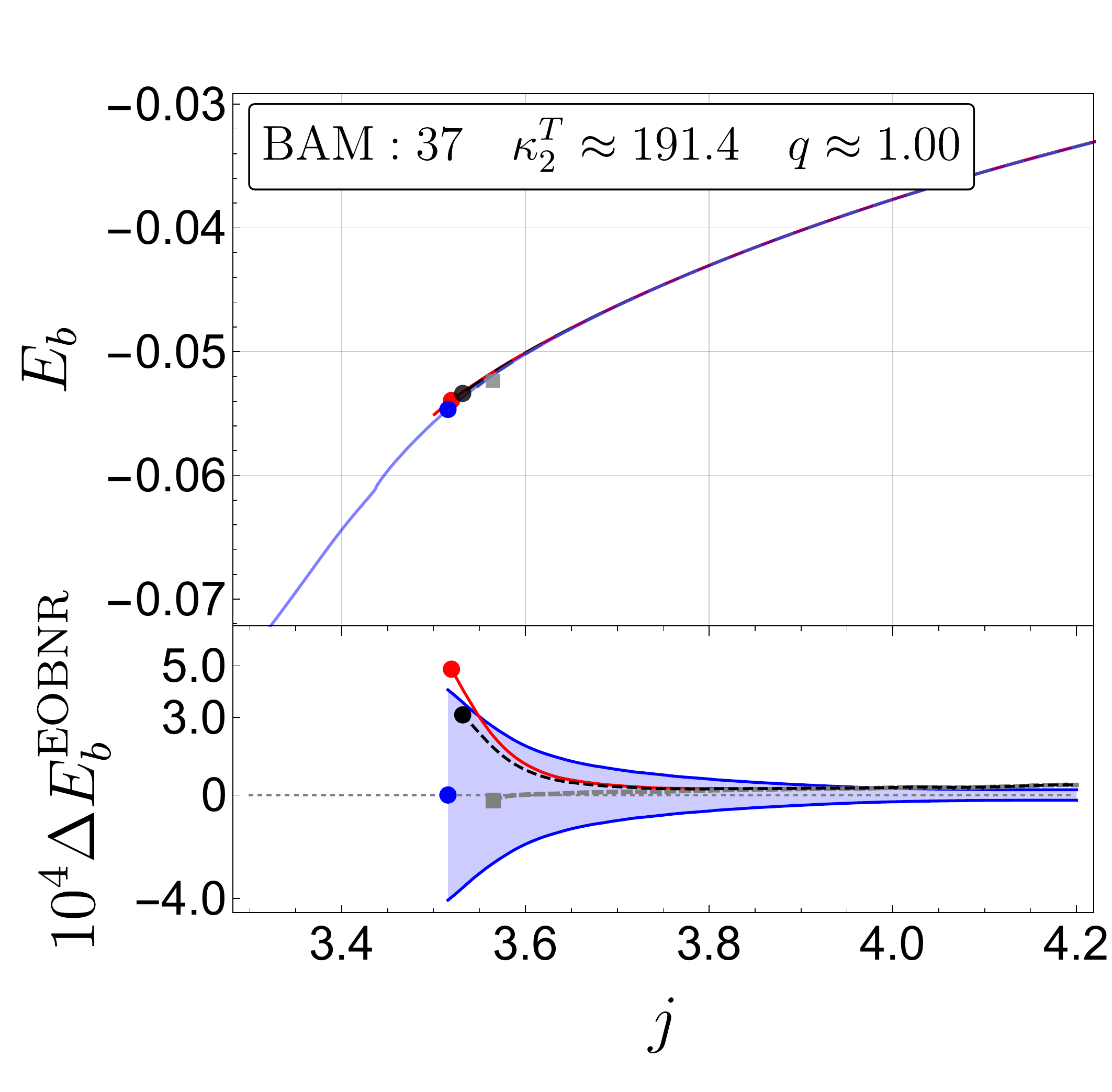} 
  \hspace{-1mm}
  \includegraphics[width=0.325\linewidth]{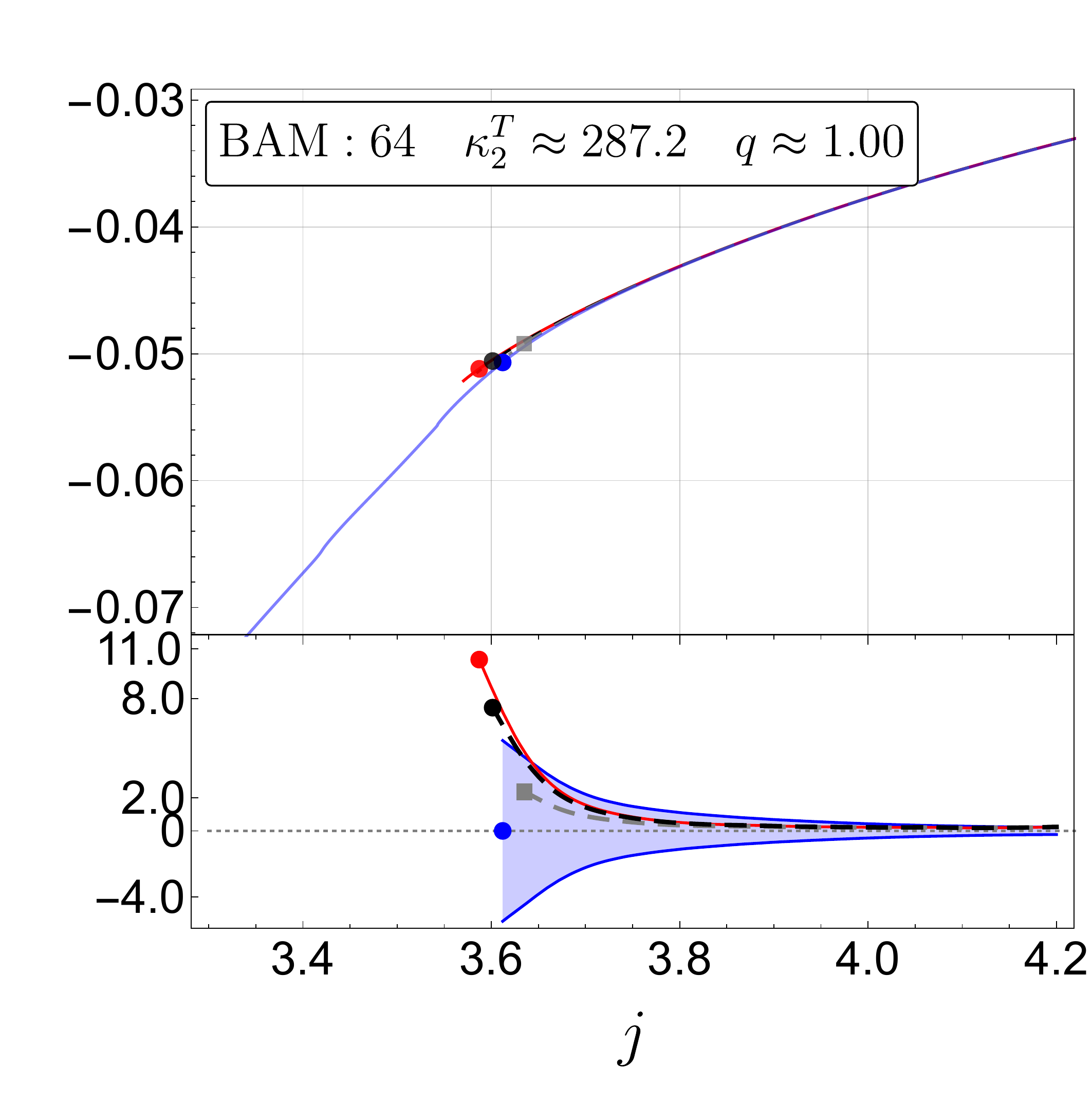}
  \hspace{-2mm}
  \includegraphics[width=0.325\linewidth]{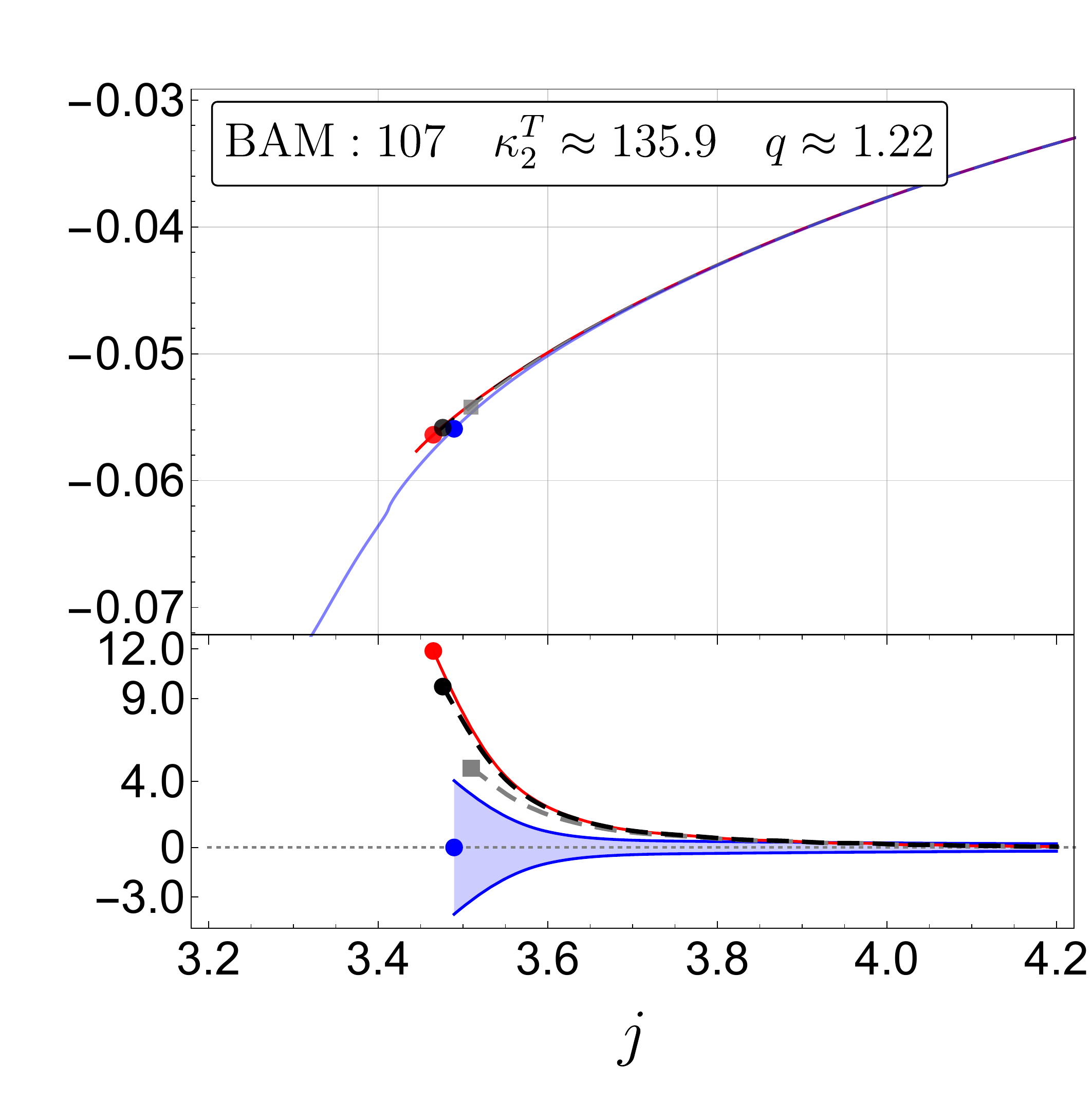}
  \vspace{-4mm}
  \caption{\label{fig:Eb_J_q1} EOB-NR comparison in terms of binding
    energy as a function of angular momentum for the $q=1$ binaries of
    Table~\ref{table:BAM_runs} and the $q\approx 1.22$ case. In each subfigure, the upper panel
    shows the \TEOBResum{} $E_b(j)$ curves for three different models listed in Table~\ref{table:opt}
    (solid red, dashed black and dashed gray)
    along with the corresponding NR data represented by the solid blue curves.
    The dots and the squares mark the peak orbital frequency of each run with the corresponding color.
    The values for NR $E_b$ and $j$ at the merger are listed in Table~\ref{table:BAM_merger_data}.
    Lower panel of each subfigure shows the difference between EOB and
    NR results, $\Delta E_b^\text{EOBNR}\equiv
    E_b^\text{EOB}-E_b^\text{NR}$, with the shaded regions representing our estimation of the NR error. The blue error regions are more reliable because they come from convergent simulations, while the pink ones are obtained from differences between the two highest NR resolutions, thus
    are less certain (cf. Sec.~\ref{sec:nrar}).
    Note that we amplify the error region by a factor of $10^4$ to improve its visibility.    
    The oscillations in the panel for \BAMc{\tt 0058} are due to residual eccentricity.}
\end{figure*}

\begin{figure*}[t]
  \centering
  %
  \includegraphics[width=0.34\linewidth]{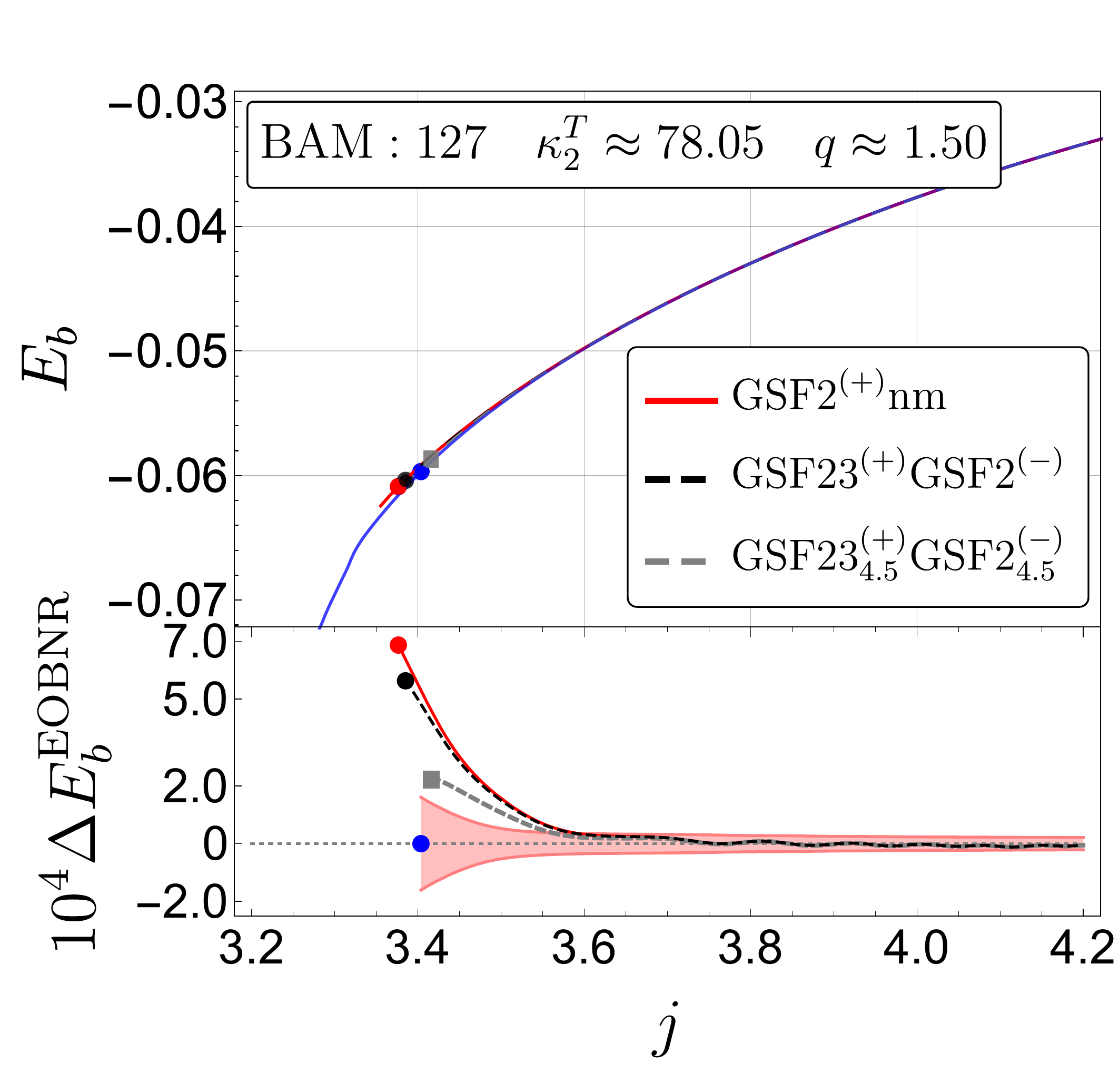}
  \hspace{-1mm}
  \includegraphics[width=0.325\linewidth]{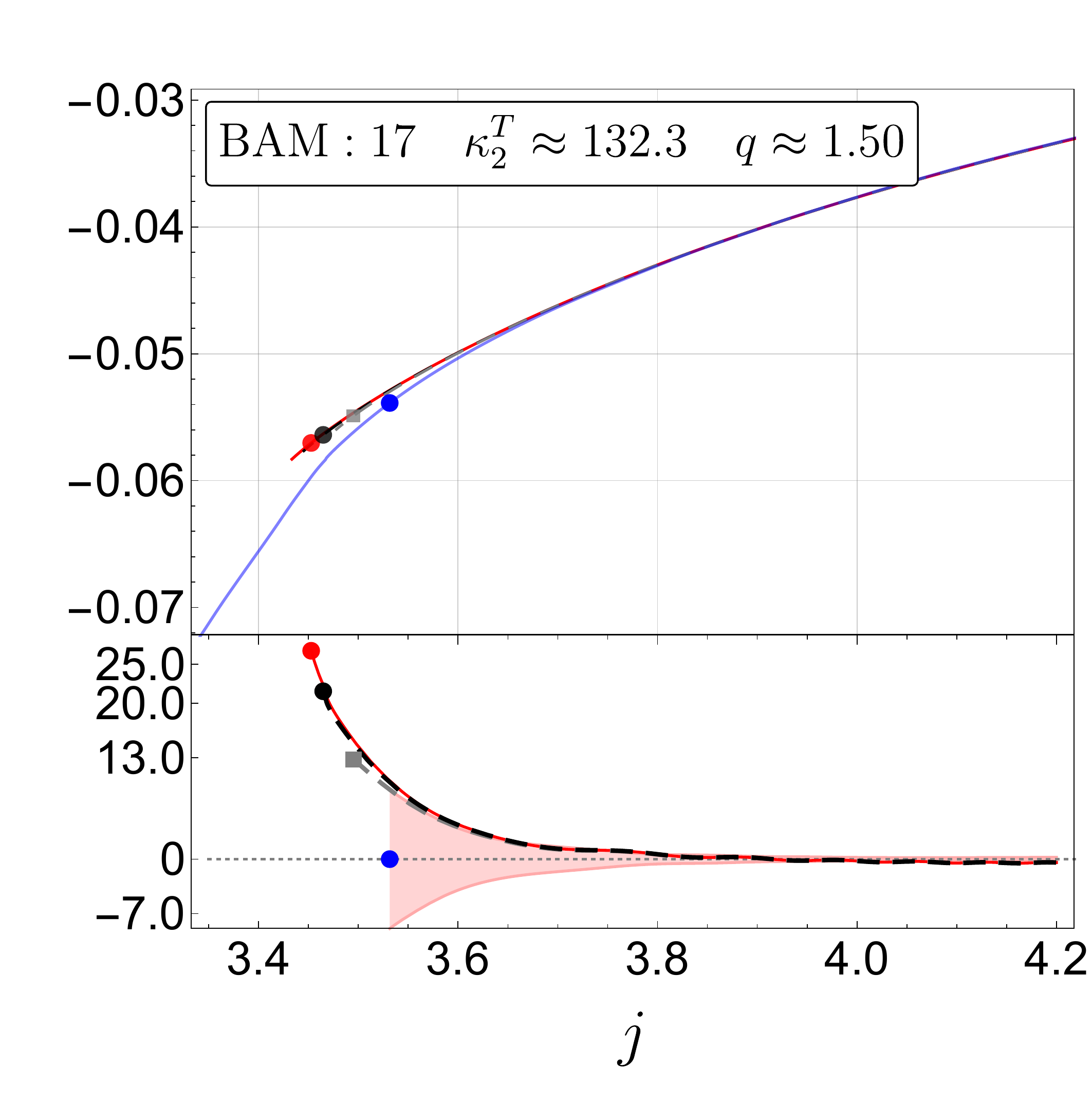}  
  \hspace{-2mm}
  \includegraphics[width=0.325\linewidth]{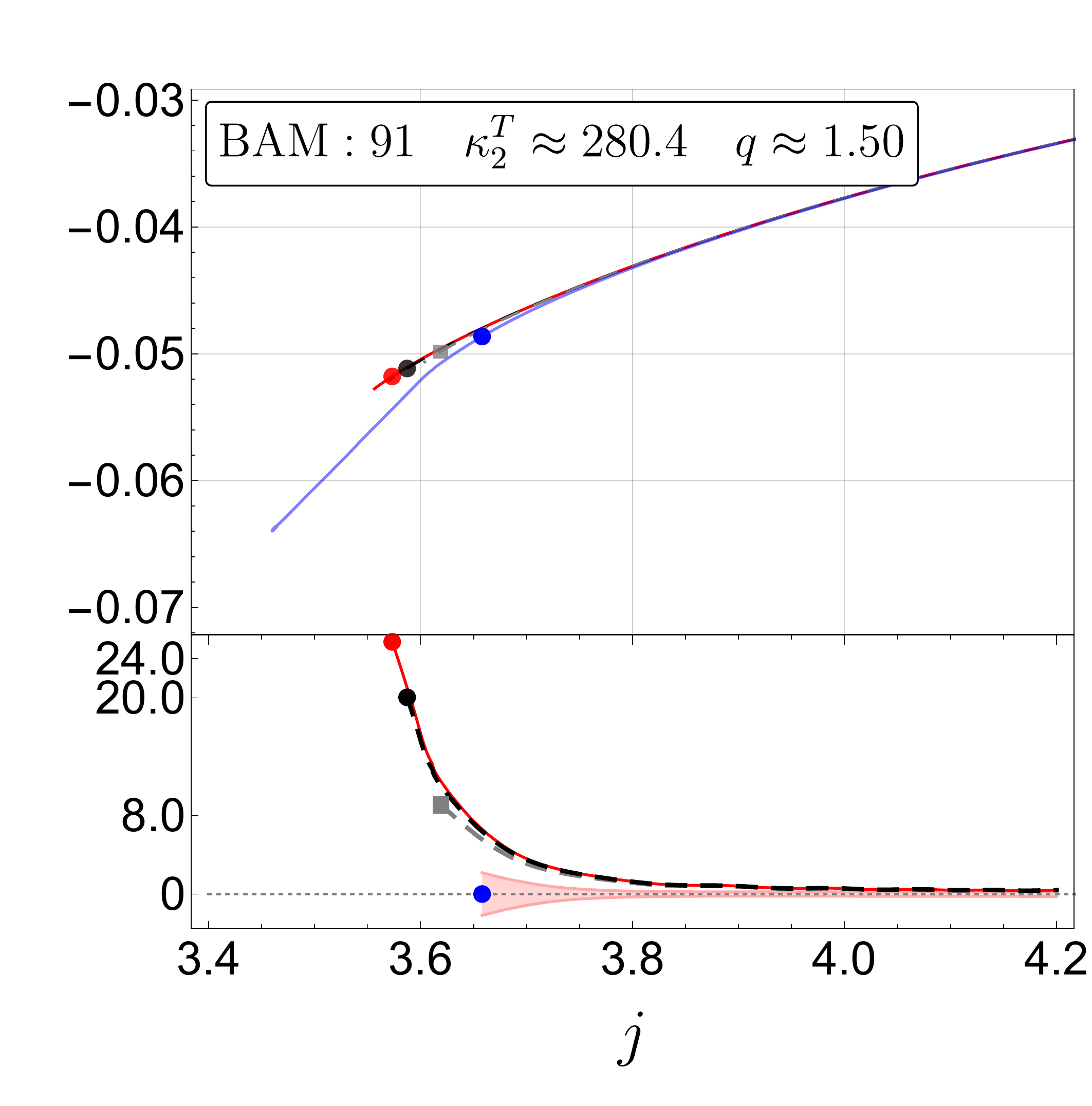}\\
  \vspace{-2mm}
  \includegraphics[width=0.34\linewidth]{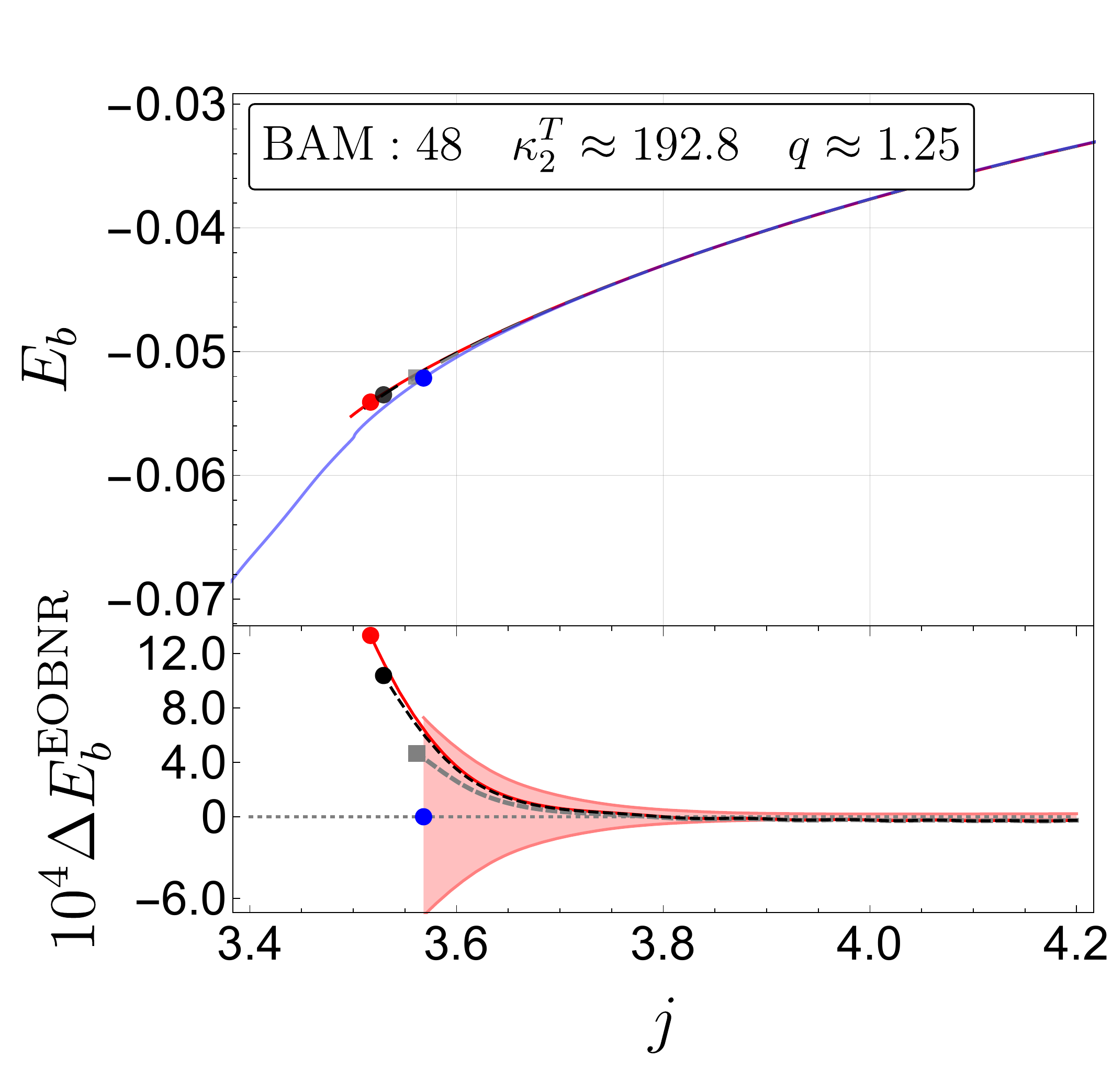} 
  \hspace{-1mm}
  \includegraphics[width=0.325\linewidth]{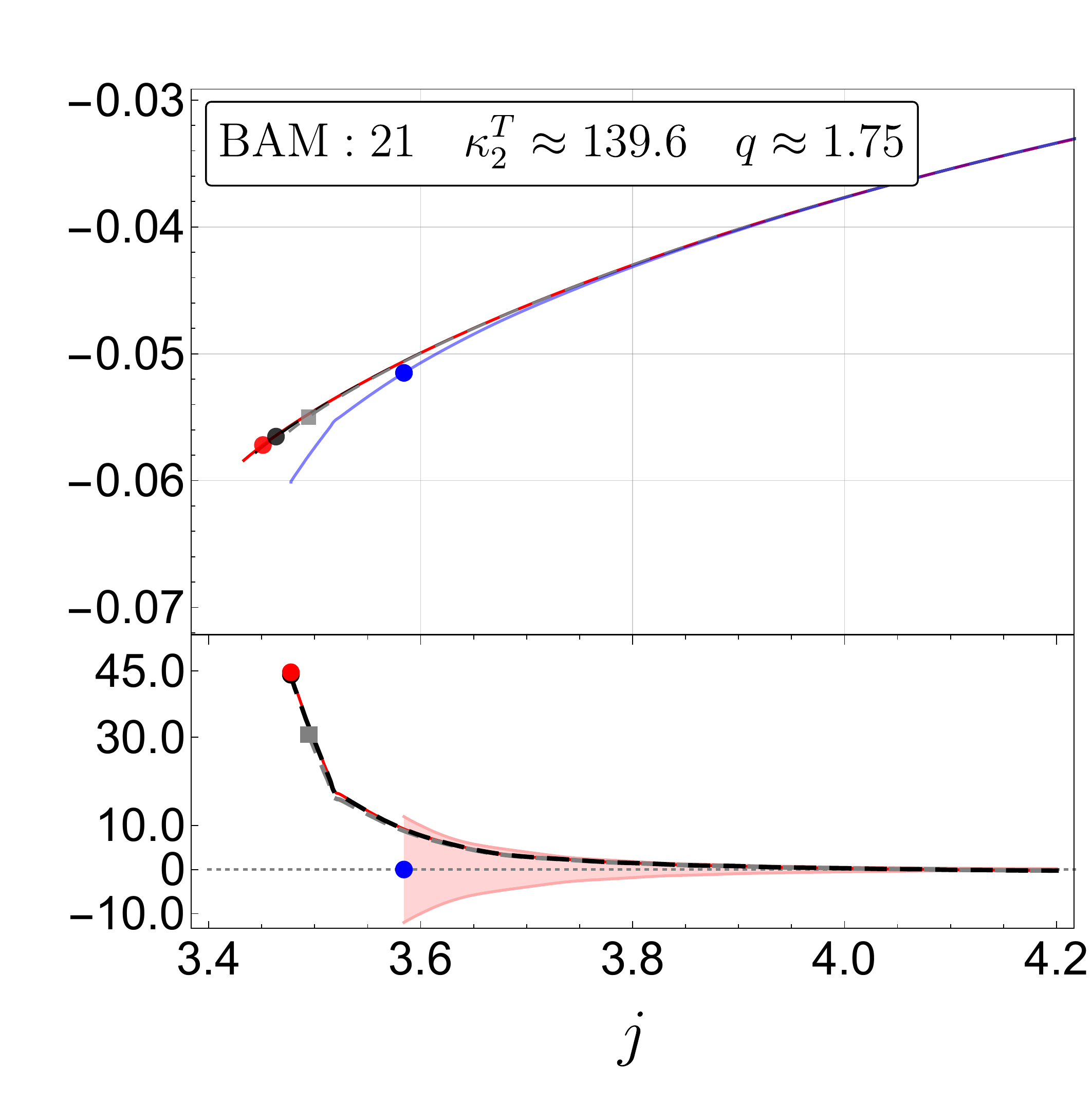}
  \hspace{-2mm}
  \includegraphics[width=0.325\linewidth]{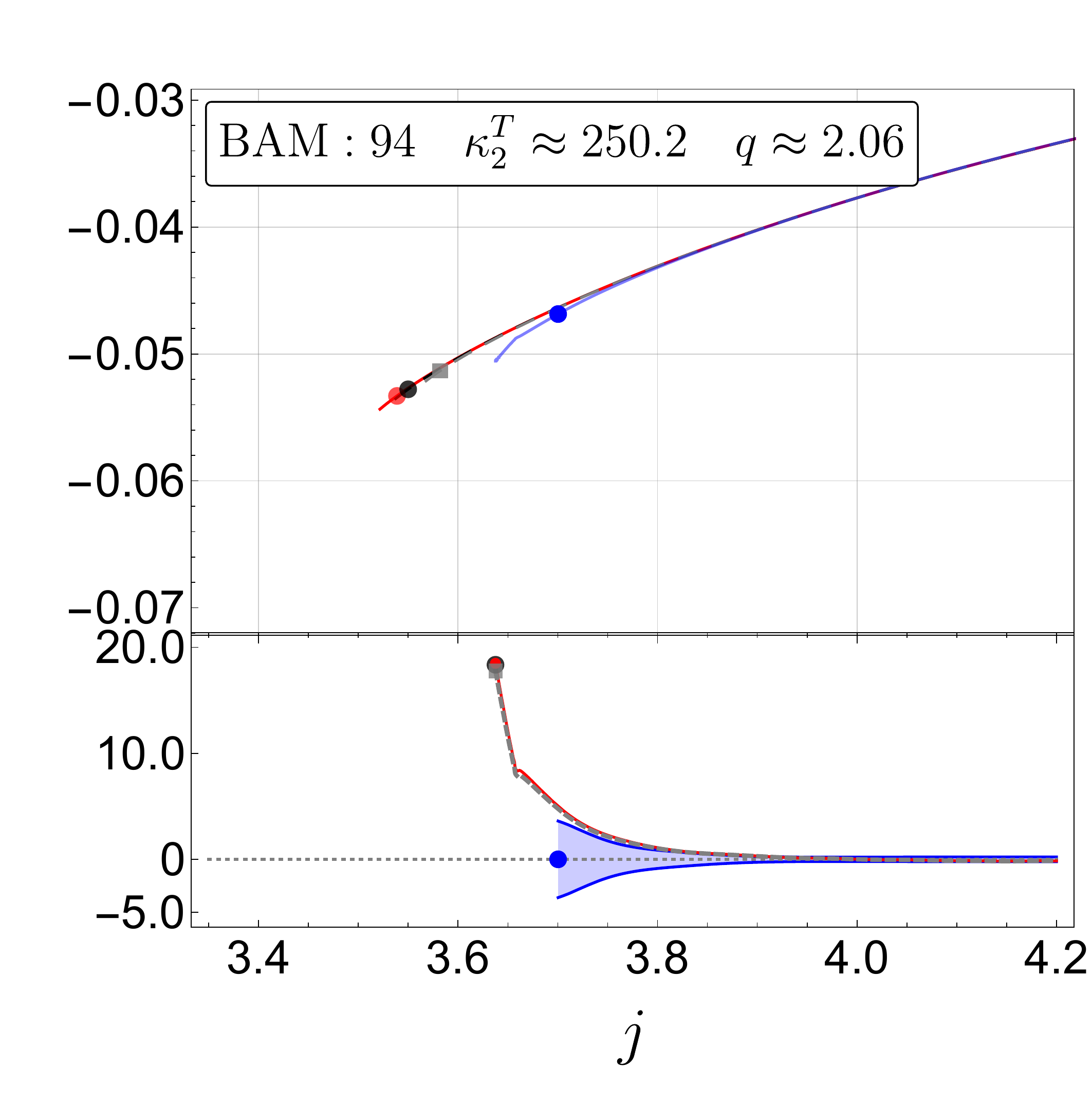}
  \vspace{-4mm}
  \caption{\label{fig:Eb_J_q_ne_q1} Same as Fig.~\ref{fig:Eb_J_q1} but for binaries with $q\gtrsim 1.25$
  with the top panels showing the $q\approx 1.5$ cases.
    The error regions for {\tt BAM:0017, 0021, 0048, 0091, 0127} are less certain thus have been shaded in light pink.}
\end{figure*}

\begin{figure*}[t!]
  \centering
    \includegraphics[width=.48\textwidth]{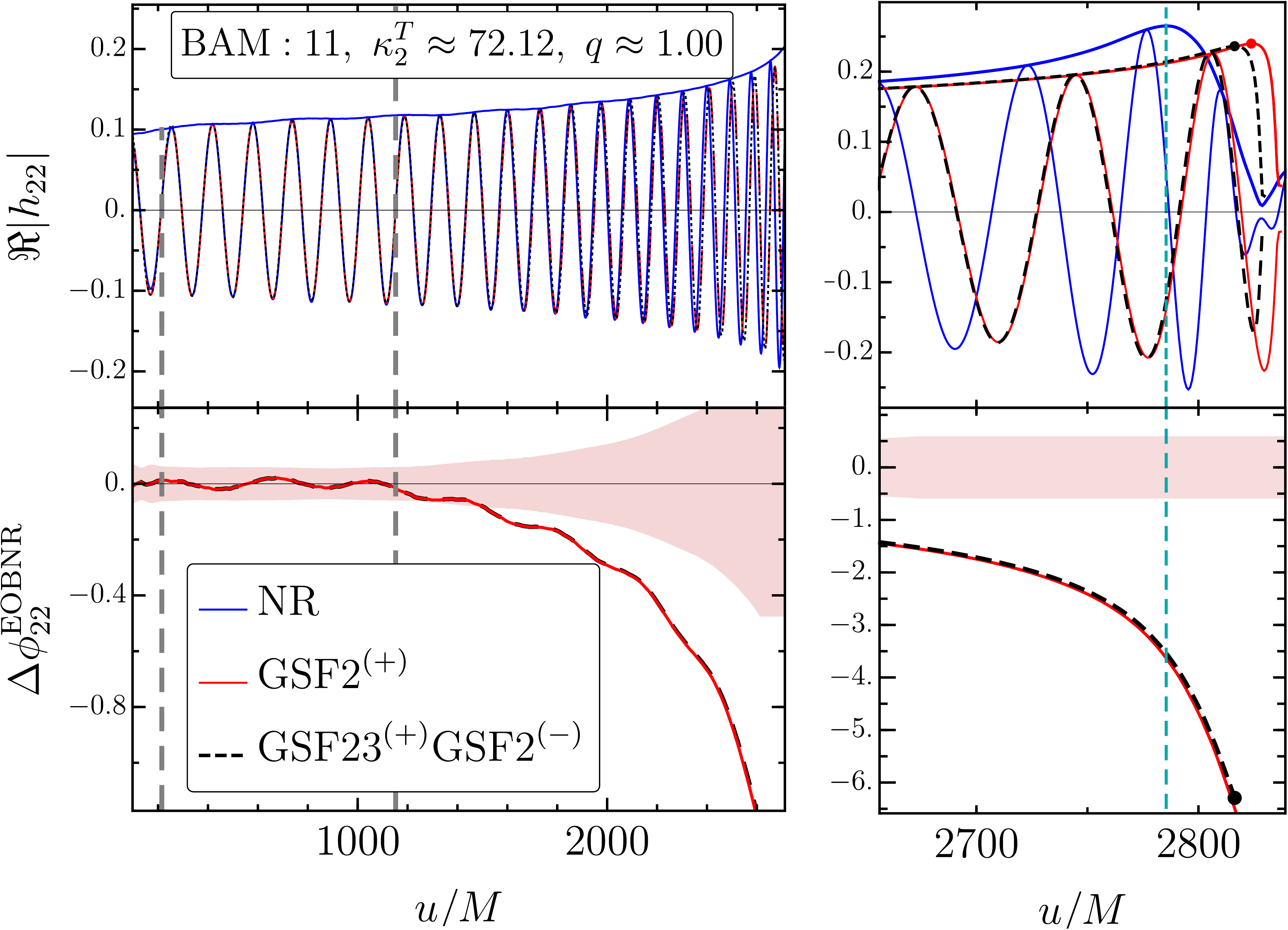} \hspace{0.5cm}
   \includegraphics[width=.48\textwidth]{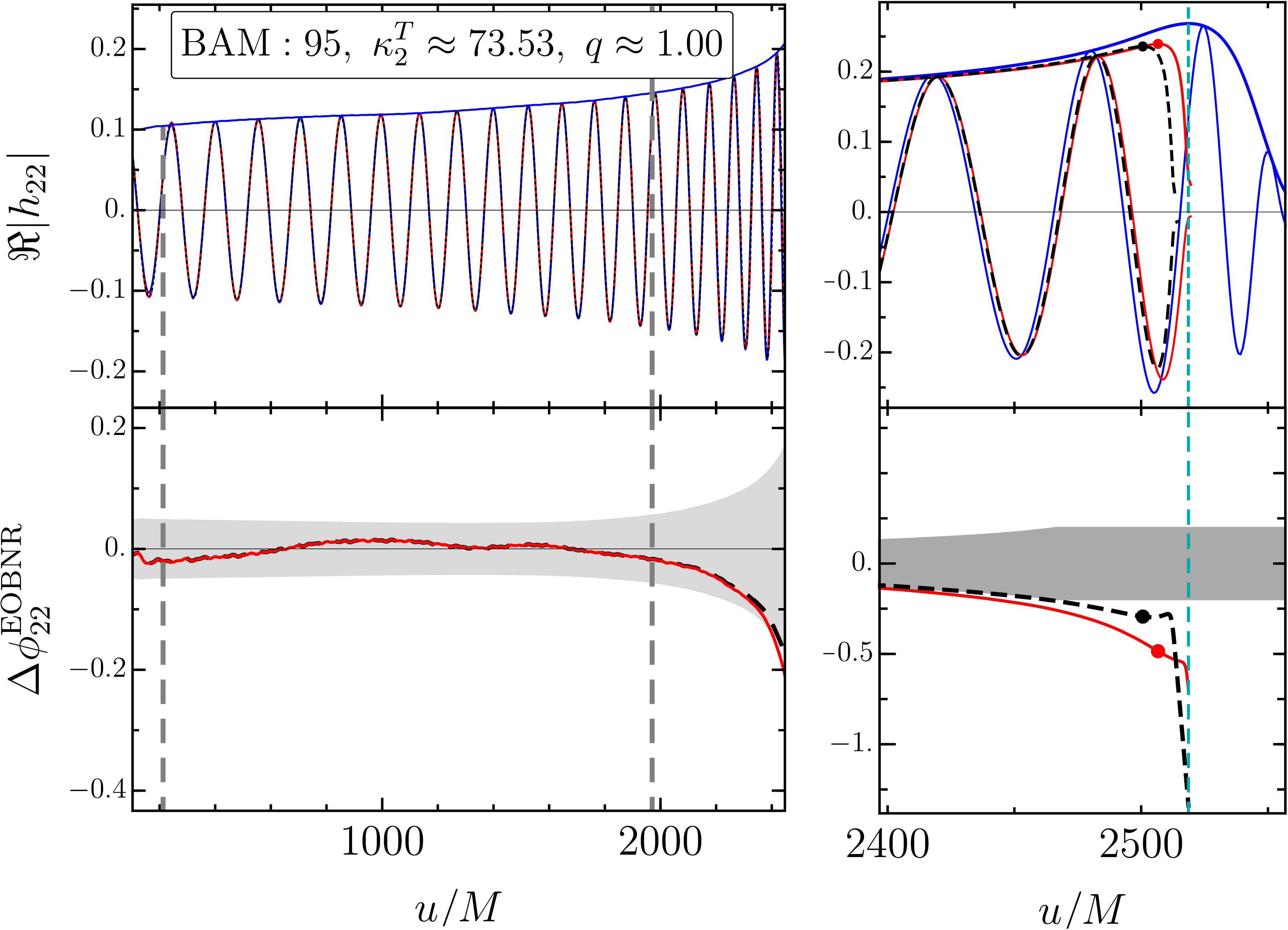}
   \\
   \vspace{1mm}
   \includegraphics[width=.48\textwidth]{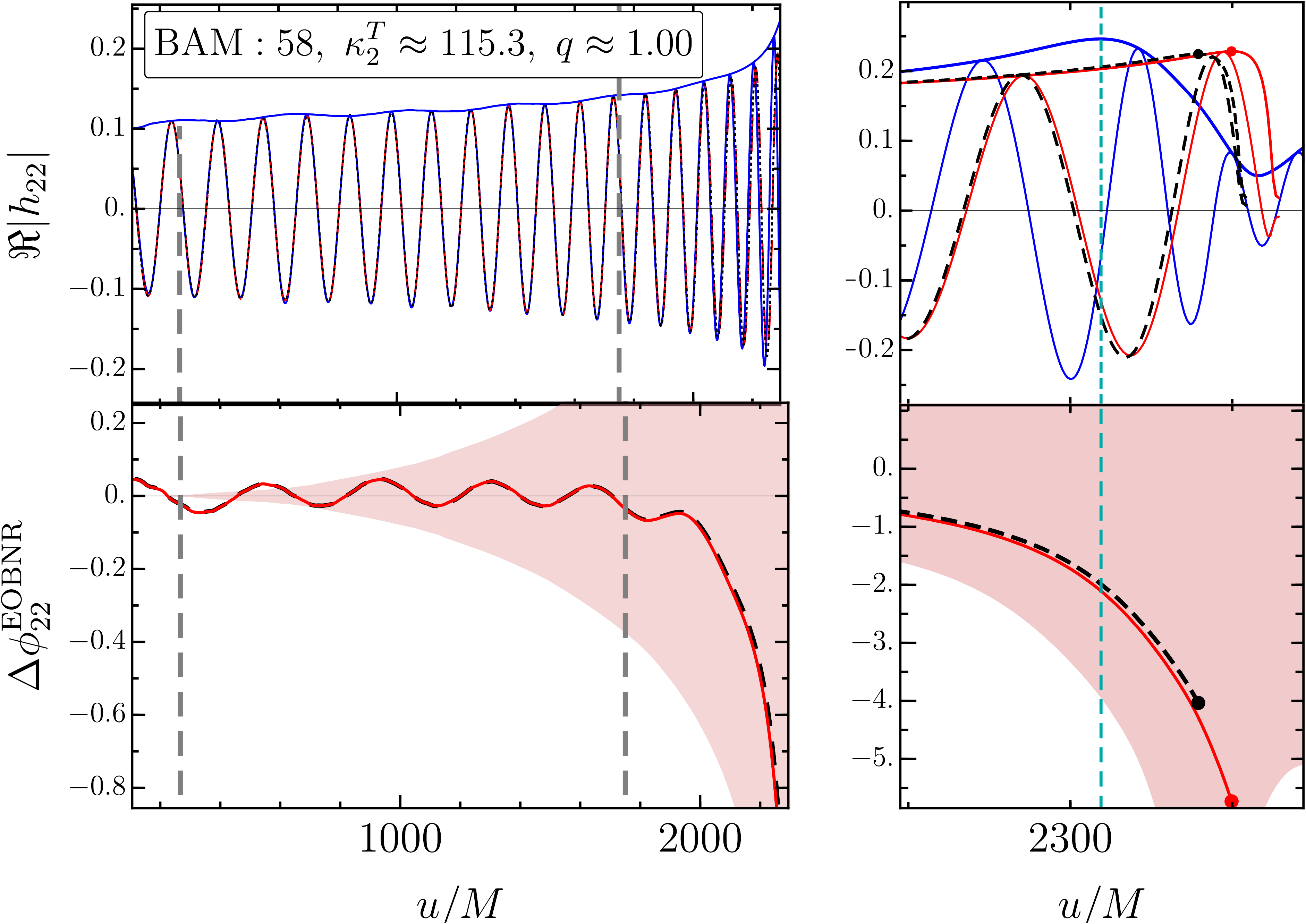} \hspace{0.5cm}  
   \includegraphics[width=.48\textwidth]{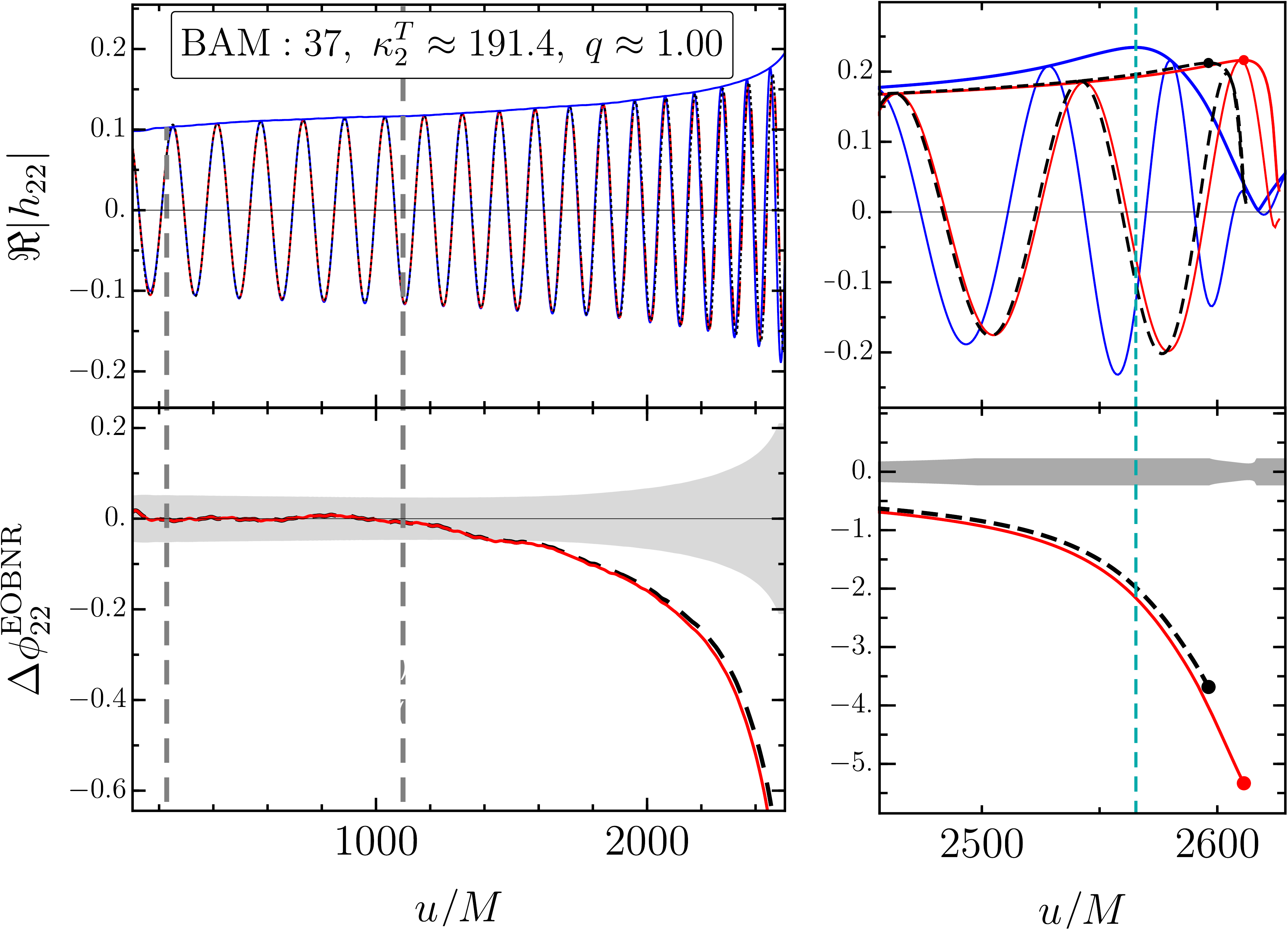}
   \\
   \vspace{1mm}
   \includegraphics[width=.48\textwidth]{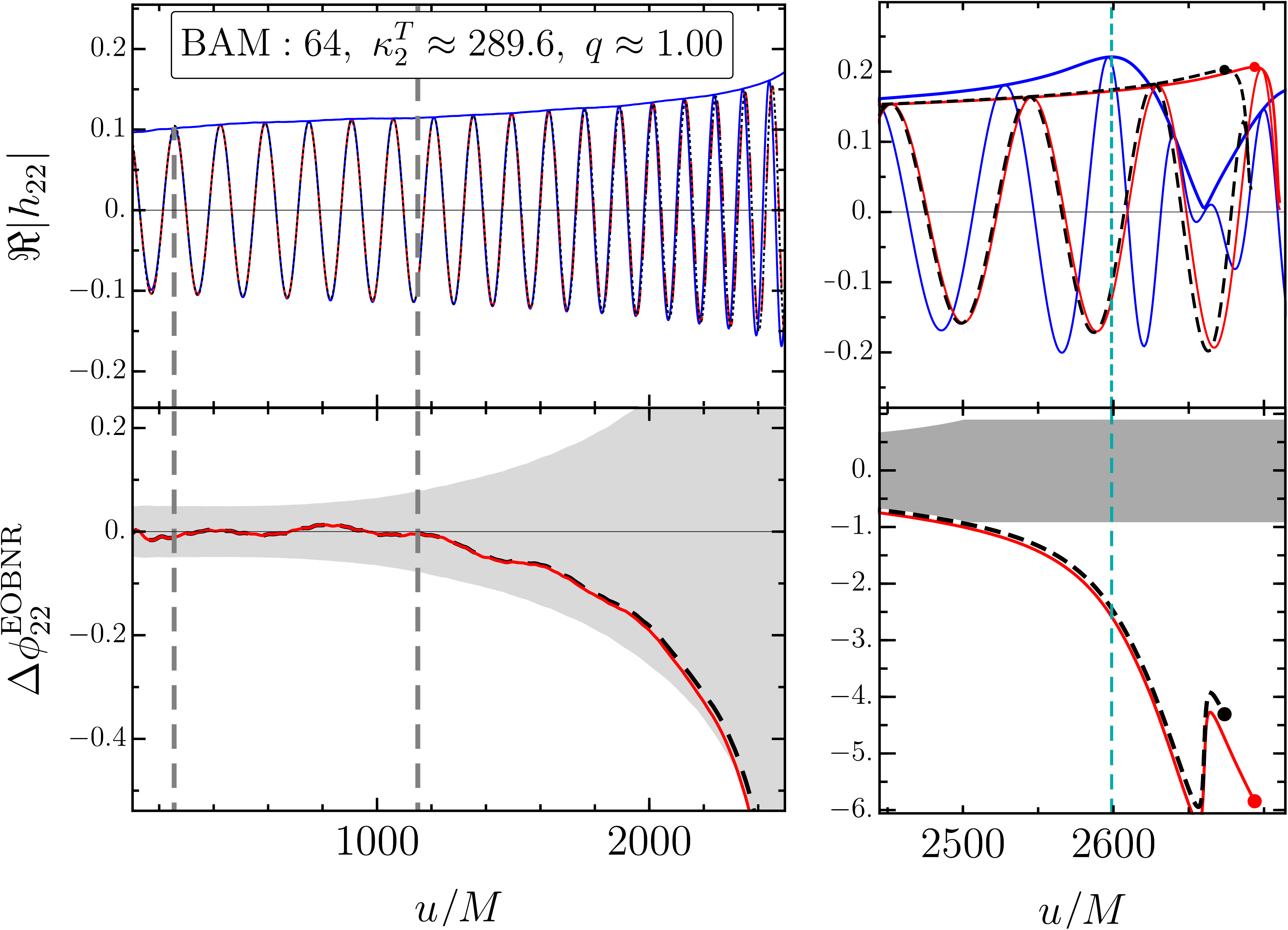} \hspace{0.5cm} 
    \includegraphics[width=.48\textwidth]{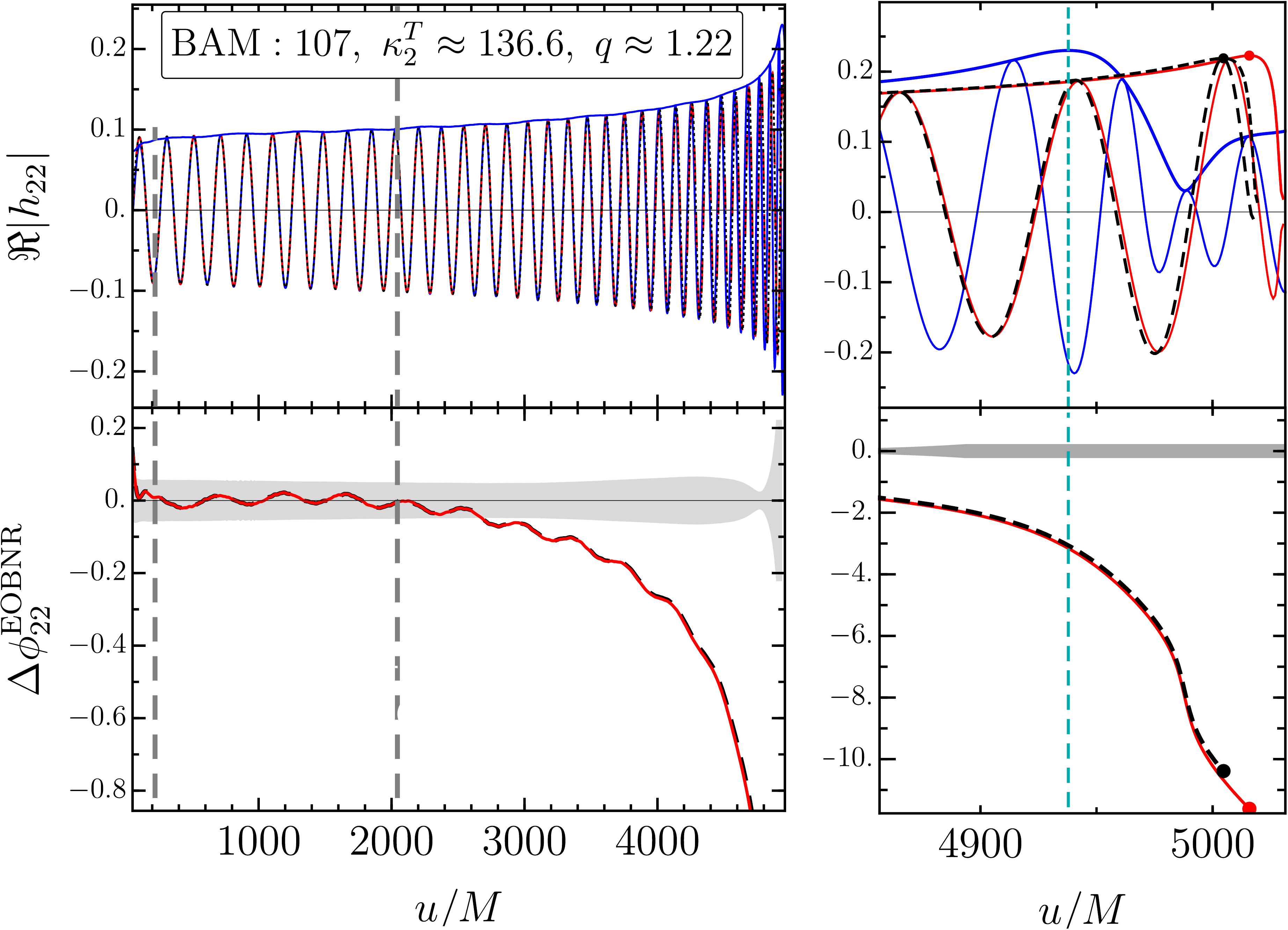}
   \caption{\label{fig:Phasing_fig1} Dephasing between \BAM{} NR simulations and two \TEOBResum{} variants for $q\approx1$ in terms of increasing $\kappa_2^T$. We also included the $q\approx1.224$ \BAMc{\tt 0107} EOBNR comparison here.
   The NR waveforms and amplitudes are plotted as solid blue curves.
   The \TEOBResum{} variants plotted are:~\GSF{2}{+} (red) and~\GSF{23}{+}\GSF{2}{-} (dashed black).
In each subfigure, upper-left panels show the waveforms starting from $\hat{\omega}\sim 0.03-0.04$ corresponding roughly
to ($M_\odot/M$)\,kHz. Upper-right panels show roughly the last cycle before and after the NR merger.
The lower panels display the phase disagreement $\Delta\phi_{22}^\text{EOBNR}\equiv \Delta\phi_{22}^X - \Delta\phi_{22}^\text{NR}$ with $X$ representing the two \TEOBResum{} variants.
The shaded (pink or gray) regions represent our estimated NR phase error.
The vertical cyan dashed lines mark the peak of NR waveform amplitude. The red, green, blue dots respectively represent the same for the three \TEOBResum{} variants listed above.
The vertical, dashed gray lines mark the waveform alignment interval $I_\omega=(\hat\omega_L,\hat\omega_R)$ introduced in Sec.~\ref{sbsec:EOBNR_phasing}.
}
\end{figure*}

\begin{figure*}[t]
\centering
   
   \includegraphics[width=.48\textwidth]{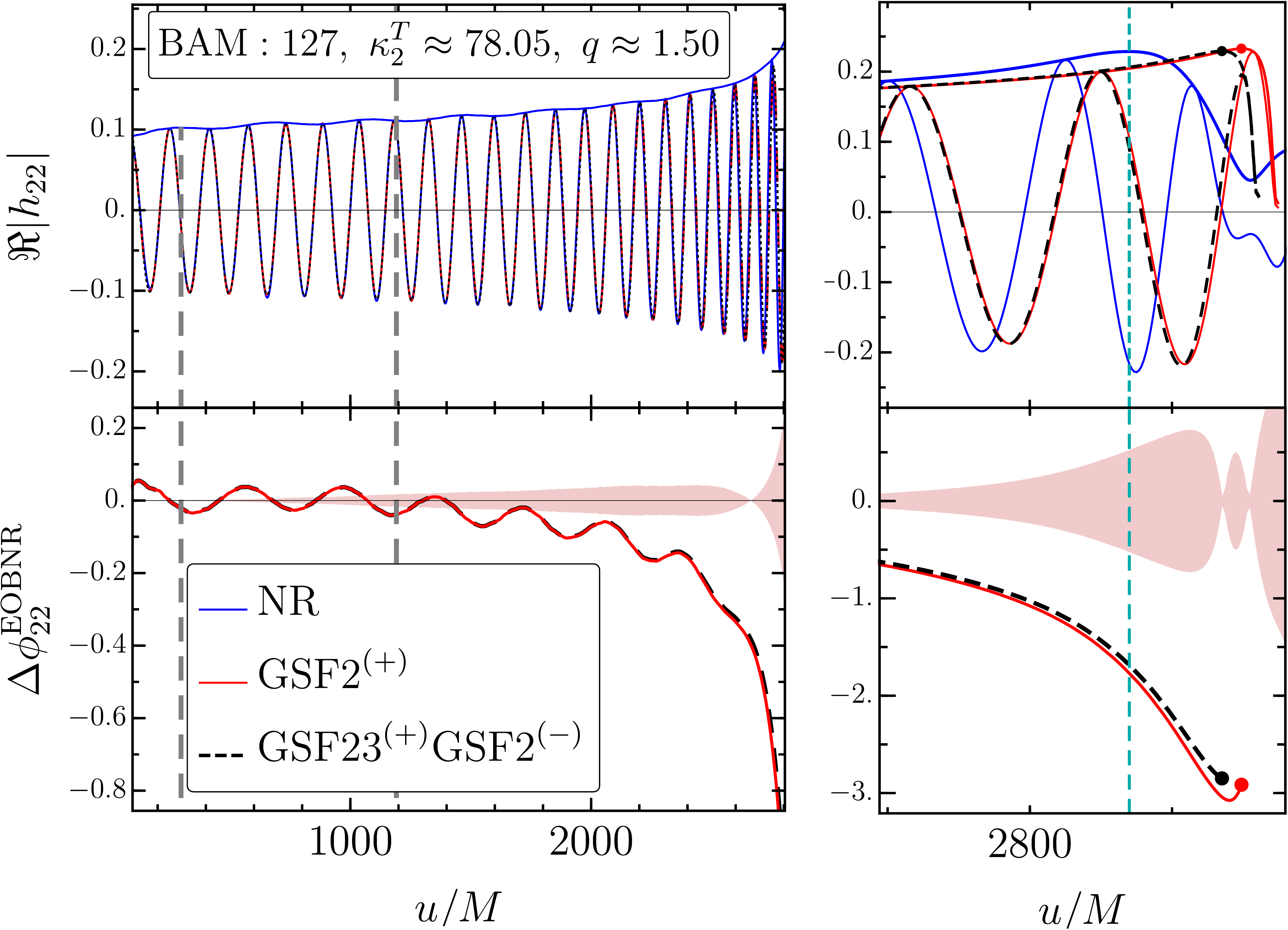} \hspace{0.5cm} 
   \includegraphics[width=.48\textwidth]{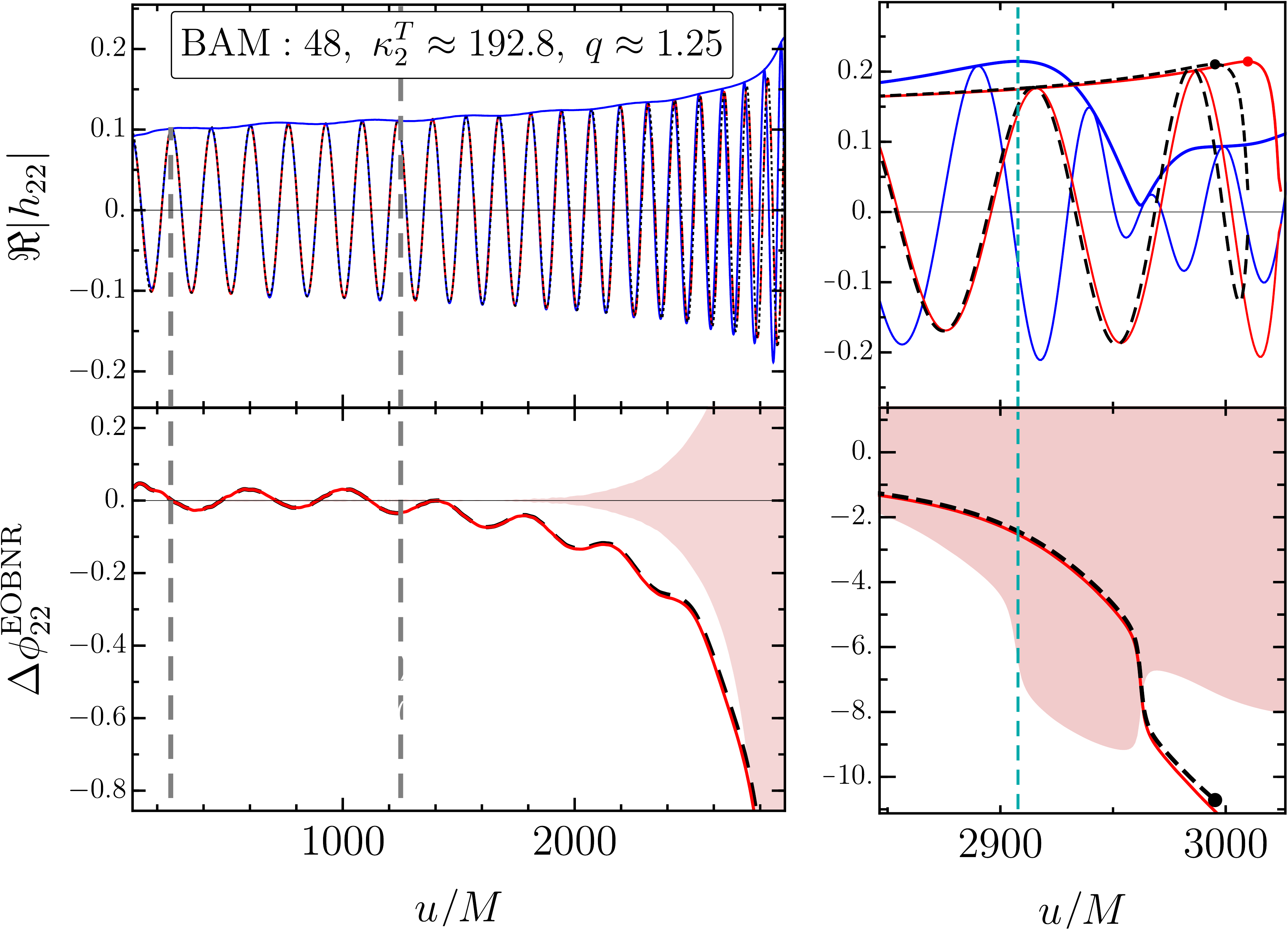}\\
   \vspace{1mm}
   \includegraphics[width=.48\textwidth]{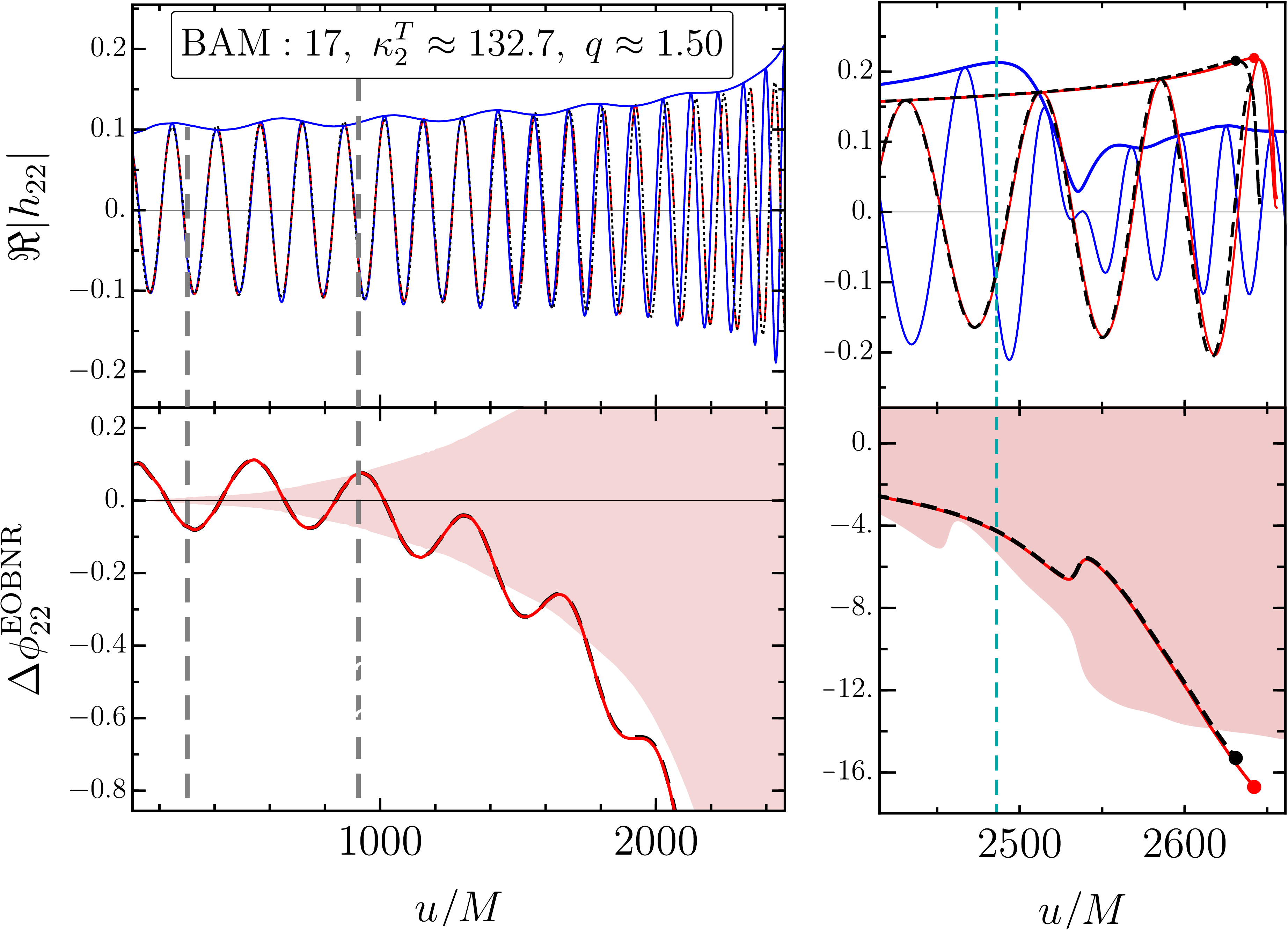}\hspace{0.5cm} 
   \includegraphics[width=.48\textwidth]{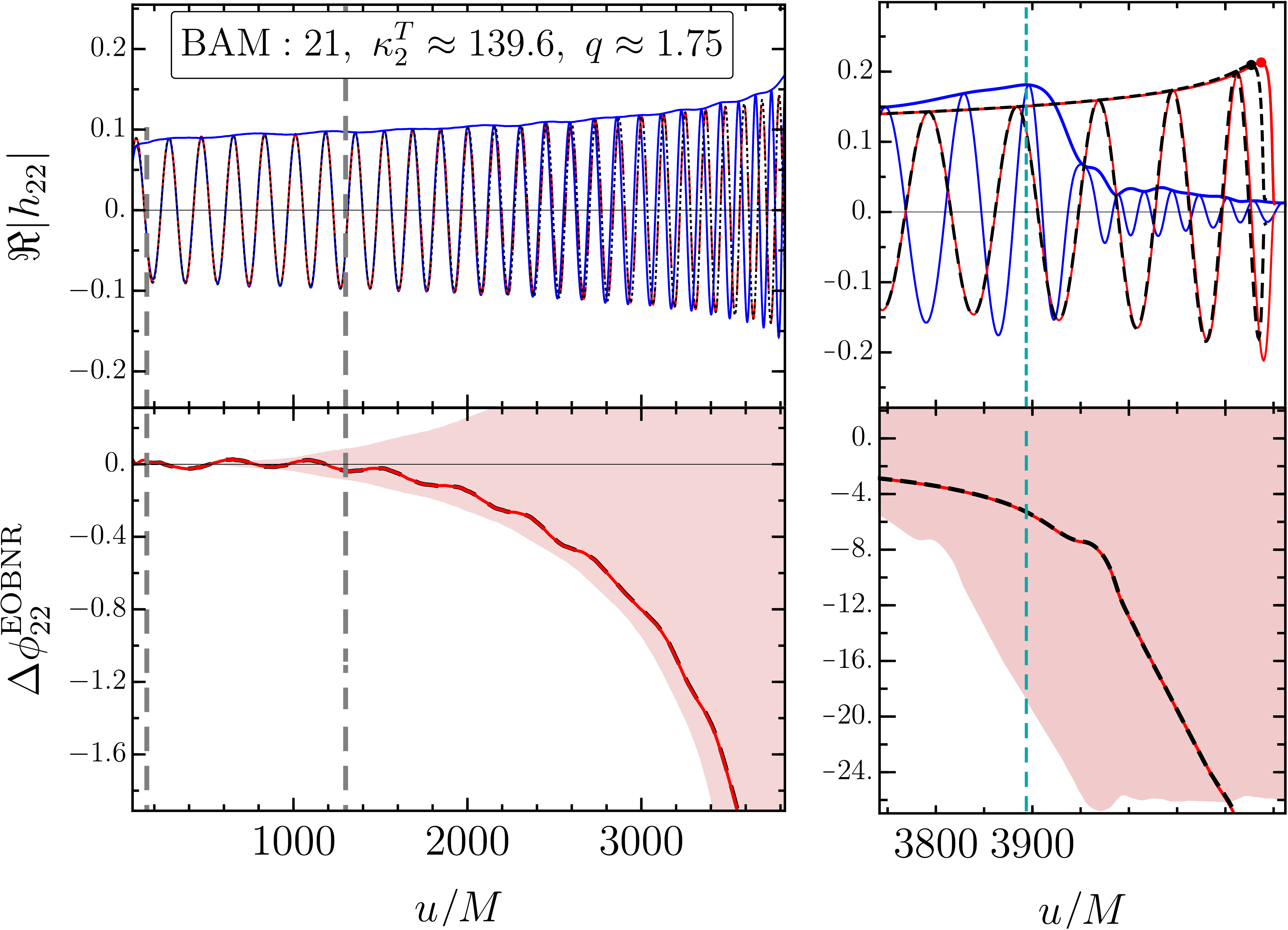}\\
   \vspace{1mm}
   \includegraphics[width=.48\textwidth]{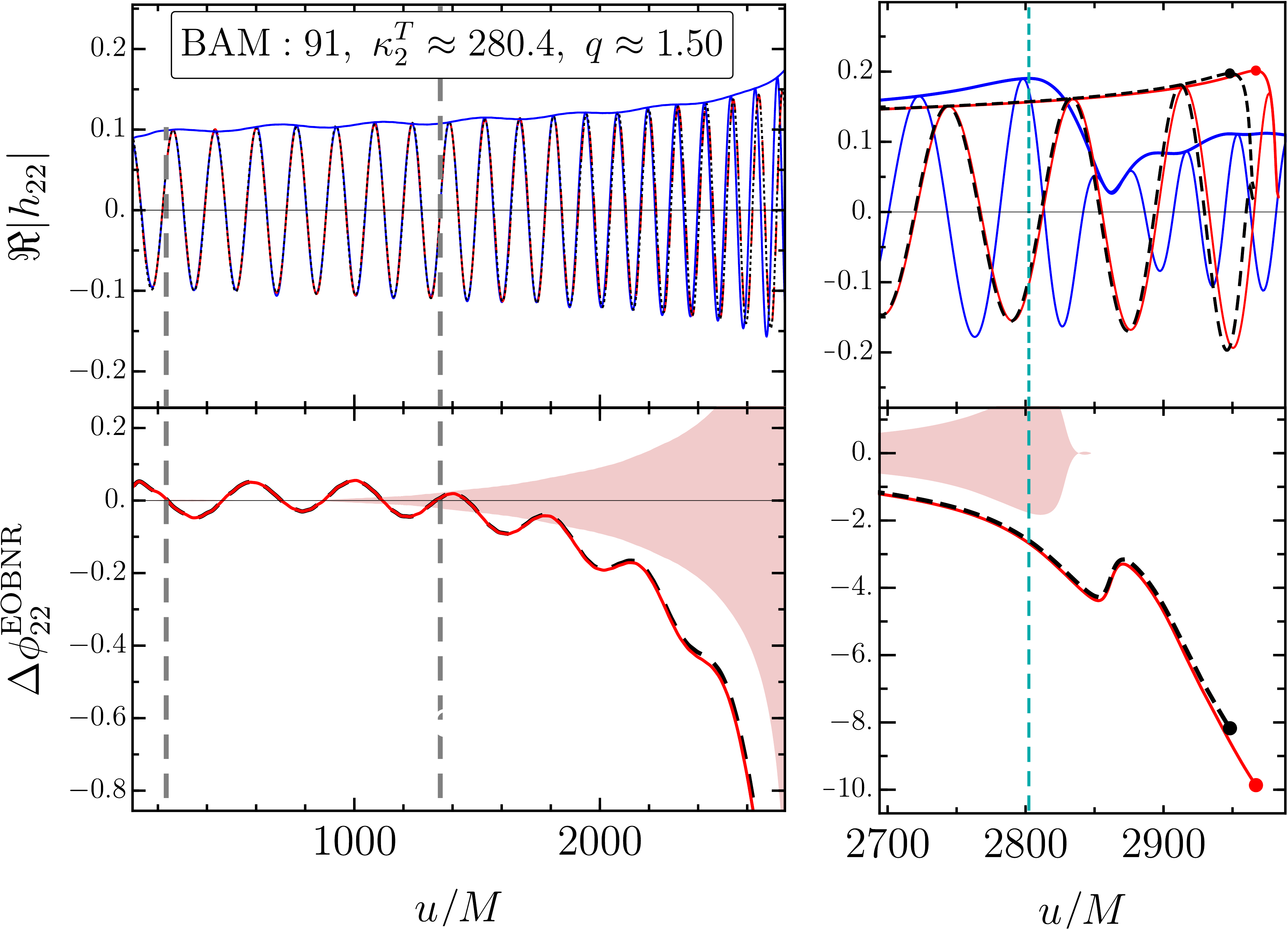}\hspace{0.5cm} 
   \includegraphics[width=.48\textwidth]{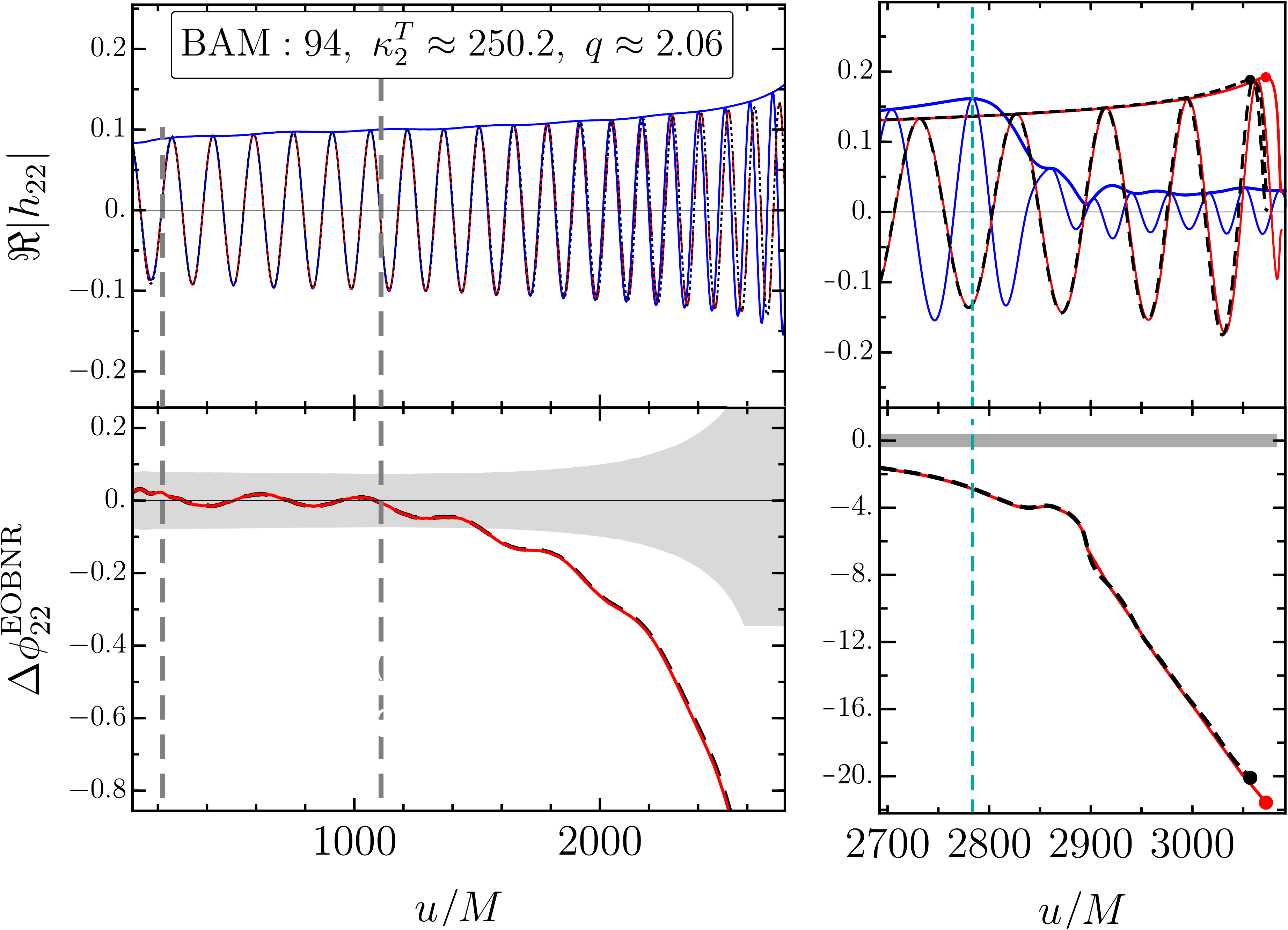}
\caption{\label{fig:Phasing_fig2} Same as Fig.~\ref{fig:Phasing_fig1}, but for the $q\gtrsim1.25$ cases. 
The figures in the left column correspond to $q\approx 1.50$ in terms of $\kappa_2^T$ increasing downward.
See the caption of Fig.~\ref{fig:Phasing_fig1} for details.}
\end{figure*}

\begin{figure*}[t]
  \centering
    %
  \includegraphics[width=0.325\textwidth]{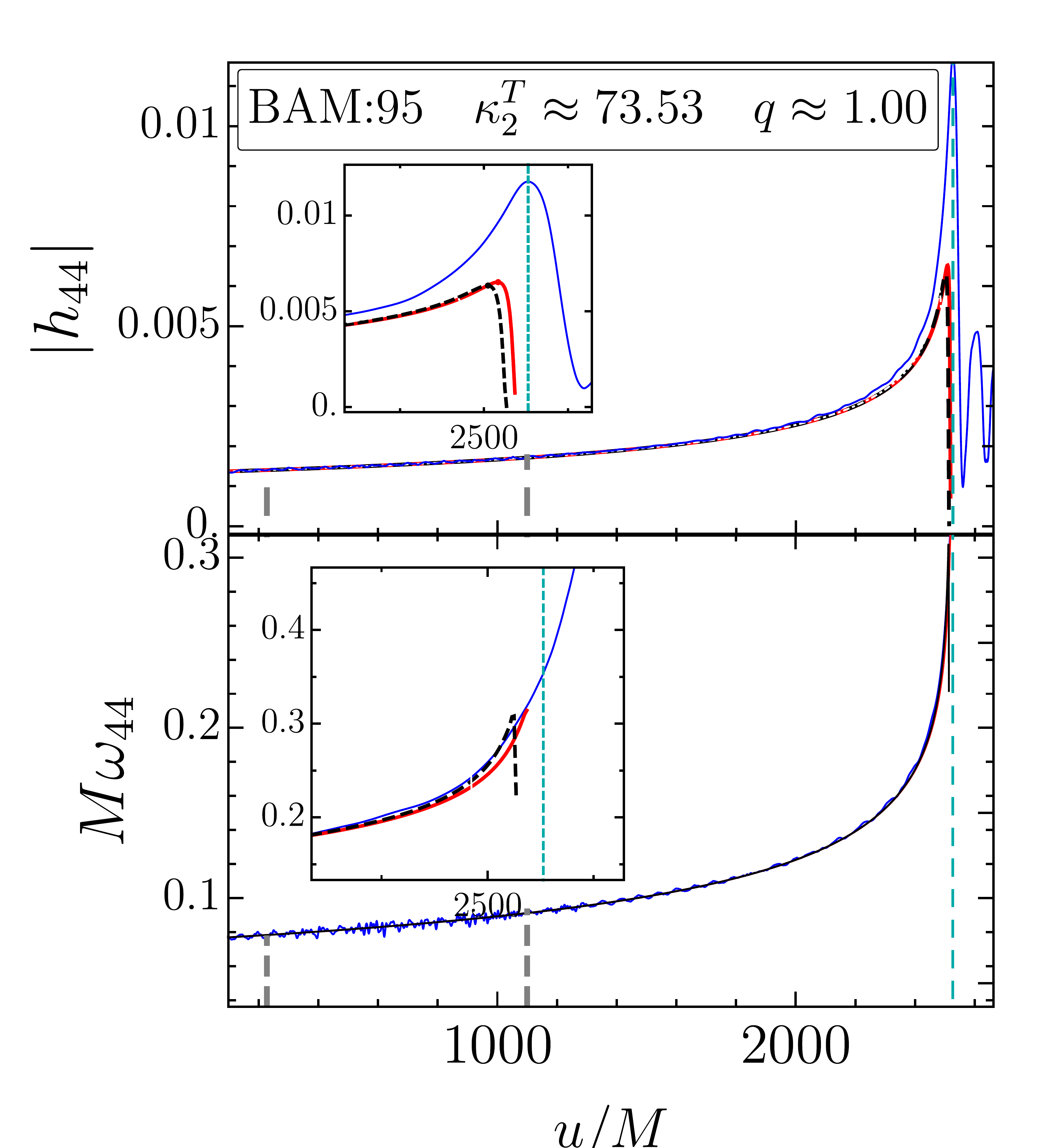}
  \includegraphics[width=0.325\textwidth]{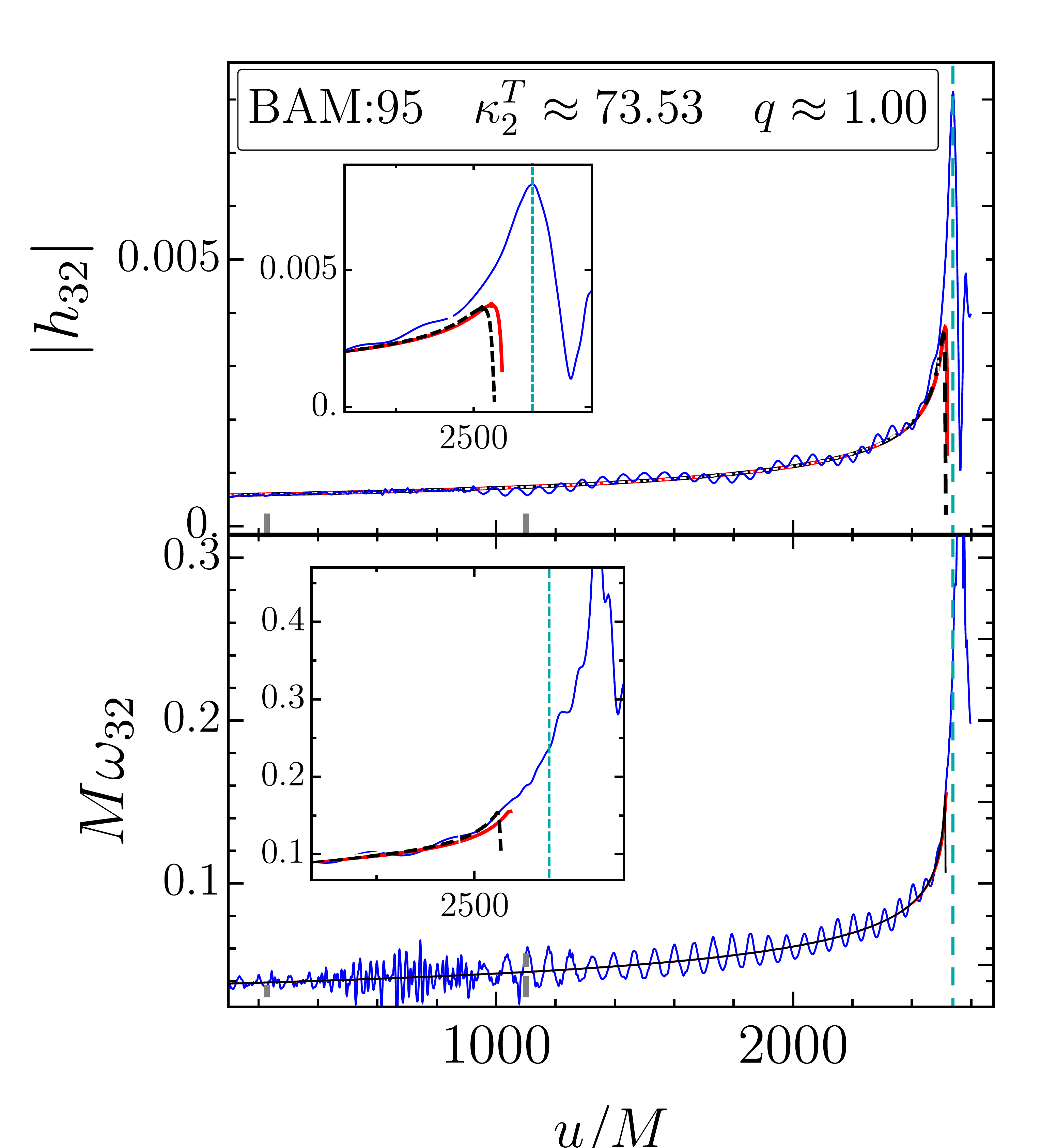}
  \includegraphics[width=0.325\textwidth]{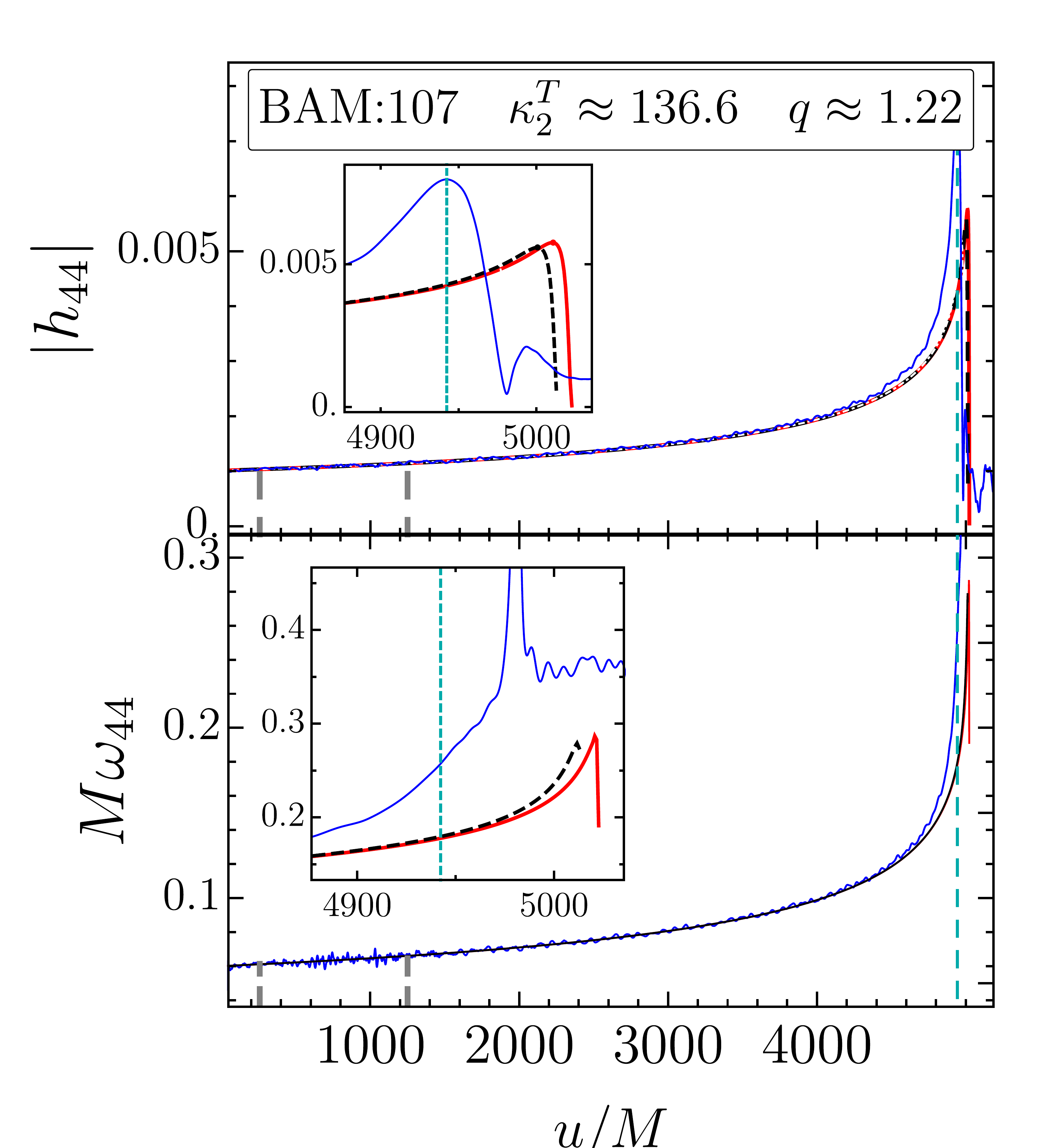}  \\
  \includegraphics[width=0.325\textwidth]{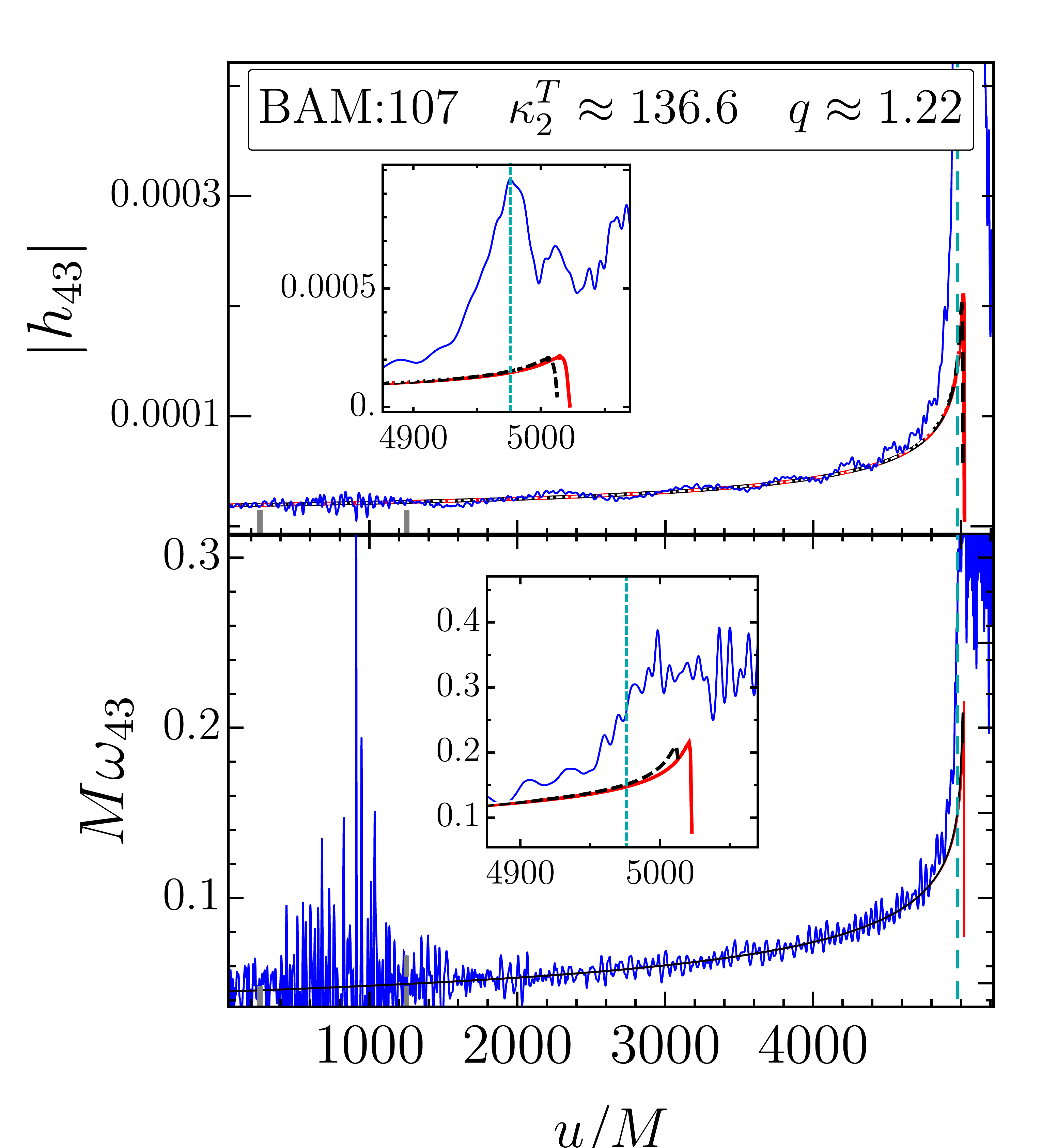}
  \includegraphics[width=0.325\textwidth]{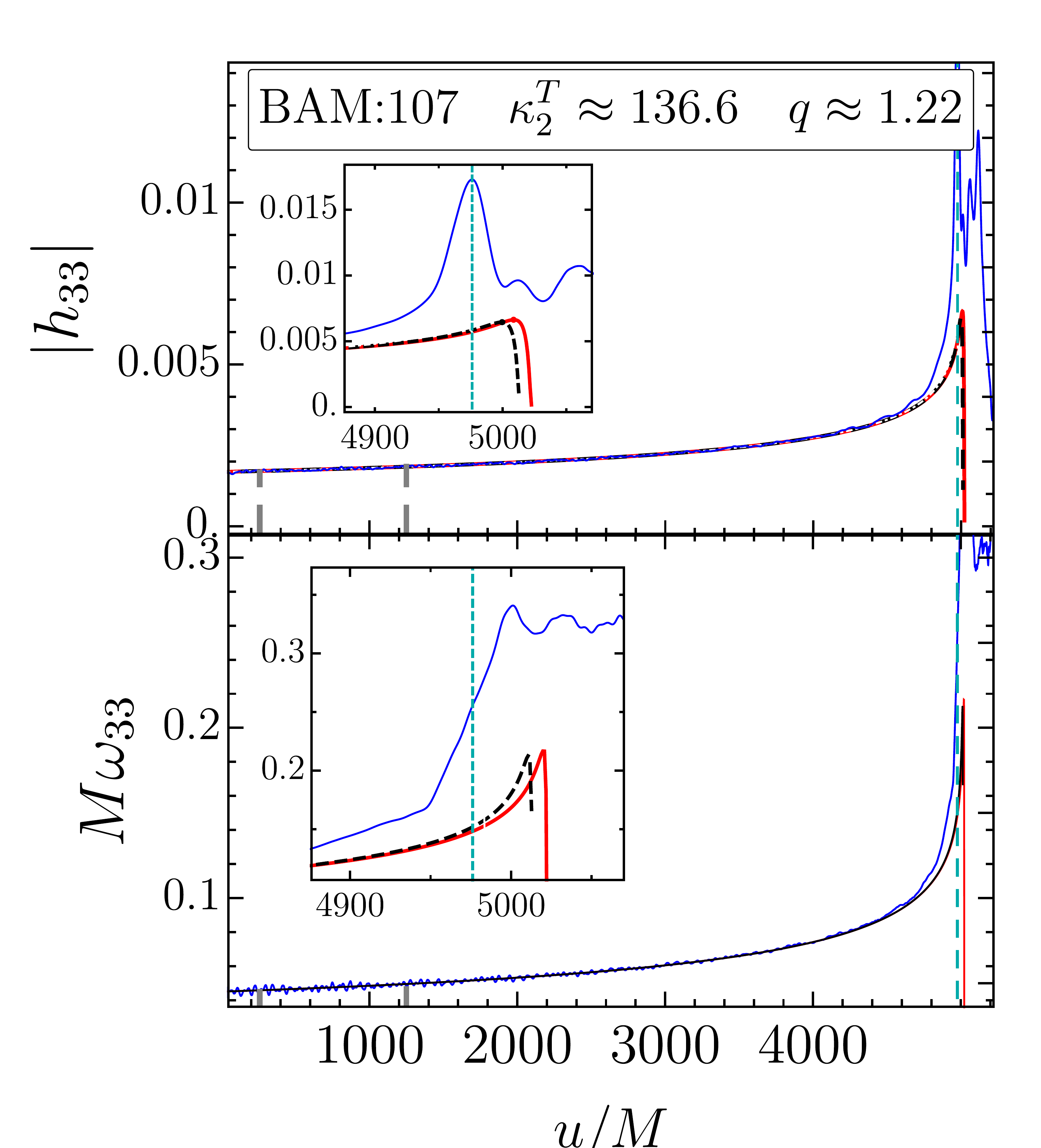}
  \includegraphics[width=0.32\textwidth]{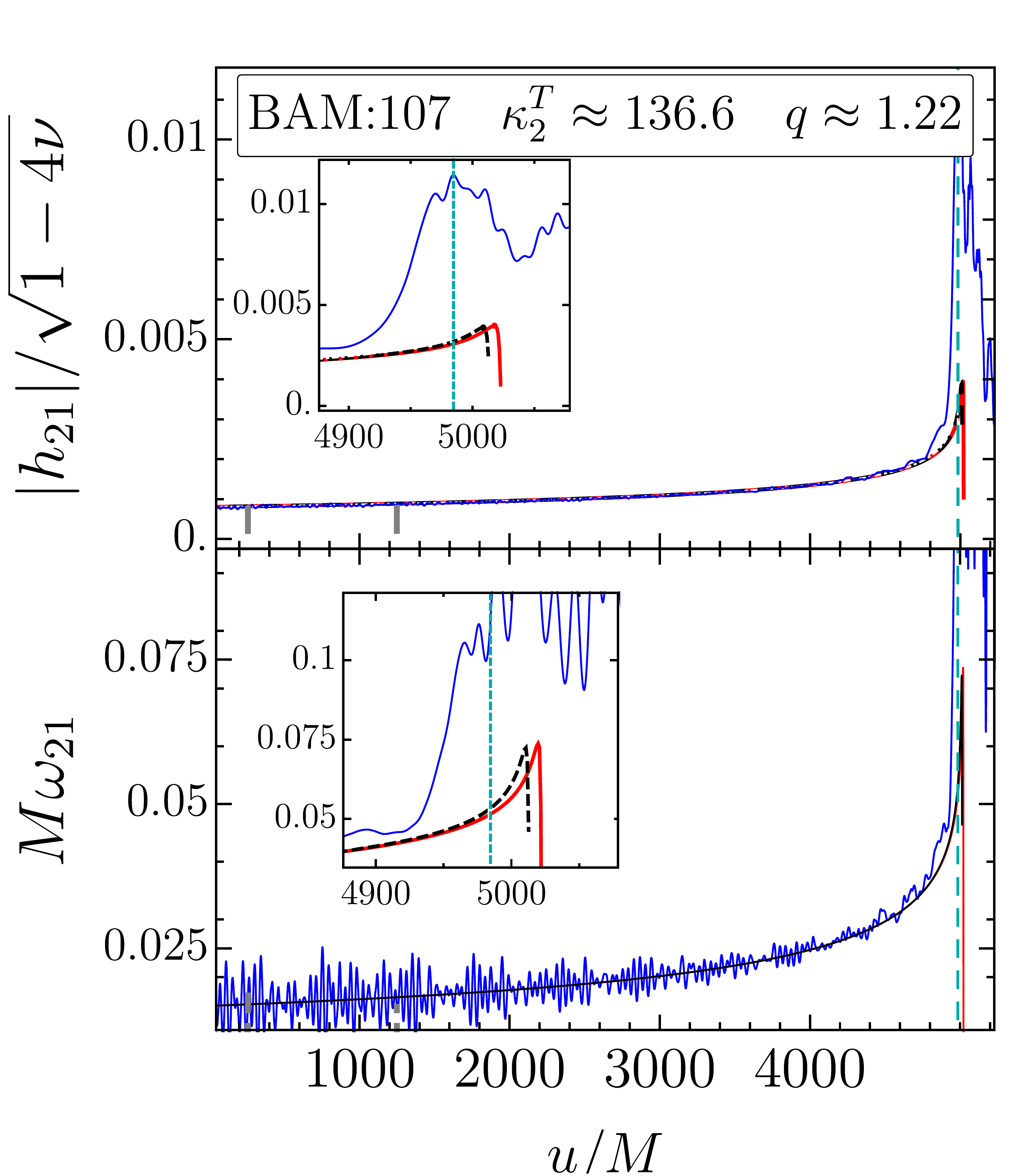}

    \caption{\label{fig:EOBNR_phase_higher_modes} 
    EOB-NR comparisons for higher multipoles. See Fig.~\ref{fig:Phasing_fig1} for legend.
    Upper panels of each subfigure show the amplitudes of the aligned EOB variants against the corresponding NR waveform amplitude. 
    The lower panels show the dimensionless frequencies $M \omega_{\ell m}$ from the NR and the EOB data.
    The insets show data near the NR merger.
    For modes with the smallest amplitudes, the noise in the NR data is clearly visible.
   }
\end{figure*}

We assess the new analytical results against NR data from the
public database\footnote{\url{www.computational-relativity.org}.} 
of the {\tt CoRe} collaboration~\cite{Dietrich:2018phi}. 
The employed datasets are summarized in Table~\ref{table:BAM_runs} and
cover a relevant range of EOS, masses and mass ratios.
Most of the NR data consist of the eccentricity-reduced,
error-controlled waveforms computed in Ref.~\cite{Dietrich:2017aum}. 
Note that much of the data employed here are of higher quality than those
employed in Ref.~\cite{Bernuzzi:2014owa} to verify the performance of
\TEOBResum{}\footnote{In particular, Ref.~\cite{Bernuzzi:2014owa}
  compared \GSF{2}{+}nm with \PN{+}.}. As a consequence, the newest data
enable a more detailed assessment of the analytical EOB model than
previously done.  
We also include simulations from Refs.~\cite{Dietrich:2015pxa,Dietrich:2016hky,Dietrich:2017feu} in order to
explore mass ratios significantly different from $q=1$. 
Roughly half of our chosen data sets show clear convergence with grid resolution
and allow us to compute consistently the error budget~\cite{Bernuzzi:2011aq,Bernuzzi:2016pie}.
The rest, specifically 
the {\tt BAM:0011, 0017, 0021, 0048, 0058, 0091, 0127} runs, 
do not show robust convergence and do not allow us to
compute consistent error bars for the phase.
Following the above references, the error from these data sets
is estimated as the difference of the two highest resolutions
and shown using pink shading in Figs.~\ref{fig:Eb_J_q1}~through~\ref{fig:Phasing_fig2}.
Therefore, the comparisons with these
data sets cannot be considered conclusive. 
Nonetheless, we present EOB-NR comparisons for all twelve cases 
with the double aim of (i) suggesting possible limitations of the analytical model and
(ii) indicating a possible direction for improving current NR simulations.
\subsection{Energetics}
\label{sbsec:Eb_vs_J}

EOB and NR dynamics are compared by considering the gauge-invariant
relation between binding energy per reduced mass, $E_b=(E-M)/(M\nu)$
and orbital angular momentum $j\equiv p_\varphi$~\cite{Damour:2011fu,Bernuzzi:2012ci,Bernuzzi:2014owa}.
We recall that in the EOB case $E$ is just the Hamiltonian function
computed along the EOB dynamics. For the NR configurations, $(E^{\rm NR},j^{\rm NR})$ 
is obtained as detailed in 
\cite{Damour:2011fu,Bernuzzi:2012ci}.
Figure~\ref{fig:Eb_J_q1} collects several $q=1$ configurations with
increasingly larger tidal interaction, as well as a $q=1.22$ case.
We display $1.25 \lesssim q \lesssim 2$ cases in Fig.~\ref{fig:Eb_J_q_ne_q1}.
In each subfigure, the bottom panels show $\Delta E_b^\text{EOBNR}\equiv E_b^\text{EOB}-E_b^\text{NR}$
with the shaded region representing our estimated NR error. We recall that
the blue-shaded regions come from convergent simulations, while the pink-shaded
regions are obtained as difference between the two highest resolutions.
For the purposes of relating our results to that of Ref.~\cite{Nagar:2018zoe},
we show \GSF{2}{+}nm as the solid red curves. On both the NR and EOB curves,
the markers indicate the conventional {\it merger points}, i.e., the values
corresponding to the peak of the amplitude of the $\ell=m=2$ waveforms.
The black dashed curves terminating at the black dots represent \GSF{23}{+}\GSF{2}{-},
while the dashed gray curves terminating at the gray squares represent \GSFp{23}{+}\GSFp{2}{-}
to illustrate the sensitivity of this quantity on the choice of the
value of the exponent $p$.
%
%

The performances of the analytical models are in broad agreement with NR within
their errors, but agreement in the $j$-interval corresponding to the last few
cycles up to merger depends on the value of $\kappa_2^T$.
Let us focus first on Fig.~\ref{fig:Eb_J_q1}. As a general statement, the, already
good, EOB/NR agreement yielded by the \GSF{2}{+} model is even improved when
the  $(3+)$-GSF resummed physical information is considered, i.e. with
the \GSF{23}{+}\GSF{2}{-} variant. The latter predicts a conventional merger
point occurring at slightly lower values of $j$ than the previous case though,
especially when $\kappa_2^T$ is increased, it gets closer to the NR prediction. This
seems to be a robust conclusion driven by inspecting the lower panels of Fig.~\ref{fig:Eb_J_q1},
where it was possible to obtain robust error bars for NR run. Same conclusion holds true,
for the same configurations, for the values of $E_b^{\rm mrg}$ (see especially {\tt BAM:0064} and {\tt BAM:0107}).
By contrast, the variant \GSFp{23}{+}\GSFp{2}{-} systematically predicts values of
the angular momentum at conventional merger that are systematically larger than the
NR ones.

Figure~\ref{fig:Eb_J_q1} ($q\approx1$) shows that our $p=4$
GSF-resummed tidal models go from slightly overestimating the tidal
interaction to slightly underestimating it as 
$\kappa_2^T$ grows. This means that there is a certain region, 
$100\lesssim \kappa_2^T\lesssim 200$, where the energetics yielded by
\GSF{23}{+} agrees rather well with the NR one.
This region corresponds to moderately stiff EOS with $500\lesssim \Lambda \lesssim 1000$
which translates to ${500\lesssim \tilde\Lambda \lesssim 2200}$ using, e.g., the
low-spin prior inferred mass ratio, $q\in [1,1.37]$ of GW170817~\cite{Abbott:2018wiz}.
Our region has some overlap with the LIGO-Virgo constraint
of $\Lambda \lesssim 800$~\cite{TheLIGOScientific:2017qsa, Abbott:2018exr, De:2018uhw, Abbott:2018wiz}
and the one from electromagnetic counterpart, $\tilde\Lambda \gtrsim 400$~\cite{Radice:2017lry}. 

We see a similar pattern in the top panels of Fig.~\ref{fig:Eb_J_q_ne_q1} corresponding to $q\approx1.5$
where the \TEOBResum{} models overshoot the NR merger with increasing $\kappa_2^T$.
However, as was the case with $q=1$, there might be a similar region of good agreement, but for $\kappa_2^T\lesssim 80$.
It seems that \GSFp{23}{+}\GSFp{2}{-} may be the most suitable model for when $q\gtrsim 1.25$.
This could be indicative of this 
variant effectively accounting for the increased NS deformability of the $q>1$ situations. 
In order to draw more definitive conclusions, we require a larger set of NR data with robust errors.
A good agreement between energetics should probably be obtained with a value of 
$p\sim 9/2$ for large values of $\kappa_2^T$ and slightly smaller than 4 for smaller value 
of $\kappa_2^T$. Since a meaningful assessment of the effective value of $p$ would require more error-controlled NR simulations,
we leave such exploration to future work.

%
%
%
%
%
%
%
\subsection{GW Phasing}
\label{sbsec:EOBNR_phasing}
We compare the EOB and NR multipolar waveforms by using a standard
(time and phase) alignment procedure in the time domain ~\cite{Baiotti:2010xh}. Relative time and phase shifts are determined by
minimizing the $L^2$ distance between the EOB and NR phases integrated on a time interval corresponding to the dimensionless frequency interval 
$I_\omega= (\hat\omega_L,\hat\omega_R)\approx (0.04, 0.06)$. 
Such a choice for $I_\omega$ allows one to average out the phase
oscillations linked to the residual eccentricity.
As a consistency check, we employed two separate codes using different alignment routines.
The waveforms we show in Figs.~\ref{fig:Phasing_fig1}, \ref{fig:Phasing_fig2} were agreed on by both codes.

In Fig.~\ref{fig:Phasing_fig1} we show several EOB $(2,2)$ waveforms aligned with NR ones for five $q\approx 1$ cases along with $q\approx 1.22$. For this comparison, we opted to include the following \TEOBResum{} variants:~\GSF{2}{+} (red) and~\GSF{23}{+}\GSF{2}{-} (dashed black).
In each subfigure, the upper-left panels show the waveforms in the late inspiral stage
with the upper-right panels showing the merger and the last few cycles before the merger.
The lower panels display the phase disagreement between EOB and NR defined as $\Delta\phi_{22}^\text{EOBNR}\equiv \Delta\phi_{22}^X - \Delta\phi_{22}^\text{NR}$ with $X$ representing the different \TEOBResum{} variants.
The shaded (gray or pink) regions represent our estimated NR phase error.

Looking at Fig.~\ref{fig:Phasing_fig1}, one notices that \GSF{23}{+}\GSF{2}{-} 
behaves very similar to \GSF{2}{+}, but merges slightly earlier due to
increased tidal attraction.
Within the moderate range of $100\lesssim \kappa_2^T\lesssim 200$, \GSF{23}{+}\GSF{2}{-} runs seem to terminate closer to the NR merger and yield marginally smaller $\Delta\phi_{22}^\text{EOBNR}$ than \GSF{2}{+}.

These trends appear to carry on to the $q\gtrsim 1.25$ cases, albeit with greater $\Delta\phi_{22}^\text{EOBNR}$
as can be seen from Fig.~\ref{fig:Phasing_fig2}.
In all these cases shown, the EOB models seem to overshoot the NR merger indicating that they underestimate
the tidal attraction.
The $q\approx 1.25$ case is consistent with $q\approx 1.22$ case with $\Delta\phi_{22}^\text{EOBNR}\approx -3\,$radians at the NR merger. Additionally, we see in the $q\approx 1.5$ comparisons that
as $\kappa_2^T$ increases, EOB models diverge from the NR phase rather significantly at the merger.
%
Overall, there is an indication that the tides might be stronger for larger $\kappa_2^T$,
which could be mimicked by $p>4 $ as in the model \GSFp{23}{+}\GSFp{2}{-}.

\subsection{GW higher multipoles}
As an additional comparison, we took the best-quality subset of our NR
data, namely $\{\tt{BAM:0037, 0064, 0091, 0094, 0095, 0107} \}$,
and compared the NR waveforms to EOB ones for higher multipolar
modes beyond the quadrupole.
This is an extension of the work of Ref.~\cite{Bernuzzi:2012ci} where
they made one comparison for the $(4,4)$ mode and another for $(3,2)$
in the $q=1$ case. For $q=1$, only $(3,2),(4,2),(4,4)$ modes
are nonzero due to symmetry. 
Figure~\ref{fig:EOBNR_phase_higher_modes} presents the $(3,2)$ and $(4,4)$
modes for  {\tt BAM:0095} ($q=1$) as well as  $(2,1),(3,3),(4,3),(4,4)$
for  {\tt BAM:0107} ($q\approx 1.22$). Among all our NR datasets, we chose,
for illustrative purposes, the two where the most important higher modes
are better resolved. The other NR modes (e.g., the (4,2)) are omitted as
they are too noisy to allow for a meaningful comparison with
the analytical models.
We consider the two main \TEOBResum{} avatars of above, ~\GSF{2}{+}
and \GSF{23}{+}\GSF{2}{-}. The relative time shift used is the one
determined on the $\ell=m=2$ mode as above. For definiteness, the figure only
reports the waveform amplitude and frequency. Note that NR errorbars are
omitted from the plots for clarity. 

For {\tt BAM:0095}, which is probably the most reliable among our
NR simulations, one finds an excellent consistency between both
the EOB and NR amplitude and frequency essentially up to the
conventional merger time as shown by the insets in the top-left panels of
Fig.~\ref{fig:EOBNR_phase_higher_modes} representing the $(4,4)$ and $(3,2)$ modes.

%
%
For these cases, we computed $\Delta\phi^\text{EOBNR}_{\ell m}$ with respect to the NR merger.
The dephasing of the various EOB variants for these modes is roughly consistent with the dephasing
of the $(2,2)$ mode shown in Fig.~\ref{fig:Phasing_fig1} for {\tt BAM:0095, 0107}.
Note that the NR data is somewhat noisy for the $(2,1),(3,2),(4,3)$ modes; 
more accuracy in the NR multipoles would be necessary for further assessments. 
Overall, we find a robust agreement between current NR data and EOB
waveforms up until the last few cycles before the merger, that 
corresponds to the GW frequencies currently observed.
Additionally, despite the noise in the NR data, the various EOB waveforms are consistent with the NR ones, 
thus deliver a reliable description of the multipolar amplitudes up to a few orbits before the merger.

\section{Conclusions}
\label{sec:end}


In this article, we have investigated analytical improvements to the
tidal sector of \TEOBResum{} \cite{Bernuzzi:2014owa,Nagar:2018zoe} for
the description of quasicircular binary neutron star waveforms valid up to
merger. Our main findings are summarized in the following.

\paragraph*{New resummed gravitoelectric terms in the EOB $A$ potential.}
The GSF-resummation of the leading order (LO) gravitoelectric $\ell=3$ term in the
tidal EOB potential gives the largest effect on the GW phasing. For
various binaries, the dephasing accumulated from $10\,$Hz is 
$-\Delta\phi_{22}\sim $ {0.5-3}~radians.

\paragraph*{New resummed gravitomagnetic terms in the $A$ potential.}
The $\ell=2$ LO gravitomagnetic, either in PN or GSF-resummed
form, term gives a smaller contribution to the GW phasing than the $\ell=3$
gravitoelectric term. In the most relevant case (stiff EOS) we find that
$\Delta\phi_{22}\lesssim {0.1}$\,radian from $10\,$Hz up to merger, cf. Fig.~\ref{fig:DeltaPhi_EOB}.
The effect on the phasing is larger by of a factor $\sim\,${two} and has the opposite sign if we assume 
that the gravitomagnetic interaction is parameterized by static Love numbers.
The inclusion of gravitomagnetic terms in the Taylor
F2 approximant is found to be negligible for GW data analysis of
LIGO-Virgo data~\cite{Jimenez-Forteza:2018buh}. Our results 
seem to support this conclusion, but we leave for the future a detailed
assessment using \TEOBResum{} waveforms.

\paragraph*{Tidal correction to the $B$ potential.}
The LO tidal correction to the EOB $B$ potential computed in
Ref.~\cite{Vines:2010ca} is positive leading to a (small) repulsive
effect. 
Its impact on the GW is rather small and it is quantified in
Fig.~\ref{fig:Delta_phi_B_tidal} for a sample of
binaries with $|\Delta\phi_{22}|\sim 0.03$ at most. 

\paragraph*{Tidal corrections in multipolar waveform and flux.}
The inclusion of the gravitoelectric and magnetic terms of
Ref.~\cite{Banihashemi:2018xfb} in the multipolar waveform and the dynamics
(via the flux) has a subleading contribution as shown by the brown dashed curves in Fig.~\ref{fig:DeltaPhi_EOB}
with $-\Delta\phi_{22}\sim 0.03$ at most, roughly equal and opposite to the contribution to the $B$ potential.

\paragraph*{Effective light-ring pole.}
We have investigated the effect of two values for the free parameter $p$ describing the
order of the 2GSF pole at the light ring, cf. Eq.~(\ref{eq:Ahat_two}).
Expected to be in the range $p\in[4,6]$ \cite{Bini:2014zxa}; NR comparisons
suggest that the effective value of $p=4$ (as in Ref.~\cite{Bernuzzi:2014owa})
is a simple and sufficient choice to yield good agreement (within NR errors) between the EOB and the NR waveforms.
We briefly explored, at the level of energetics, the sensitivity of the
analytical models to varying $p$ by considering the value $p=9/2$.
We stress that the light-ring pole in \TEOBResum{} is always 
``dressed'' in the sense that for all the possible neutron star
binaries, the EOB dynamics terminate at larger radii, roughly given by $r_\text{peak}\sim({1.35-1.4})\,r_{\rm LR}$, than the GSF
pole at $r_\text{LR}$ (cf. Fig.~\ref{fig:rPeak_rLR}).
On the other hand, the light-ring pole is a gauge artefact, resulting, in particular, from
working in the Damour-Jaranowski-Sch\"{a}fer gauge~\cite{Damour:2008qf}.
It was shown in Ref.~\cite{Akcay:2012ea} that the LR pole is a coordinate singularity
in EOB phase space, which was eliminated via a canonical transformation in Ref.~\cite{Steinhoff:2016rfi}.
Recent approaches based on the post-Minkowskian expansion employ a different gauge
with no LR singularity in the $A$ potential including the GSF, $X_A\ll 1$, limit~\cite{Damour:2017zjx}.

\paragraph*{Inspiral-merger BNS waveforms.}
We find that the new analytical waveform information improves the agreement
between \TEOBResum{} and high-resolution NR simulations. 
We require more high-quality NR data to fully assess the potential
benefits of the $(3+)$-GSF resummed tidal models with $p > 4$.
The binding energy vs. angular momentum plots of Figs.~\ref{fig:Eb_J_q1}, \ref{fig:Eb_J_q_ne_q1} 
are the most telling of our comparisons made in this article since they contain plots of gauge-invariant quantities, thus enabling unambiguous EOB-NR comparisons. 
The subset of NR data with robust errors (shaded blue regions) in these figure carry the most weight in 
judging the faithfulness of EOB models. 
For this reason, \GSF{2}{+} supplied with either PN or GSF $(2-)$ tides should be taken as the current most faithful \TEOBResum{} variant.
%
%
%
%
%
%
%
%
%
\paragraph*{Higher multipoles.}
We also presented EOB-NR waveform comparisons for multipoles beyond
the leading-order quadrupole. In
Fig.~\ref{fig:EOBNR_phase_higher_modes}, we showed a small sample of
various modes up to $(4,4)$ showing good phase alignment between \TEOBResum{}
and NR up to frequencies corresponding to the last one-two orbits before the merger.
Our results indicate that the \TEOBResum{} {\it multipolar} waveform can be
accurately used in current GW parameter estimation studies.
At the analytical level, more information on the amplitudes
would be desirable to verify the match of the NR waveform amplitudes
up to merger. \\

\begin{figure}[t]
  \centering
 \includegraphics[width=0.45\textwidth]{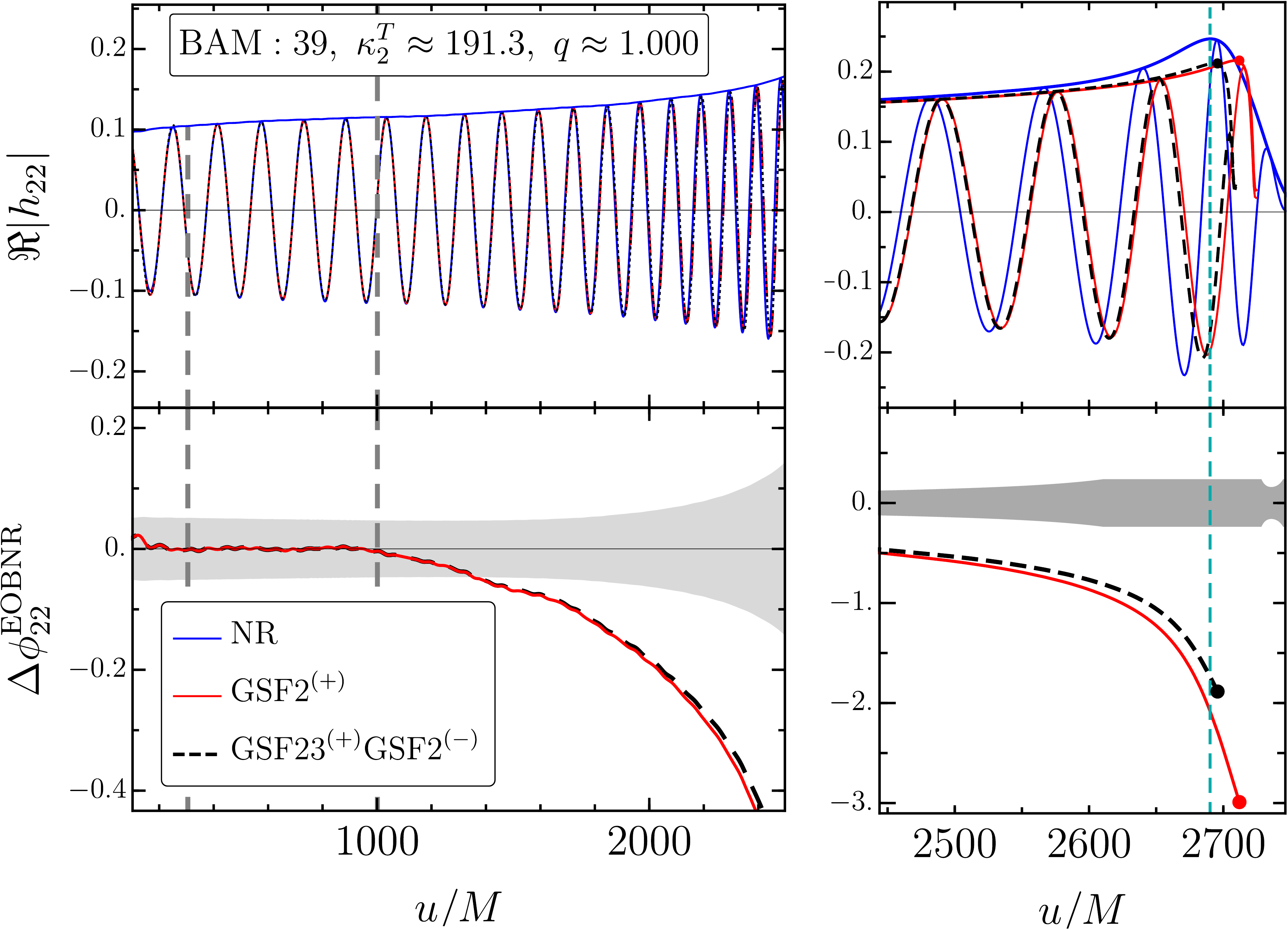}
     \caption{Phasing comparison for the case involving BNSs with spins, 
  specifically {\tt BAM:0039} with $q=1$, $\Lambda=1001.8$ and dimensionless spins $\chi_1=\chi_2= 0.14$.
  See the caption of Fig.~\ref{fig:Phasing_fig1} for details.}
  \label{fig:spin_case}
\end{figure}

The improvements in the tidal sector presented in this paper carry over
to spinning binaries. We show in Fig.~\ref{fig:spin_case}, as a
preliminary example, a comparison between the spin-accommodating \TEOBResumS{} and {\tt
  BAM:0039}, a high-quality BNS waveform with $q\approx1$, $\Lambda\approx1001.8$, and
dimensionless spins equalling $0.14$. 
The new GSF resummation of the gravitoelectric LO term seems to reduce the gap to NR data. We will present
elsewhere a detailed comparison with NR binary neutron star waveforms
that include spin effects. The reason is that we are currently improving the spinning vacuum
sector of \TEOBResumS{} with the new waveform resummation presented in
Refs.~\cite{Nagar:2016ayt, Messina:2018ghh} and with a resummed expression for self-spin terms that
include the NLO PN terms \cite{Nagar:2018zoe, Bohe:2015ana}.
It will be also interesting to incorporate more spin-tidal couplings~
\cite{Pani:2015hfa,Pani:2015nua, Landry:2017piv, Gagnon-Bischoff:2017tnz, Landry:2018bil,Abdelsalhin:2018reg}, 
albeit their effect is likely to be negligible for realistic spins \cite{Jimenez-Forteza:2018buh}.

\TEOBResumS{} has been used for a recent analysis of GW170817
\cite{Abbott:2018exr} within the rapid parameter estimation approach
of Ref.~\cite{Lange:2018pyp}. 
Parameter estimation with direct use of
\TEOBResum{} (or \TEOBResumS{}~\cite{Nagar:2018zoe})
waveforms might be possible by generating
the waveform using the post-adiabatic (PA) approximation as
pointed out in Ref.~\cite{Nagar:2018gnk}.
The procedure and performance for BNSs are discussed in detail in
Appendix~\ref{sec:PA}. We find that BNS waveforms from $10$~Hz can be
generated in about $\sim0.06$~s in the PA approximation while their
require $1.26$~s solving the ODE on an adaptive grid. The relative
phase difference accumulated between the PA approximation at 8th order
and the ODE runs is below $10^{-5}$~rad, thus practically negligible.
Fast waveform evaluation can usually be
performed by constructing surrogate models based on reduced order
models~\cite{Lackey:2016krb}. 
The current implementation of \TEOBResum{}
(as well as \TEOBResumS{}~\cite{Nagar:2018zoe})
proves competitive with these approaches. 
In addition, \TEOBResum{}
can be used as a key building block for the
construction of closed-form frequency-domain
approximants~\cite{Dietrich:2017aum,Kawaguchi:2018gvj,Dietrich:2018uni}.

A public implementation of our C code is available at
\begin{center}
{\footnotesize \url{https://bitbucket.org/account/user/eob_ihes/projects/EOB}}
\end{center}

\begin{acknowledgments}
We thank Paolo Pani, Justin Vines, and Philippe Landry for helpful discussions about gravitomagnetic Love numbers. We thank Tim Dietrich for sharing with us the highest resolution {\tt BAM:0011} data. S.~A., S.~B., and N.~O. acknowledge support by the EU H2020 under ERC Starting Grant, no.~BinGraSp-714626.
\end{acknowledgments}

\appendix

\section{Derivation of the GSF-resummed (3+) potential}
\label{sec:A3plus}
We follow the formalism and notation of Ref.~\cite{Bini:2014zxa}
(henceforth BD). For more details, see their work.
Setting the NS label $A=1$ we have
\begin{align}
\hat{A}^{(3+)}_1 &=\hat{A}_1^{(3+)\text{0GSF}}  
+ \hat{A}_1^{(3+)\gsf} X_1 +  \hat{A}_1^{(3+)\text{2GSF}} X_1^2 
\label{eq:A3hatGSF_AppA}.
\end{align}
%

To obtain the explicit expression for $\hat{A}_1^{(3+)\text{1GSF}}$ we start with Eq.~(6.11) of BD
\be
\hat{A}^{(3+)}_1(u)=\sqrt{F(u;\nu)}\, \Gamma^{-1}[y(u)]\, \f{J_{3+}[y(u)]}{J^\text{Newt}_{3+}(u)} \label{eq:A3hat_generic0},
\ee
where $\Gamma(y)$ is the usual redshift factor and $y\equiv (m_2\Omega)^{2/3}$ is the GSF inverse separation.
$F(u;\nu)$ is a function of the circular-orbit, ``bare'' potential $A(u)$ and its derivative, and $J^\text{Newt}_{3+}= 90 m_2^2/r^8_\text{EOB} $ which becomes $ J^\text{Newt}_{3+}(u) = 90 X_2^8 u^8/m_2^6$ using $r_\text{EOB}= M/u$. This results in
\be
\hat{A}^{(3+)}_1(u)=\f{\sqrt{F(u;\nu)}}{(1-X_1)^8}\, \Gamma^{-1}[y(u)]\, \f{m_2^6 J_{3+}[y(u)]}{90 u^8} \label{eq:A3hat_generic},
\ee
which is the $(3+)$ version of BD Eq.~(7.3).
Note that Eq.~(\ref{eq:A3hat_generic0}) is a general expression that holds for all order of $X_1$, but current GSF knowledge limits us to $\ord(X_1)$. Additionally, 
the $\ord(X_1)$ difference between
the EOB inverse separation $u$ and the GSF inverse separation $y$ needs to be accounted for [cf. Eqs.~(2.18, 2.19) of BD].

Combining the work of Ref.~\cite{Nolan:2015vpa} and BD App.~D we have that $J_{3+} = K_{3+} + \tfrac{1}{3} J_{\dot{2}+}$ 
where the latter are given as a series in $q\equiv X_1/X_2\ll 1$
\begin{align}
K_{3+} &= K_{3+}^\bk\left[1 + q\left( \hat{\delta}k_{3+}+2 h_{uu}\right)\right],\label{eq:K_3plus}\\
J_{\dot{2}+} &= J_{\dot{2}+}^\bk\left[1+ q\,\Delta j_{\dot{2}+}\right]\equiv J_{\dot{2}+}^\bk\left[1+ q\,(\hat{\delta}j_{\dot{2}+}+3h_{uu})\right]\label{eq:J_2dot},
\end{align}
where numerical values for $\hat{\delta}k_{3+}$ are given in Table~V of Ref.~\cite{Nolan:2015vpa} and $\Delta j_{\dot{2}+} $ can be obtained from Ref.~\cite{Nolan:2015vpa} Eqs.~(2.44, 2.45) in terms of Ref.~\cite{Dolan:2014pja}'s redshift and spin-precession invariants. 
$\hat{\delta}j_{\dot{2}+}$ is given as PN series in Appendix~D of BD. $h_{uu} \equiv 2 \Delta U/U_0$ where $U_0=(1-3y)^{-1/2}$.

The background, i.e., 0GSF terms in Eqs.~(\ref{eq:K_3plus}, \ref{eq:J_2dot}) can be extracted
from Ref.~\cite{Nolan:2015vpa} or Appendix~D of BD. They read
%
\begin{align}
K_{3+}^\bk &= 6 y^8(1 - 2 y) \f{(42 y^2 - 46 y + 15)}{(1 - 3 y)^2}, \label{eq:K_3plus0GSF}\\
J_{\dot{2}+}^\bk &= \frac{18 y^9 (1-2 y)^2}{(1-3 y)^2}. \label{eq:J_2dot0GSF}
\end{align}

\begin{figure}[t!]
\center
\includegraphics[width=0.46\textwidth]{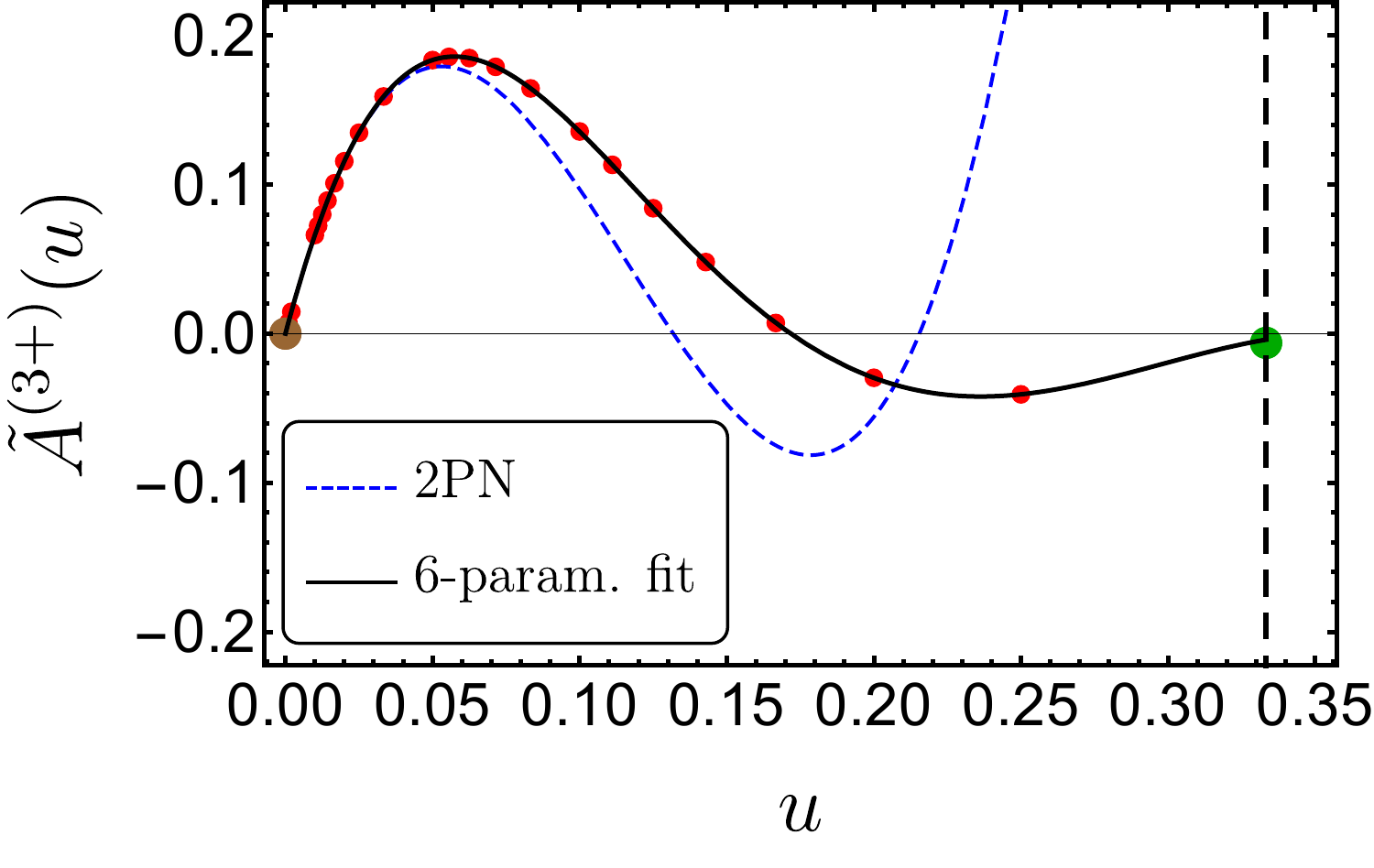}
  \caption{\label{fig:A3tilde}  $\tilde{A}^{(3+)}\equiv(1-3u)^{7/2}\hat{A}^{(3+)\text{1GSF}}$ data in red dots with the PN series as the blue dashed curve. 
  The values for the $\tilde{A}^{(3+)}(u)$ data set are obtained using the numerical data for the tidal invariants of Refs.~\cite{Dolan:2014pja, Nolan:2015vpa}.
  The black curve is our six-parameter fit given by Eqs.~\eqref{eq:A3tilde_fit}-\eqref{eq:A3tilde_fit_coefs}.
  The green dot marks the light-ring limit and the brown dot the $u\to 0$ limit. 
  The vertical dashed black line marks the position of the light ring.
}
   	\end{figure}

With the above equations and the numerical data of Refs.~\cite{Dolan:2014pja,Nolan:2015vpa} we can now calculate the 1-GSF contribution to $\hat{A}^{(3+)}$. We performed several checks on our result:
\begin{enumerate}
\item 0-GSF limit: Simply taking the $q\sim X_1\to 0$ limit of our expression for $\hat{A}^{(3+)}$ yields
\be
\hat{A}^{(3+)\text{0GSF}} = (1-2u) \left( 1 + \f{8}{3} \f{u^2}{(1-3u)}\right) \, .\label{eq:A3hat_0GSF}
\ee
This agrees with the test-mass limit result given by Eq.~(6.45) of Ref.~\cite{Bini:2012gu}.
\item Weak-field limit: Using BD's PN series expansions for $\hat{\delta}k_{3+}$ and $\hat{\delta}j_{\dot{2}+}$ and Ref.~\cite{Kavanagh:2015lva}'s series for $\Delta U$ we straightforwardly obtain the PN series for $\hat{A}^{(3+)}(u)$ 
\be
\lim_{u\to 0} \hat{A}^{(3+)\text{1GSF}}= \f{15}{2}u-\f{311}{24}u^2 +\ord(u^3)\,
\ee
which agrees with the $\ord(X_1)$ part of Eq.~(\ref{eq:alpha_3s}).
Our numerical data is also consistent with this as can be seen in Fig.~\ref{fig:A3tilde}.
\end{enumerate}
	 
%

We next investigate the light-ring (LR) limit. BD provide ample explanations on how to ascertain the singular behaviour of $\hat{A}^{(\el\pm)}$ as $u\to 1/3$ and how to obtain the LR limit of singularity-factored potentials $\tilde{A}^{(2\pm)}\equiv (1-3u)^{7/2}\hat{A}^{(2\pm)\text{1GSF}}$. Following the same analysis, we straightforwardly establish that 
\begin{align}
\lim_{u\to \tfrac{1}{3}}\hat{A}^{(3+)\text{1GSF}}&= -\f{\zeta}{162} (1-3u)^{-7/2}\nn \\
&= \f{8}{27}\left(-\f{\zeta}{48} (1-3u)^{-7/2}\right), \label{eq:A3hat_LRlimit}
\end{align}
where the last quantity in parentheses is the LR limit of $\hat{A}^{(2\pm)\text{1GSF}}$. 

%
Accordingly, we now introduce the LR rescaled function
\be
\tilde{A}^{(3+)}(u) \equiv (1-3u)^{7/2} \hat{A}^{(3+)\text{1GSF}}\!(u) \label{eq:A3tilde}
\ee
whose PN series expansion 
\be \tilde{A}^{(3+)}(u\ll 1) = \f{15}{2}u \left[1 - \f{2201}{180}u + \ord(u^2)\right]
\ee
hints a cubic strong-field fit to the data of the form $\tfrac{15}{2}u(1+ C_1 u + C_2 u^2)$.
However, after much experimenting we settled on the following best fit to the data
\begin{align}
\tilde{A}^{(3+)}(u)&\approx \tilde{A}^{(3+)}_\text{fit}(u)\label{eq:A3tilde_fit} \\
&= \frac{15}{2}u(1+ C_1 u + C_2 u^2+C_3 u^3)\, \f{1+C_4 u+C_5 u^2}{1+C_6 u^2} , \nn
\end{align}
where
\begin{align}
C_1 & = -3.682095, \qquad	C_2 =\ 5.171003,\nn\\	
C_3 & = -7.639164,\qquad	C_4 = -8.632781,\nn\\	
C_5 & = 16.36009,\qquad	\ \ C_6 = 12.31964 \, . \label{eq:A3tilde_fit_coefs}
\end{align}
	%
%
%
   	%
   	%
This fit and a 2PN expression for $\tilde{A}^{(3+)}$ are shown as the black and blue curves in
Fig.~\ref{fig:A3tilde}, respectively.
Although our fitting procedure excluded the data point at the
    light ring, our fit nearly crosses it anyway (see Fig.~\ref{fig:A3tilde}). 
Additionally, the fit approximates every one of the 23 data points to a relative difference of $< 5\times 10^{-4}$ with the exception of one point with 1\% mismatch and another 0.1\%. The norm of the relative disagreement over the entire data is
\be
|| 1-\tilde{A}^{(3+)}_\text{fit}/\tilde{A}^{(3+)\text{1SF}}_\text{num}|| \approx 0.0118\, . \label{eq:A3tilde_rel_diff}
\ee
Putting everything together, we arrive at
\begin{align}
\hat{A}_A^{(3+)}\,=\ & (1-2u) \left( 1 + \f{8}{3} \f{u^2}{(1-3u)}\right)\nn \\
&+ X_A\,\f{\tilde{A}^{(3+)}_\text{fit}(u)}{(1-3u)^{7/2}}
+ X_A^2\, \f{110}{3}\f{u^2}{(1-3u)^{p_{3+}}}. \label{eq:A3plus_GSF_all_App}
\end{align}

\section{Post-adiabatic dynamics}
\label{sec:PA}

Within \TEOBResumS{}, the dynamics of a (non-precessing) binary system
is usually determined by numerically solving four of Hamilton's
equations. The time needed to solve these four ODEs 
is the main contribution to the waveform evaluation
time. Using our publicly available C code (see main text) a typical
time-domain BNS waveform requires $\sim 1$~sec to be generated
starting from a GW frequency of 10~Hz and employing standard
Runge-Kutta integration routines with adaptive timestep.  
Thus, ODE integration cannot be used in parameter estimation runs that
require the generation of $~ 10^7$ waveforms. 
Ref.~\cite{Nagar:2018gnk} pointed out a way of reducing the evaluation
time by making use of the PA approximation to compute the system
dynamics. While the approach was then restricted to the inspiral
phase, we here present, for the first time, results that include the
full evolution up to merger.

\begin{figure}[t]
\vspace{2mm}
\centering
\includegraphics[width=0.48\textwidth]{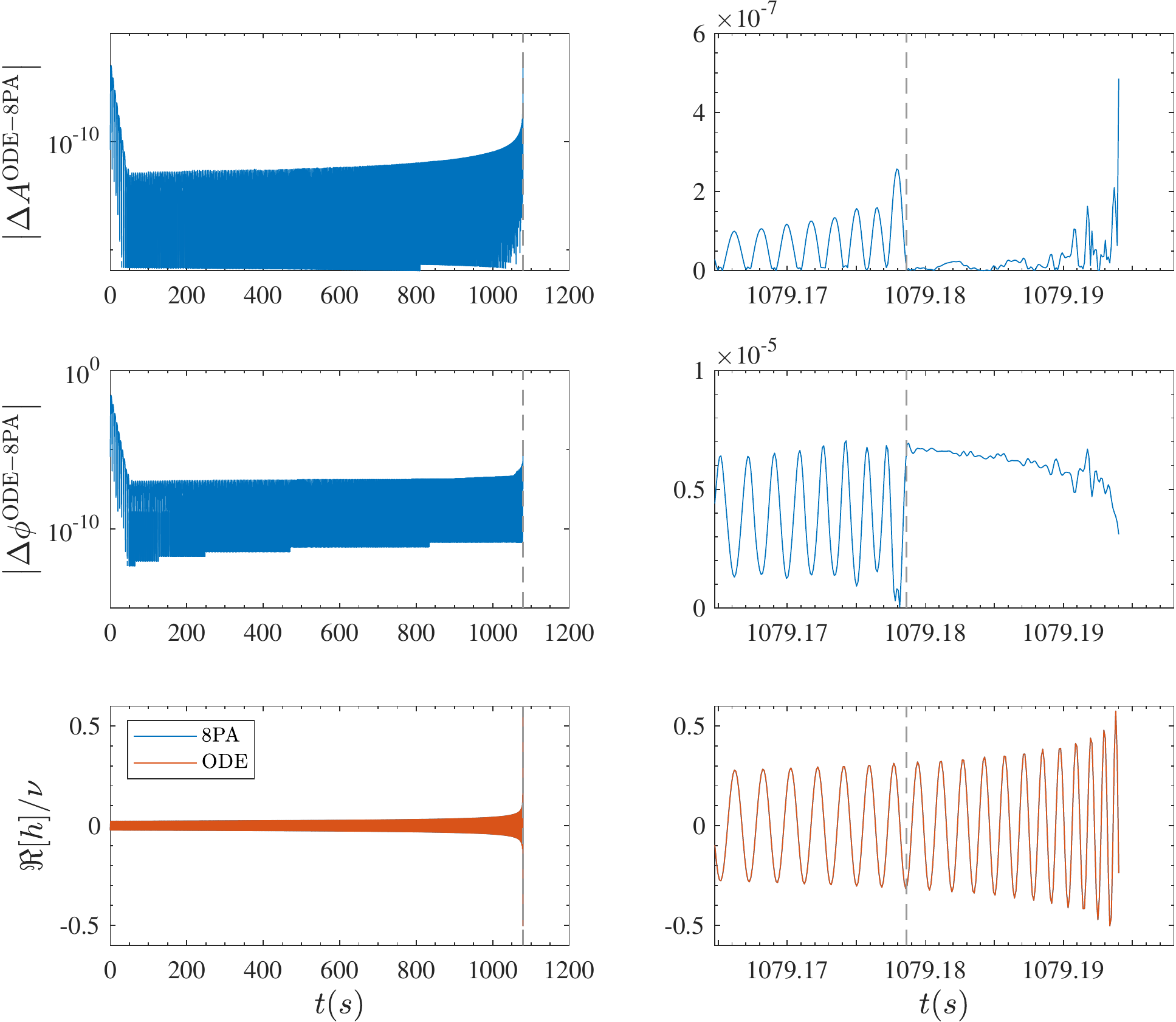}
\caption{
	\label{fig:PA}
	Comparison between the waveforms computed solving the
	ODEs with the \texttt{GSL} \texttt{rk8} routine and
	adaptive timestep, and the PA waveform completed with
	the same ODE solver after $r<r_\text{min}$ for a
	non-spinning BNS system with $1.35M_\odot+1.35M_\odot$
	and SLy EOS starting at $10$~Hz. 
	The PA parameters used are the ones described in the second row of Table~\ref{table:PAtime}.	
	The dashed grey line marks the stitching point, $r_\text{min}$, between the PA and ODE-based dynamics.
	Having written the waveform strain as $h/\nu \equiv A
	e^{\rm -i \phi}$, we defined the phase difference as
	$\Delta\phi^{\rm ODE-8PA} \equiv \phi^{\rm ODE} -
	\phi^{\rm 8PA}$ and the fractional amplitude
	difference as $\Delta A^{\rm ODE-8PA} \equiv (A^{\rm
		ODE} - A^{\rm 8PA})/A^{\rm ODE}$. 
	The higher differences at the start of the evolution are due to the fact that the complete ODE is currently started using only 2PA data.}
\end{figure}

We start by briefly summarizing the procedure described in Ref.~\cite{Nagar:2018gnk}.
The PA approximation is an extension of the one introduced in
Refs.~\cite{Buonanno:1998gg,Buonanno:2000ef} (and expanded in
Refs.~\cite{Damour:2007yf,Damour:2012ky}) and is currently used
to determine the initial conditions of \TEOBResumS{}. 
Using this approximation, it is possible to analytically compute the
radial and angular momentum of a binary system, under the assumption
that the GW flux is small. 
This is obviously true in the early inspiral phase and progressively
loses validity when the two objects get close. The approach starts by
considering the conservative system, when the flux is null, and then
computes the successive corrections to the momenta. We denote with
$n$PA the $n$-th order iteration of this procedure.
\begin{table}[h!]
\centering
\begin{tabular}{ccccccc}
\hline\hline
$f_0$ [Hz] & $r_0$ & $r_{\rm min}$ & $N_r$ & $\Delta r$ & $\tau_{\rm 8PA}$ [sec] & $\tau_{\rm ODE}$ [sec] \\
\hline
20 & 112.80 & 12 & 500 & 0.20 & 0.04 & 0.53 \\
10 & 179.01 & 12 & 800 & 0.21 & 0.06 & 1.26 \\
\hline\hline
\end{tabular}
\caption{
	\label{table:PAtime}
	Performance of the \TEOBResumS{} C code for a non-spinning BNS system with $1.35M_\odot+1.35M_\odot$ and SLy EOS.
	$f_0$ and $r_0$ denote the initial GW frequency and radial separation (in units of $(GM)/c^2$).
	The 8PA dynamics is computed on a grid with $N_r$ points and grid separation $\Delta r$
        that ends at $r_{\rm min}$ and then completed by the standard ODE one. The evaluation
        times $\tau$ are determined using a standard Intel Core i7, 1.8GHz and 16GB RAM.
	The code is compiled with the GNU gcc compiler using O3 optimization.}
\end{table}
Practically, to compute the PA dynamics, we first build a uniform
radial grid from the initial radius $r_0$ to an $r_\text{min}$ up until
which we are sure the approximation holds. 
We then analytically compute the momenta that correspond to each radius at a chosen PA order.
Finally, we determine the full dynamics recovering the time and orbital
phase by quadratures.
From $r_\text{min}$ we can then start the usual ODE-based dynamics
using the PA quantities as initial data as it is usually done (at 2PA
order) in \TEOBResumS{}. The benefits of using this method come from
the fact that we can avoid the numerical solution of two Hamilton's
equations and that we can integrate the other two on a very sparse
radial grid.
\begin{table}[t]
\centering
\begin{tabular}{cccc}
\hline\hline
$f_0$ [Hz] & $r_0$ & $\tau_{\rm 8PA}^{\rm int}$ [sec] & $\tau_{\rm ODE}^{\rm int}$ [sec] \\
\hline
20 & 112.80 & 0.54  & 1.06  \\
10 & 179.01 & 3.2 & 4.4 \\
\hline\hline
\end{tabular}
\caption{
	\label{table:PAtimeint}
	Performance of the \TEOBResumS{} C code when the final waveform is interpolated on a time
        grid sampled at 1/(4096~Hz).
	We use a standard, non-optimized, GSL interpolation routine.
	The considered system coincides with the one of Table~\ref{table:PAtime}.}
\end{table}

With the initial radius is fixed, there are three parameters that can be chosen at will in the PA procedure.
These are the PA order, the number of grid points (or, equivalently, the grid step), and $r_{\rm min}$.
We use the 8PA order, a grid separation $\Delta r \sim 0.2$, and $r_{\rm min} \sim 12$
(note the latter value can be tuned depending on the BNS spin).

\begin{figure}[t]
\vspace{2mm}
\centering
\includegraphics[width=0.48\textwidth]{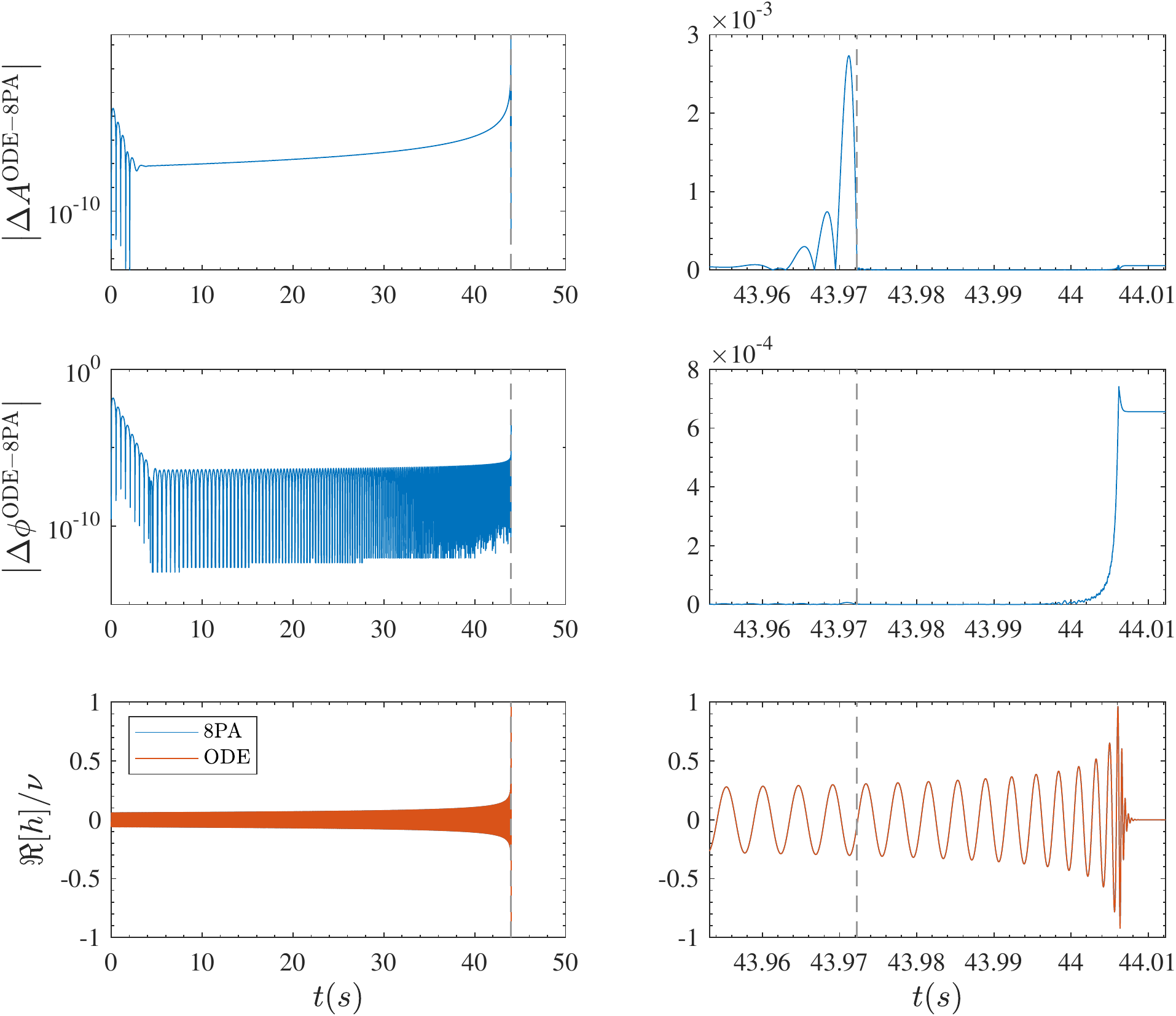}
\caption{
	\label{fig:PA_bbh}
	Comparison of Fig.~\ref{fig:PA} in a case of a BBH system with $m_A = m_B = 3 M_\odot$ and $\chi_A = \chi_B = -0.99$.
	The evolution is started at a GW frequency $f_0 = 20$ Hz, which corresponds to an initial radius $r_0 = 66.34$.
	The 8PA dynamics is computed using a grid separation $\Delta r = 0.2$ and then stitched to the ODE-based one at $r_\text{min} = 13$.}
\end{figure}

This is a conservative choice of parameters that guarantees a
remarkable agreement with the dynamics computed by solving the ODEs.
We show in Fig.~\ref{fig:PA} the waveform fractional-amplitude difference
(top panel) and phase difference (medium panel)
for a non-spinning BNS system with $1.35M_\odot+1.35M_\odot$ and SLy EOS.
The vertical dashed line marks the stitching point between the PA evolution
and the ODE evolution for the last orbits where the PA approximation brakes
down. Table~\ref{table:PAtime} highlights the performances of the C code for such a case.
Here, the initial radius is determined by solving the circular Hamilton's equations
instead of relying on Kepler's law, as discussed in Sec. VI of Ref.~\cite{Nagar:2018zoe}.

We can see that the waveform computed using the PA dynamics (completed with the ODE for the last few orbits)
only takes around 60 milliseconds to be evaluated. Such a time is competitive with respect to the surrogate models
that are currently being constructed in order to reduce waveform evaluation times~\cite{Lackey:2016krb}.
Finally, Table~\ref{table:PAtimeint} illustrates the performance of \TEOBResumS{} when the waveform,
which is obtained on a {\it nonuniform} temporal grid, is interpolated on an evenly spaced time grid,
sampled at $\Delta t^{-1}=4096$~Hz. Note that the interpolation routine is not optimized, and as such,
it by far makes the dominant contribution to the global computational cost.

\subsection{Binary black hole case}
For completeness, we also show in Fig.~\ref{fig:PA_bbh} a case of a binary black hole
(BBH) system, completed with the postmerger and ringdown phase. We consider
an equal-mass black-hole binary with $m_A=m_B=3M_\odot$ and nearly extremal anti-aligned spins,
$\chi_A=\chi_B=-0.99$. We do not want to discuss these cases in detail here.
It suffices to note that the main conclusions do not change when we take into
account BBH systems.


\bibliography{refs.bib}
 
\end{document}